\pdfoutput=1

\documentclass[11pt]{article}

\usepackage{graphicx,pstricks,pst-node,psfrag,amsthm,amssymb,amsmath,dsfont,bbm,mathrsfs}
\usepackage[round]{natbib}
\usepackage[a4paper]{geometry}
\usepackage{float} 
\usepackage{subcaption} 
\usepackage{url}
\usepackage{multirow}
\usepackage{fancybox}
\usepackage{lastpage}

\usepackage{xr} 

\makeatletter
\newcommand*{\addFileDependency}[1]{
\typeout{(#1)}
\@addtofilelist{#1}

\IfFileExists{#1}{}{\typeout{No file #1.}}
}\makeatother

\newcommand{\bzero}{\mbox{\boldmath $0$}}

\newcommand{\bdelta}{\mbox{\boldmath $\delta$}}

\newcommand{\bSigma}{\mbox{\boldmath $\Sigma$}}

\newcommand{\bPsi}{\mbox{\boldmath $\Psi$}}

\newcommand{\ba}{\mbox{\boldmath $a$}}

\newcommand{\be}{\mbox{\boldmath $e$}}

\newcommand{\bm}{\mbox{\boldmath $m$}}

\newcommand{\bu}{\mbox{\boldmath $u$}}

\newcommand{\by}{\mbox{\boldmath $y$}}

\newcommand{\bs}{\mbox{\boldmath $s$}}

\newcommand{\bA}{\mbox{\boldmath $A$}}
\newcommand{\bB}{\mbox{\boldmath $B$}}
\newcommand{\bD}{\mbox{\boldmath $D$}}
\newcommand{\bE}{\mbox{\boldmath $E$}}
\newcommand{\bG}{\mbox{\boldmath $G$}}
\newcommand{\bI}{\mbox{\boldmath $I$}}
\newcommand{\bL}{\mbox{\boldmath $L$}}
\newcommand{\bM}{\mbox{\boldmath $M$}}
\newcommand{\bP}{\mbox{\boldmath $P$}}
\newcommand{\bS}{\mbox{\boldmath $S$}}

\newcommand{\bU}{\mbox{\boldmath $U$}}
\newcommand{\bV}{\mbox{\boldmath $V$}}

\newcommand{\bY}{\mbox{\boldmath $Y$}}
\newcommand{\bX}{\mbox{\boldmath $X$}}
\newcommand{\bZ}{\mbox{\boldmath $Z$}}

\newcommand{\sumt}{\sum_{t=1}^T}

\newcommand{\sumv}{\sum_{v=1}^V}

\DeclareMathOperator{\EX}{\mathbb{E}}
\DeclareMathOperator{\VX}{\mathbb{V}}

\definecolor{turquoise}{RGB}{112,212,193}

\begin{document}


\title{Leveraging population information in brain connectivity via Bayesian ICA with a novel informative prior for correlation matrices}
\author{AMANDA F. MEJIA$^{1,\ast}$, DAVID BOLIN$^2$, DANIEL A. SPENCER$^1$, ANI ELOYAN$^4$\\[4pt]
\textit{$^1$Department of Statistics, Indiana University, Bloomington, IN, USA}\\
\textit{$^2$Statistics Program, Computer, Electrical and Mathematical Sciences and}\\
\textit{Engineering Division, KAUST, Saudi Arabia}\\
\textit{$^3$Department of Biostatistics, Brown University, Providence, RI, USA}
\\[2pt]
{afmejia@iu.edu}}

\maketitle

\begin{abstract}
    {Brain functional connectivity (FC), the temporal synchrony between brain networks, is essential to understand the functional organization of the brain and to identify changes due to neurological disorders, development, treatment, and other phenomena. Independent component analysis (ICA) is a matrix decomposition method used extensively for simultaneous estimation of functional brain topography and connectivity. However, estimation of FC via ICA is often sub-optimal due to the use of ad-hoc estimation methods or temporal dimension reduction prior to ICA. Bayesian ICA can avoid dimension reduction, estimate latent variables and model parameters more accurately, and facilitate posterior inference. In this paper, we develop a novel, computationally feasible Bayesian ICA method with population-derived priors on both the spatial ICs and their temporal correlation, i.e. FC. For the latter we consider two priors: the inverse-Wishart, which is conjugate but is not ideally suited for modeling correlation matrices; and a novel informative prior for correlation matrices. For each prior, we derive a variational Bayes algorithm to estimate the model variables and facilitate posterior inference. Through extensive simulation studies, we evaluate the performance of the proposed methods and benchmark against existing approaches. We also analyze fMRI data from over 400 healthy adults in the Human Connectome Project. We find that our Bayesian ICA model and algorithms result in more accurate measures of functional connectivity and spatial brain features. Our novel prior for correlation matrices is more computationally intensive than the inverse-Wishart but provides improved accuracy and inference. The proposed framework is applicable to single-subject analysis, making it potentially clinically viable.}{functional MRI, variational Bayes, independent component analysis, neuroimaging}
\end{abstract}

\section{Introduction}

The widespread availability of functional magnetic resonance imaging (fMRI) technology for obtaining high-resolution images of brain function non-invasively has made it possible to study the function and organization of the human brain in ever-increasing detail. One of the techniques that has proven most useful in this context is independent component analysis (ICA; \citeauthor{hyvarinen2002independent}, \citeyear{hyvarinen2002independent}). First introduced for blind source separation \citep{mckeown1998analysis}, ICA is a matrix factorization method used for estimating latent spatial signals based on the blood oxygen level dependent (BOLD) signals from fMRI data \citep{beckmann2005investigations}. The spatial independent components (ICs) from fMRI are believed to represent brain ``networks'', or collections of locations that tend to activate in a coordinated manner. By estimating these networks we can identify, for instance, differences in brain network localization between populations of interest. 

While early ICA studies in fMRI focused primarily on the spatial ICs, ICA can also be used to study brain functional connectivity (FC), i.e. the temporal synchrony between networks. FC is based on the correlation between columns of the ``mixing matrix'' containing the time courses corresponding to each independent component \citep{van2004functional}. Compared with anatomical or functional brain atlases traditionally used in FC analyses, ICA deconstructs connectivity into between- and within-network components, providing a richer and more nuanced picture of brain organization \citep{joel2011relationship}. In addition, ICA has greater flexibility to capture the overlapping, continuous nature of brain functional topography, unlike deterministic atlases with hard boundaries \citep{bijsterbosch2018relationship}.  ICA has been used in a range of scientific investigations to study FC, advancing understanding of autism \citep{nebel2014precentral}, Alzheimer's disease \citep{li2012analysis}, major depression \citep{greicius2007resting}, and others. More methodological research in this area is needed, however, since existing ICA methods do not focus on estimating dependence among the columns of the mixing matrix.

One of the historical limitations of non-parametric \cite[e.g.][]{hyvarinen2002independent} and likelihood-based \citep{eloyan2013likelihood, guo2013hierarchical} ICA methods is the assumption of a square and invertible mixing matrix. With few exceptions \citep[e.g.,][]{risk2019linear}, most ICA methods require dimension reduction via principal component analysis (PCA) prior to model fitting. The full temporal activation profile associated with each IC is obtained by projecting back to the full temporal dimension, which may not provide optimal estimation.  

Bayesian ICA techniques can address this limitation by avoiding the need to invert the unmixing matrix, thereby allowing for rectangular mixing matrices and avoiding dimension reduction \citep{lawrence2000variational}. Bayesian approaches for ICA additionally facilitate inference on the spatial and/or temporal brain measures based on the posterior distributions of the model parameters. However, fitting hierarchical Bayesian ICA models to high-dimensional fMRI data may be computationally demanding or even impossible. Computationally efficient Bayesian ICA techniques are needed to provide more accurate FC estimation versus standard ICA methods that require a square mixing matrix, while being feasible and practical for fMRI analysis.

We previously proposed a computationally efficient Bayesian ICA approach for fMRI data using known population brain networks and subject-specific topographic deviations modeled as latent variables \citep{mejia2020template}. Here, we extend that model by incorporating multivariate priors on the temporal activity, whose correlation represents the FC between brain networks. Building on our earlier work, wherein we used population-derived priors to inform the spatial configuration of brain networks, here we use population data to build an informative prior on the correlation to encode expected patterns of FC. Thus, in this work we use multiple sources of prior population information (spatial and temporal) to inform the ICA model.

A key scientific advantage of the proposed methods is the ability to estimate unique subject-level functional brain features with high accuracy. The conventional use of population-level atlases can produce biased FC estimates, due to spatial misalignment with the functional topography of a subject's brain \citep{bijsterbosch2018relationship}. Such biases may compromise the validity of observed associations between FC and behavior, given that topographic differences are themselves biologically relevant \citep{kong2019spatial}. Our proposed Bayesian ICA approach can disentangle functional topography and FC to accurately depict functional brain organization and its relationship to disease, development, and treatment.  Importantly, the techniques developed here perform well when applied to typical 10-15 minute fMRI experiments, avoiding the need to collect hours of data \citep{laumann2015functional} to overcome high noise levels.

In the remainder of this manuscript, in Section \ref{sec:methods} we present the proposed Bayesian ICA model, describe our novel informative prior for correlation matrices, and derive two variational Bayes algorithms for posterior estimation and inference. In Section \ref{sec:simulation} we present extensive realistic simulation studies to evaluate the performance of the proposed methods versus existing approaches. In Section \ref{sec:DA} we analyze test-retest fMRI data to further evaluate our algorithms' performance and feasibility in real fMRI studies. We conclude with a discussion in Section \ref{sec:discussion}.

\section{Methods}
\label{sec:methods}

For a single subject, let $\bY\in\mathbb{R}^{T\times V}$ be the observed fMRI data, where $V$ is the number of brain locations (voxels or vertices) and $T$ is the number of time points. We assume $\bY\in\mathbb{R}^{T\times V}$ is centered across $t=1,\dots,T$. A probabilistic ICA model is given by $\bY = \bA\bS + \bE$, where $\bS\in\mathbb{R}^{Q\times V}$ contains the spatially independent components, the unmixing matrix $\bA\in\mathbb{R}^{T\times Q}$ contains the temporal activity of each IC, and $\bE$ is Gaussian white noise. As it is necessary to constrain the scale of $\bA$ or $\bS$ for identifiability \citep{hyvarinen2002independent,eloyan2013likelihood}, we constrain the columns of $\bA$ to have unit variance.  Let $\ba_t'$ represent the $t$th row of $\bA$, and let $\bs_v$ represent the $v$th column of $\bS$.  Of interest are the spatial ICs $\bS$ and their functional connectivity (FC) $Cor(\bA)$, defined as the correlation among the columns of $\bA$. We refer to the proposed Bayesian ICA model as \textit{FC template ICA}, and to the model of \cite{mejia2020template} wherein $\bA$ is treated as a parameter, as \textit{template ICA}.  The first level of the proposed model is
\begin{equation}\label{eqn:model1}
    y_{tv} = \ba_t^\top \bs_{v} + e_{tv}, \mbox{ where } e_{tv} \sim N(0, \tau^2).
\end{equation}
At the second level, we model the unmixing matrix $\bA$ and the ICs $\bS$ using population-derived priors. Following our previous work \citep{mejia2020template}, we model $\bS$ as
\begin{equation}\label{eqn:model2}
    s_{qv} = s^0_{qv} + \delta_{qv},\quad\delta_{qv} \sim N(0, \sigma^2_{qv}), \quad  q=1,\ldots, Q,\ v = 1,\ldots, V,
\end{equation}
where $s^0_{qv}$ and $\sigma^2_{qv}$ are estimated a-priori (Appendix \ref{app:priorS}), and the $\delta_{qv}$ are mutually independent. Here, we further treat $\bA$ as random via a multivariate prior with covariance $\bG$. 
The full model is 
\begin{align*}
    y_{tv} &= \ba_t^\top \bs_{v} + e_{tv}, \mbox{ where } e_{tv} \sim N(0, \tau^2), \tau^2 \sim IG(\alpha_0, \beta_0) \\
    s_{qv} &= s^0_{qv} + \delta_{qv}, \mbox{ where } \delta_{qv} \sim N(0, \sigma^2_{qv}) \\
    \ba_t &\sim MVN(\mathbf{0}, \bG), \mbox{ where } \bG \sim p(\bG) ,
\end{align*}
where the inverse Gamma (IG) parameters $\alpha_0$ and $\beta_0$ are set to be uninformative.  The FC prior $p(\bG)$ is described below. Note that $\bG$ is a correlation matrix due to the unit variance constraint. 

\subsection{Population-Derived Prior on $\bG$}
\label{ss:hyperparameters}

No conjugate prior exists for correlation matrices. Existing priors for correlation matrices include the LKJ prior \citep{lewandowski2009generating} and restricted inverse-Wishart \citep{wang2018equivalence}. These are designed for uniform sampling from the space of positive-definite correlation matrices, resulting in an uninformative prior. They therefore do not fulfill our goal of incorporating population information through an informative prior. There is a lack of informative priors for correlation matrices in the literature \citep{merkle2023opaque}.

Here, we consider two choices of informative prior for $\bG$, a correlation matrix: (1) a conjugate prior for covariance matrices, the inverse Wishart (IW), centered at a correlation matrix, and (2) a novel informative prior for correlation matrices based on permuted Cholesky factorizations. During model estimation, we constrain the posterior mean of $\bA$ to have unit variance to satisfy the identifiability constraint.  Note that $\bG$ is treated as a hyperparameter in the prior on $\bA$ and $p(\bG)$ as a hyperprior; we therefore do not update the distribution of $\bG$ based on $\bY$. This is an important distinction, since updating $\bA$ based on the prior on $\bG$, rather than its posterior based on data from a single subject, is in accord with the goal of incorporating population information into estimation of $\bA$ by shrinking $Cor(\bA)$ toward the population mean FC. 

To estimate the prior parameters, we require test-retest FC estimates from a set of $n$ training subjects (see Appendix \ref{app:priorS}). In some cases, we may have access to an appropriate training set from a large, publicly available fMRI database \citep[e.g.,][]{van2013wu, casey2018adolescent}. The training sample should represent a similar population as the focal subject. Alternatively, a holdout portion of the focal study can also be used for training, which has been done successfully in previous work \citep{gaddis2022psilocybin, derman2023individual}. As a last resort, one can take an empirical Bayes approach and use the same subjects for parameter estimation and model estimation, recognizing the associated risks of over-fitting and under-estimation of posterior variances.

\subsubsection{Inverse Wishart (IW) Prior.}
\label{sec:IW}
A convenient choice of prior on $\bG$ is the IW, since it is conjugate for the multivariate Gaussian prior on $\ba_t$. Let $\bG \sim IW(\bPsi_0, \nu_0)$, where $\bPsi_0 \in \mathbb{R}^{Q\times Q}$ and $\nu_0 > Q + 1$.   Using the FC estimates $\bX_i$, $i=1,\dots, n$ from the training set, we first estimate the element-wise population mean and variance of $\bG$, $\bar{x}(q,q')$ and $s^2_x(q,q')$.  While for the spatial IC priors we use between-subject variance, here $s^2_x(q,q')$ refers to total variance, since FC is known to vary both between and within subjects.  We then estimate the IW parameters $\bPsi_0$ and $\nu_0$ via a constrained method of moments (MoM) approach, motivated by a well-known limitation of the IW prior: $\nu_0$ controls the variance in an omnibus fashion for all elements of $\bG$. The standard MoM estimator of $\nu_0$ would likely give rise to an element-wise prior variance smaller than the empirical population variance for some elements. We therefore require that the element-wise prior variance not be less than the empirical variance to avoid an overly informative prior for any element of $\bG$.  

\subsubsection{Permuted Cholesky (pChol) Prior.} 
A limitation of assuming an IW prior for a correlation matrix is that the diagonal elements have nonzero variance, so samples from $IW(\bPsi_0, \nu_0)$ are unlikely to be correlation matrices. Furthermore, the IW distribution assumes a monotonic relationship between the element-wise mean and variance, which may be unrealistic. 

Here, we develop a novel informative prior for correlation matrices that can be used to sample correlation matrices that mimic the element-wise mean and variance in the training sample. This approach utilizes Cholesky decompositions, which have played an important role in modeling correlation and covariance matrices, since ensuring positive definiteness only requires positive diagonal elements \citep{pourahmadi2007simultaneous, ghosh2021bayesian, merkle2023opaque}. Our proposed permuted Cholesky (pChol) prior is presented in detail in Appendix \ref{app:CP_prior} but is briefly described here. First, consider the naive approach of constructing univariate priors based on the element-wise training set mean and variance. Univariate sampling from these priors would fail to satisfy positive definiteness. However, the Cholesky factorization of correlation matrices, $\bX = \bL\bL^\top$, only requires (1) $diag(\bL)>0$ and (2) row sum of squares $= 1$. Consequently, the diagonal and off-diagonal elements are respectively constrained to $[0,1]$ and $[-1,1]$. 

This suggests a simple approach: for each FC estimate $\bX_i$ in the training sample, compute the Cholesky factor $\bL_i$, transform each element to $\mathbb{R}$, and sample from univariate priors designed to match the empirical mean and variance of each lower triangular element. After applying the reverse transformations to the sampled values, we only need to rescale to satisfy the sum-of-squares constraint. However, this univariate Cholesky approach has two flaws, which we now address. The first flaw is that strong dependencies between the elements of $\bL$ are not accounted for through univariate priors. To address this, we first perform PCA on the elements of $\bL$ from the training set. We then assume univariate priors on the PC scores (see Appendix \ref{app:CP_prior}). Sampling from those priors produces a sample Cholesky factor $\bL_*$ and correlation matrix $\bX_* = \bL_*\bL_*^\top$. 

However, examining the element-wise mean and variance of the resulting correlation matrix samples reveals a second flaw. While the mean of the $\bX_*$ is very close to $\bar{\bX}$, their variance exhibits systematic bias, with some elements' variances being overestimated and others underestimated. The correlation elements that are a function of fewer elements of $\bL$ (i.e. the upper-left part of $\bX$) have higher variance, while those that are a function of {more} elements of $\bL$ (i.e. the lower-right part of $\bX$) have lower variance. This reflects an undesirable ordering effect. To address this, we randomly permute the rows and columns of the training samples $\bX_i$ prior to Cholesky factorization,  then reverse-permute the samples to obtain sample correlation matrices with the original ordering. In our setting, we find $100$ permutations sufficient to mitigate this bias. 

Samples from this pChol prior are generated based on this procedure to obtain a set of $K$ samples from $p(\bG)$. In our experiments, we sample $500$ matrices for each of $100$ random permutations, resulting in $K=50,000$. As we describe in the following section, these prior samples will be used to (1) obtain the approximate posterior mean and covariance of $\ba_t$ and (2) to produce posterior samples of $\bA$ to facilitate inference on the FC matrix $Cor(\bA)$.

\subsection{Model Estimation}
\label{sec:model_estimation}

Our previous work relied on expectation-maximization (EM) to estimate the model parameters and conditional posterior moments of the latent variables \citep{mejia2020template, mejia2023template}.  Previously $\bA$ was considered as a parameter, whereas here both $\bS$ and $\bA$ are random.  As a result, the E-step in EM is not straightforward, involving unknown joint posterior moments of $\bS$ and $\bA$. One option is to use Gibbs sampling at the E-step, but this would be computationally intensive since the sampler must be run at each iteration. Further, many samples may be required to accurately estimate the required high-dimensional moments. Therefore, we consider an alternative approach using variational Bayes (VB) to relax the posterior dependence between $\bS$ and $\bA$. 

Using VB, we assume that the joint posterior factorizes over $\bS$, $\bA$, and $\tau^2$. That is, if $\bZ=\{\bS,\bA,\tau^2\}$ represents all the latent variables, we assume that $q(\bZ|\bY)=q(\bS|\bY)q(\bA|\bY)q(\tau^2|\bY)$. This factorization assumption is fundamental to VB and results in approximate inference. The convergence of the estimators has been studied in various settings, e.g. see \cite{wang2006convergence} for convergence properties of the estimators for a Gaussian mixture model. 

We derive two VB algorithms based on the two different priors for $\bG$ described above. Derivation details are given in Appendix \ref{app:VB}. The first algorithm, VB1, is based on the IW prior for $\bG$. VB1 is very computationally efficient, since it benefits from conjugacy for every approximate posterior. The second, VB2, is associated with our novel pChol prior. It uses a pre-determined set of sample correlation matrices from $p(\bG)$ to update $\bA$. For both VB1 and VB2, the approximate posteriors for $\bS$ and $\tau^2$ have analytical solutions, given as follows.

\begin{itemize}
    \item $q(\bS|\bY)$ factorizes across $v$, and $q(\bs_{v}|\bY)$ is Normal with mean and covariance given by $
\hat\bs_{v} = V(\bs_{v}) \left(\frac{1}{\hat\tau^2}\hat\bA^\top\by_v + \bD_v^{-1}\bs_v^0 \right)$ and $
V(\bs_{v}) = \left( \frac{1}{\hat\tau^2}\EX[\bA^\top\bA] +\bD_v^{-1}  \right)^{-1}$

\item $q(\tau^2|\bY) \sim IG(\alpha,\hat\beta)$ with mean $\hat\tau^2 = (\alpha - 1)^{-1}\hat\beta$, where $\alpha = \alpha_0 + \frac{TV}{2}$ and \\
$
 \hat\beta = \beta_0 + \frac{1}{2}\sumv\sumt y_{tv}^2 - \sumv\sumt y_{tv}\hat\ba_t^\top\hat\bs_v + Tr\left\{ \sumt \EX\Big[\ba_t\ba_t^\top\Big] \sumv \EX \Big[\bs_v\bs_v^\top\Big] \right\}
$
\end{itemize}

\noindent Turning now to $\bA$, using the approximate posterior requires integrating over $\bG$. However, the two choices of prior for $\bG$ necessitate different integration strategies, which we now describe.

\begin{itemize}
\item For VB1, $\bG \sim IW(\bPsi_0,\nu_0)$, so the marginal prior $p(\ba_t) = \int_{\bG}p(\ba_t|\bG)p(\bG)d\bG$ follows a multivariate $t$ distribution with $\nu_a = \nu_0 + 1 - Q$ degrees of freedom. This can be represented as a scale-mixture of Normals: conditional on $u\sim Gamma(\tfrac{\nu_a}{2}, \tfrac{\nu_a}{2})$, $\ba_t$ is Normal with mean zero and covariance $(u\nu_a)^{-1}\bPsi_0$  a-priori. Hence, the  approximate posterior $q(\bA|\bY,u)$ is Normal and factorizes over $t$, with mean $\EX[\ba_t|u] = \VX(\ba_t|u)\left(\hat\tau^{-2}\hat\bS\by_t\right)$ and covariance $\VX(\ba_t|u) = \Big( \hat\tau^{-2}\EX[\bS\bS^\top] + u\nu_a\bPsi_0^{-1}\Big)^{-1}$, where $\hat\tau^2$, $\hat\bS$, and $\EX[\bS\bS^\top]$ are approximate posterior moments. Applying laws of total expectation and covariance, we obtain:
\begin{align*}
    \EX[\ba_t] &\equiv \hat\ba_t = \EX_u\left[ \Big( \frac{1}{\hat\tau^2}\EX[\bS\bS^\top] + u\nu_a\bPsi_0^{-1}\Big)^{-1}\right]
    \left(\frac{1}{\hat\tau^2}\hat\bS\by_t\right)  \\
    \VX(\ba_t) &= \EX_u\left[ \Big( \frac{1}{\hat\tau^2}\EX[\bS\bS^\top] + u\nu_a\bPsi_0^{-1}\Big)^{-1}\right] 
     + \VX_u\left( \Big( \frac{1}{\hat\tau^2}\EX[\bS\bS^\top] + u\nu_a\bPsi_0^{-1}\Big)^{-1} \left(\frac{1}{\hat\tau^2}\hat\bS\by_t\right)\right)
\end{align*}
\noindent which are estimated via Monte Carlo using samples from $u$. To facilitate inference, we generate samples from the approximate posterior of $\bA$ using the same samples $u$: for each $u$, we draw a sample from $q(\bA|\bY,u)$, which collectively form a set of samples from $q(\bA|\bY)$.

\item For VB2, we have sample correlation matrices $\bG_k$ from the pChol prior $p(\bG)$. Conditional on $\bG$, the approximate posterior $q(\bA|\bY,\bG)$ is Normal and factorizes over $t$, with mean $\EX[\ba_t|\bG] = \VX(\ba_t|\bG)\left(\frac{1}{\hat\tau^2}\hat\bS\by_t\right)$ and covariance $\VX(\ba_t|\bG) = \Big( \frac{1}{\hat\tau^2}\EX[\bS\bS^\top] + \bG^{-1}\Big)^{-1}$. Let  $\bV_k = \Big( \frac{1}{\hat\tau^2}\EX[\bS\bS^\top] + \bG_k^{-1}\Big)^{-1}$. Applying the laws of total expectation and covariance, we obtain 
$$
    \EX[\ba_t] \equiv \hat\ba_t =\frac{1}{K} \sum_{k=1}^K \bV_k\left(\frac{1}{\hat\tau^2}\hat\bS\by_t\right) 
    \quad\text{and}\quad
    \VX(\ba_t) = \Big(\frac{1}{K}\sum_{k=1}^K \bV_k\Big) + {Cov}_k\Big( \frac{1}{\hat\tau^2}\bV_k\hat\bS\by_t\Big),
$$
\noindent  where ${Cov}_k$ refers to the covariance over the $k$ samples. Optionally, we can use an approximation for $\bV_k$ (described in Appendix \ref{app:VB}) until the final iteration. We obtain samples from the approximate posterior of $\bA$ similarly to VB1: for each sample $\bG_k$ from $p(\bG)$, we draw a sample of $\ba_t$, $t=1,\dots,T$, which collectively form a set of samples from $q(\bA|\bY)$.
\end{itemize}

To facilitate posterior inference on the FC, we first draw samples from $q(\bA|\bY)$ as described above.  For VB2, we obtain one sample of $\bA$ for each $\bG\sim p(\bG)$; for VB1, we obtain one sample of $\bA$ for each Gamma-distributed $u$. In our analyses, we draw $10,000$ samples of $u$ in VB1 or $50,000$ samples of $\bG$ in VB2. Using these samples, we can estimate and perform posterior inference on the FC matrix, i.e. ${Cor}(\bA)$. For instance, we can construct element-wise credible intervals to identify statistically significant FC pairs. While our focus is the FC matrix, these posterior samples of $\bA$ have other potential uses, such as analyzing FC dynamics and performing inference on dwell time and other metrics.

We initialize the parameter values as follows: we use standard template ICA to produce initial estimates of $\bA$ and $\bS$ and the element-wise variance of $\bS$.  We initialize the residual variance $\tau^2$ based on the template ICA residuals. We then iteratively estimate the approximate posteriors until convergence of the posterior estimates of $\bS$, $\bA$ and $\tau^2$.

\section{Simulation Study}
\label{sec:simulation}

We simulate realistic functional MRI data to test the accuracy of the proposed algorithms for FC template ICA (FC-tICA) and compare them with existing approaches. These include our previous template ICA (tICA) model, wherein the temporal mixing matrix is treated as a model parameter rather than a latent variable, and dual regression (DR), a popular ad-hoc method. For each method, we quantify the accuracy of the resulting spatial IC maps and temporal FC matrices, with respect to the unique true subject-level spatial maps and FC matrices. We also assess the sensitivity of each method to the duration of the fMRI time series.

\subsection{Data Generation}
\label{ss:sim_data}

Our simulated data was generated using realistic features derived from the Human Connectome Project (HCP) dataset, which is introduced in Section \ref{sec:DA} \citep{van2013wu}. The generation of subject-specific spatial IC maps is described in Appendix \ref{app:sim_data_generation}. Among the 25 HCP group ICA maps, we choose three visual components (ICs 1-3), one default mode network (DMN) (IC 4), and one motor (IC 5). For each IC, we randomly generate unique subject-level versions containing subtle individual features, a reasonable representation of the differences known to exist between individuals. An example subject's deviations and ICs are shown in Figure \ref{fig:sim:subjICs}.

To generate realistic IC time courses, we use the HCP minimally preprocessed, surface-projected fMRI timeseries from the left-to-right (LR) run of the first visit, which contains 1200 time points (volumes) over 14.4 minutes. Details of the HCP data processing pipelines can be found in \cite{glasser2013minimal}. We utilize observations from 1,068 subjects with valid data. To obtain IC timeseries for the simulation, we regress the high-pass filtered fMRI data against the 25 HCP group ICs, then select the time courses corresponding to our five ICs. 

Sample IC time courses are shown in Figure \ref{fig:sim:TC_FC_real}, along with their FC matrices. They show the expected patterns of high connectivity within the three visual ICs, moderate connectivity between the visual and motor regions, and lower connectivity between the DMN and visuomotor regions. Yet individual differences in FC patterns can also be clearly seen. The population mean and standard deviation of each FC pair across all subjects are shown in Figure \ref{fig:sim:template_FC}. The mean FC is similar to the patterns seen in the example subjects. The variance patterns show that all FC pairs exhibit variation across subjects, with more variance for weaker connections.

Finally, we generate synthetic fMRI data for each subject as follows.  Let $\bE_{T\times V}$ contain Gaussian white noise with standard deviation $\sigma_e$ such that the signal-to-noise ratio (SNR) $\sigma_a/\sigma_e$ is equal to $0.5$, where $\sigma_a^2$ is the average variance of IC time courses scaled by the IC spatial intensity at the peak (top 1\% of) vertices. This SNR level is chosen to be similar to real fMRI data.  The synthetic fMRI data is then given by $\bY = \bA\bS + \bE$, where $\bS_{Q\times V}$ contains the $Q$ IC maps and $\bA_{T\times Q}$ contains the corresponding IC time courses. We use 500 subjects for template estimation and 50 held-out test subjects to fit the models and evaluate model performance. For the test subjects, the first half ($T=600$, 7.2 min) of volumes is used for model estimation, while the second half is reserved to evaluate the predictive accuracy of our FC estimates. We also consider the effect of shorter scan duration, varying $T$ from $200$ to $600$ volumes.

\subsection{Prior and model estimation}

We estimate the population priors on the ICs and the FC, as described in Section \ref{ss:hyperparameters}.  Figure \ref{fig:sim:template_vs_oracle} shows the estimated prior mean and variance maps for each spatial IC, along with the true (oracle) population mean and variance maps. The prior means closely mimic the true population means. The prior variance mimics the true variance in high-variance areas but is larger in low-variance areas. This is an expected result of our biased non-negative variance estimation approach, which avoids under-estimation of the true variance in favor of a less informative prior.  

Figure \ref{fig:sim:template_FC} shows the true population mean and variance of each FC pair, along with the prior mean and variance associated with the IW and pChol priors. The prior mean is close to the true population mean for both priors. For the IW prior, the variance is monotonically related to the magnitude of the mean, in contrast to the population. Our conservative estimation of the IW prior variance therefore leads to a relatively uninformative prior, particularly for stronger connections. The variance of the pChol prior, conversely, closely mimics the population variance.

For each of the 50 test subjects, we fit our FC-tICA model using the two proposed VB algorithms, along with the existing methods tICA and DR. For tICA and FC-tICA, we run each algorithm to a tolerance of 0.001 for a maximum of 100 iterations.

\subsection{Simulation results}

Here, we summarize the performance of FC-tICA versus existing methods. Our primary metric of accuracy is median absolute error (MAE) relative to the subject-specific ground truth values across all 50 test subjects. To assess the predictive accuracy of FC estimates, we use ground truth FC values for each subject based on the held-out half of the simulated session.

Considering first the spatial maps, Figure \ref{fig:sim:results_subjICs} shows example true and estimated deviation maps for three test subjects. Recall that the spatial ICs for each subject are generated as the prior mean plus a subject-level deviation, as described in Section \ref{ss:sim_data}. Both tICA and FC-tICA identify the unique subject-level spatial features. While these features can also be seen with DR, the estimates exhibit much higher noise levels. Figure \ref{fig:sim:results_MAE_S} shows the accuracy of spatial IC maps. FC-tICA and tICA have similar accuracy, with higher error in areas of engagement and lower error in background areas, while the accuracy of DR is much worse.

\begin{figure}
    \centering
    \begin{tabular}{cccccc}
    & MAE & \%Change vs. DR & \%Change vs. tICA & \\[4pt]
    \hline \\[-8pt]
    \begin{picture}(10,70)\put(5,35){\rotatebox[origin=c]{90}{VB1}}\end{picture} & 
    \includegraphics[height=28mm, page=2, trim = 5mm 3mm 25mm 22mm, clip]{simulation/plots/MAE_FC.pdf} & 
    \includegraphics[height=28mm, page=2, trim = 5mm 3mm 25mm 22mm, clip]{simulation/plots/MAE_FC_diff_DR.pdf} & 
    \includegraphics[height=28mm, page=1, trim = 5mm 3mm 25mm 22mm, clip]
    {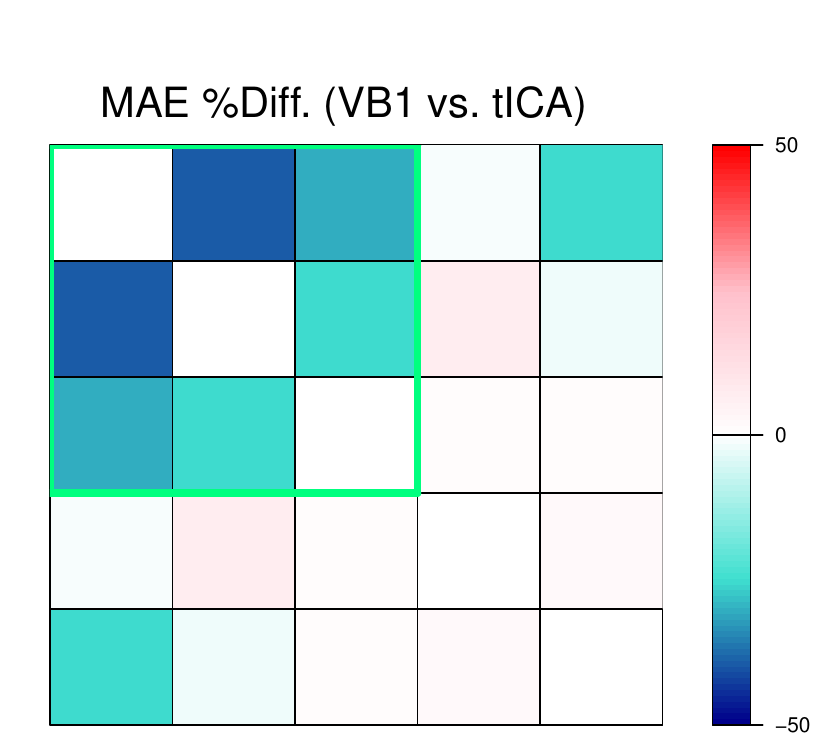} \\ 
    \begin{picture}(10,70)\put(5,35){\rotatebox[origin=c]{90}{VB2}}\end{picture} & 
    \includegraphics[height=28mm, page=3, trim = 5mm 3mm 25mm 22mm, clip]{simulation/plots/MAE_FC.pdf} & 
    \includegraphics[height=28mm, page=3, trim = 5mm 3mm 25mm 22mm, clip]{simulation/plots/MAE_FC_diff_DR.pdf} & 
    \includegraphics[height=28mm, page=2, trim = 5mm 3mm 25mm 22mm, clip]
    {simulation/plots/MAE_FC_diff_tICA.pdf} \\ 
    \begin{picture}(10,70)\put(5,35){\rotatebox[origin=c]{90}{tICA}}\end{picture} & 
    \includegraphics[height=28mm, page=1, trim = 5mm 3mm 25mm 22mm, clip]{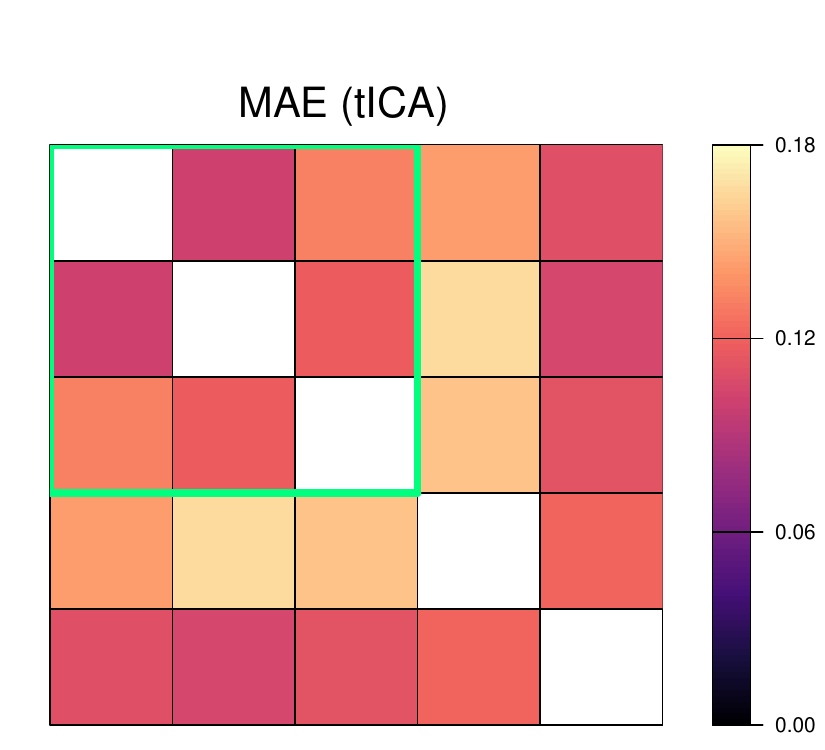} & 
    \includegraphics[height=28mm, page=1, trim = 5mm 3mm 25mm 22mm, clip]{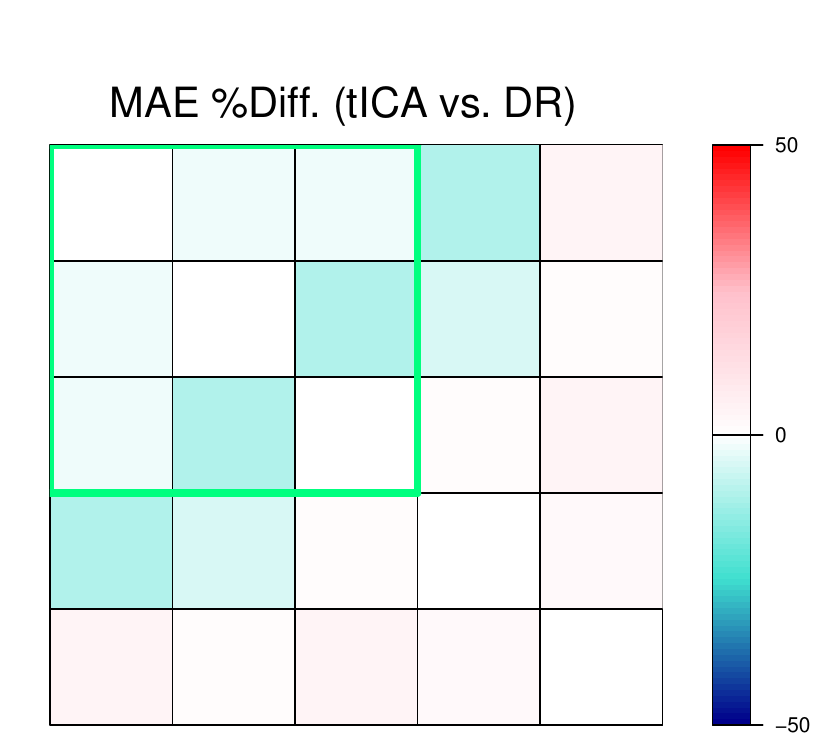} \\ 
    \begin{picture}(10,70)\put(5,35){\rotatebox[origin=c]{90}{DR}}\end{picture} & 
    \includegraphics[height=28mm, page=4, trim = 5mm 3mm 25mm 22mm, clip]{simulation/plots/MAE_FC.pdf}  &
    \multicolumn{3}{c}{{\includegraphics[height=28mm, trim = 0 5mm 0 0, clip]{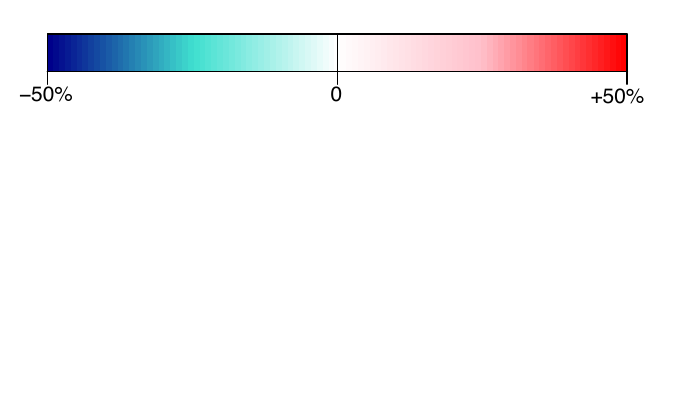}}} \\ 
    & \includegraphics[height=4mm]{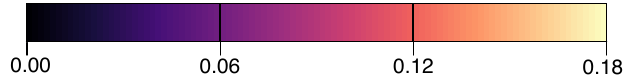} \\
    \end{tabular}
    \caption{\textit{Accuracy of FC estimates across 50 simulated test subjects.} The green box indicates the stronger connections within the visual network; the fourth row/column of each matrix corresponds to weak connections between the default mode and visual/motor regions; the fifth row/column of each matrix corresponds to connections with the motor network. Accuracy is summarized in terms of median absolute error (MAE). Columns 2-4 show the percent change in MAE compared with existing methods. Negative values (blue) indicate improved accuracy. Compared with DR and tICA, both FC-tICA VB algorithms tend to result in higher accuracy, particularly for the stronger within-network correlations (green box). VB1 and VB2 have similar performance, even though they assume different priors on the FC. }
    \label{fig:sim:MAE_FC}
\end{figure}

Turning to the FC matrices, the main focus in this work, Figure \ref{fig:sim:FC_true_est} displays true and estimated FC matrices for three randomly selected subjects. All methods produce visually similar estimates, and idiosyncrasies across subjects can be clearly seen. Figure \ref{fig:sim:MAE_FC} quantifies the accuracy of FC estimates for each method. The stronger within-network connections, generally of greater interest, are outlined in green on each image. The fourth row/column contains connections with the DMN IC, and the fifth row/column contains connections with the motor IC.  Overall, FC-tICA tends to outperform existing methods in terms of predictive accuracy of FC. The advantage of FC-tICA is predominantly in the stronger within-network connections, but is also apparent for some moderate and weak connections. Comparing VB1 and VB2, there is not a clear advantage of one over the other, though VB2 shows somewhat better performance for the stronger within-network connections, suggesting a possible advantage of the pChol prior. 


\begin{figure}
    \centering
    \includegraphics[width=0.9\linewidth]{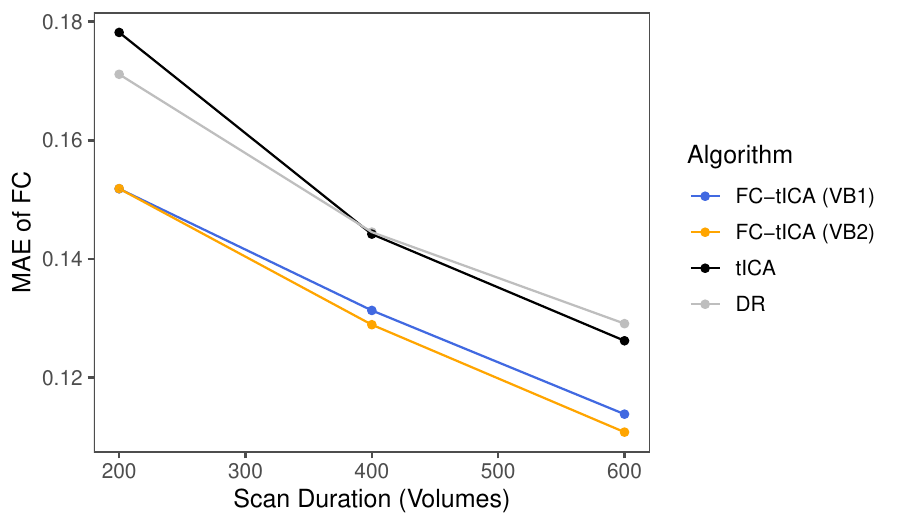}
    \caption{\textit{Accuracy of FC estimates by scan duration in simulation study.} Accuracy is summarized in terms of median absolute error (MAE), averaged over FC pairs. Appendix Figure \ref{fig:sim:FC_MAE_by_duration} shows the pair-specific MAE values by scan duration. As scan duration varies from $T=200$ volumes ($2.4$ minutes) to $T=600$ volumes ($7.2$ minutes), both VB algorithms for FC-tICA outperform the existing methods tICA and DR. VB2 slightly outperforms VB1 for longer scan durations, possibly reflecting a benefit of the permuted Cholesky prior over the inverse-Wishart for modeling correlation matrices.}
    \label{fig:sim:MAE_FC_duration}
\end{figure}

Figures \ref{fig:sim:MAE_FC_duration} and \ref{fig:sim:FC_MAE_by_duration} summarize the accuracy of FC estimates by scan duration. Both VB algorithms consistently outperform tICA and DR across scan durations, with VB2 having a slight advantage over VB1 overall.  As a preliminary investigation of asymptotic properties as $T\to\infty$, and on the influence of the prior for small $T$, we perform an additional simulation study described in Appendix \ref{app:long_duration_sim}. We observe that as $T\to\infty$, estimation of FC improves to a point, and that the data dominate the prior even for small $T$. We discuss the complex effects of $T\to\infty$. 

Finally, we perform Bayesian inference by constructing 95\% credible intervals (CIs) for the FC. We obtain posterior samples of $\bA$ as described in Section \ref{sec:model_estimation}, then compute ${Cor}(\bA)$ for each sample. For each FC pair, we construct a middle 95\% CI and evaluate their coverage. Figure \ref{fig:sim:FC_CIs} shows the CI and true FC for the first 10 subjects. We observe that the VB2 intervals include the true in-sample FC value 73\% of the time, while the VB1 intervals only do so 12\% of the time. This reflects the narrow width of the VB1 CIs (0.015 on average) compared with VB2 (0.12 on average). The risk of posterior variance underestimation, a common problem in VB \citep{bishop2006pattern}, is apparently greater in VB1. Therefore, while VB1 and VB2 produce visually similar estimates, VB2 appears to produce more valid inferences. This suggests an additional benefit of the pChol prior for correlation matrices.

The mean computation time of our two VB algorithms for FC-tICA was 2.99 seconds for VB1 and 100.77 seconds for VB2, indicating considerably higher computational burden associated with the pChol prior. For comparison, tICA converged in 2.14 seconds on average.  Since tICA is used within FC-tICA to initialize parameter values, the VB1 computations only required an additional 0.85 seconds.  FC-tICA with VB1 therefore represents a highly pragmatic model extension to tICA.

\section{Experimental Data Analysis}
\label{sec:DA}

In this section, we use fMRI data from the Human Connectome Project (HCP) to evaluate the performance of our proposed FC template ICA model over standard template ICA and to assess the computational feasibility and practicality of the proposed VB algorithms.

\subsection{Data Processing and Analysis}

We analyze resting-state fMRI from the HCP 1200-subject release (http://humanconnectome.org). The HCP fMRI data were collected using a custom multiband acquisition for high spatial and temporal resolution \citep{van2013wu} and were processed according to the HCP minimal preprocessing pipelines, including projection to the cortical surface \cite{glasser2013minimal}. We avoid spatial smoothing to preserve the functional spatial features of each individual.  To reduce the computational burden during model estimation, we resample to approximately $10,000$ vertices per hemisphere and exclude subcortical regions.

Each HCP participant underwent four resting-state fMRI runs across two visits. Each run contains $1200$ volumes $0.72$ seconds apart over approximately $15$ minutes. At each visit, the two runs were acquired using opposite phase encoding directions (LR and RL). To avoid acquisition-related spatial distortions, here we limit analysis to the two LR runs per subject. We utilize a version of the released data that has been high-pass filtered and denoised via ICA-FIX \citep{salimi2014automatic}.  We randomly select 362 subjects for template estimation and 100 subjects for analysis, analyzing in total 924 fMRI runs from 462 subjects. 

\begin{figure}
\centering
\begin{tabular}{cccc}
& Population & IW Prior (VB1) & pChol Prior (VB2)\hspace{1cm} \\[10pt]
{\begin{picture}(10,120)\put(5,60){\rotatebox[origin=c]{90}{Mean}}\end{picture}} &
\includegraphics[height=45mm, page=1, trim = 5mm 0 25mm 20mm, clip]{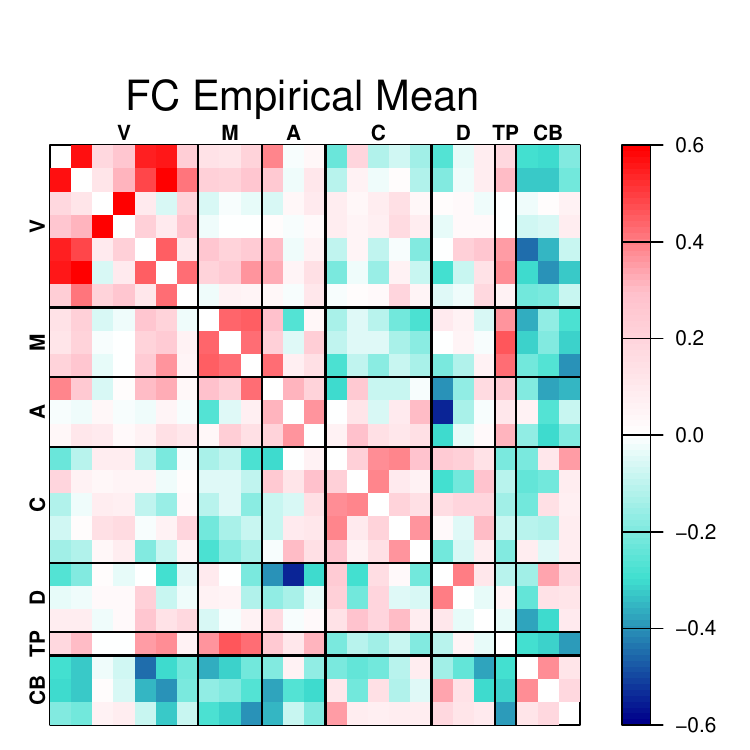} &
\includegraphics[height=45mm, page=1, trim = 5mm 0 25mm 20mm, clip]{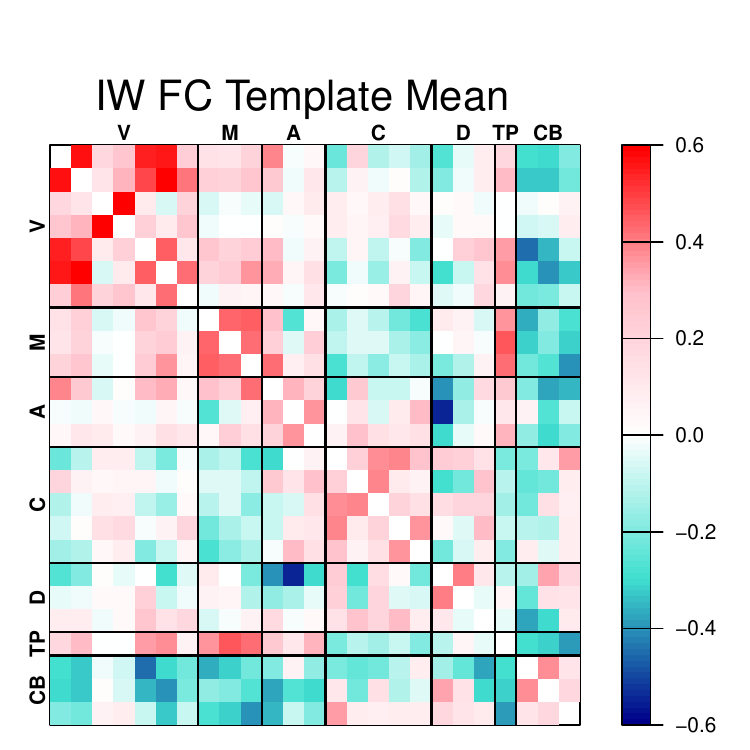} &
\includegraphics[height=45mm, page=1, trim = 5mm 0 0 20mm, clip]{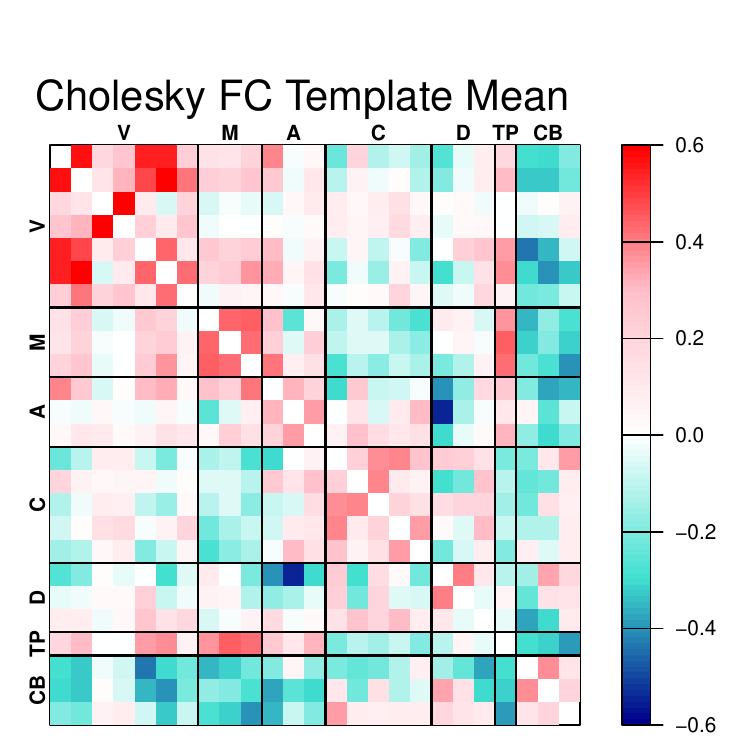} \\
{\begin{picture}(10,120)\put(5,60){\rotatebox[origin=c]{90}{Standard Deviation}}\end{picture}} &
\includegraphics[height=45mm, page=2, trim = 5mm 0 25mm 20mm, clip]{plots/GICA25/FC_empirical.pdf} &
\includegraphics[height=45mm, page=2, trim = 5mm 0 25mm 20mm, clip]{plots/GICA25/FC_template_IW.pdf} &
\includegraphics[height=45mm, page=2, trim = 5mm 0 0 20mm, clip]{plots/GICA25/FC_template_Chol.pdf} \\
\end{tabular}
    \caption{\textit{Population and prior means and standard deviations of functional connectivity (FC) in the HCP analysis.} ICs are grouped by network, with abbreviations given in Table \ref{tab:RSNs}. 
 The prior means closely mimic the population mean FC patterns. For the inverse-Wishart (IW) prior, the prior variance is generally inflated compared with the empirical population variance. This is by design, given that the IW distribution has a single parameter to control the variance across the matrix, and we wish to avoid an overly informative prior for any FC pair (see Section \ref{sec:IW}). By contrast, the element-wise variance of our novel pChol prior closely mimics the population variance.}
    \label{fig:DA:template_FC}
\end{figure}

Group ICA is a prerequisite for estimating the hyperparameters of the population-derived priors in our Bayesian ICA models, as described in Section \ref{ss:hyperparameters}. The HCP 500-subject release includes group ICA maps based on nearly 500 subjects.  We utilize the HCP 25-component group ICA (see Appendix \ref{app:IC_assignments} for IC maps and their network memberships). We estimate the prior hyperparameters for the IC maps and FC based on full-resolution test-retest data from the 362 training subjects. Figure \ref{fig:DA:template_ICs} in Appendix \ref{app:DA_results} illustrates the population-derived priors for the spatial ICs, and Figure \ref{fig:DA:template_FC} illustrates the two population-derived priors for the FC. Note that the variance of the IW prior is generally larger than the population variance. This is by design to an overly informative prior for any FC pair, as described in Section \ref{sec:IW}. By contrast, the pChol prior closely mimics the population variance.


We analyze both visits for each of the 100 test subjects using FC template ICA (FC-tICA) via each of our proposed VB algorithms (VB1 and VB2). We also apply standard template ICA (tICA) to assess the additional benefit of the population-derived priors for the FC in the model. We run all algorithms to a tolerance of $0.001$. For better computational efficiency, we set the number of nuisance ICs to estimate and remove prior to model fitting, as described in \cite{mejia2020template}, to $200$ for all methods. Our open-source templateICAr R package (version 10.0) and R version 4.4.1 are used for all analyses.

\subsection{Results}

FC-tICA requires approximately $23$ minutes per session using VB1 and $278$ minutes using VB2. By comparison, tICA requires approximately $17$ minutes. These computation times include all preprocessing steps, parameter initialization and model estimation.  This shows that FC-tICA with VB1, which uses the IW prior, represents a highly practical model extension to standard tICA.  While our pChol prior requires a much greater computational investment, it is computationally feasible in practical settings.

\begin{figure}
\centering
\begin{tabular}{cccc}
& FC-tICA (VB1) & FC-tICA (VB2) & \hspace{-1cm}tICA \\[10pt]
\begin{picture}(10,90)\put(5,45){\rotatebox[origin=c]{90}{Subject 1}}\end{picture} & 
\includegraphics[height=35mm, page=1, trim = 5mm 0 25mm 21mm, clip]{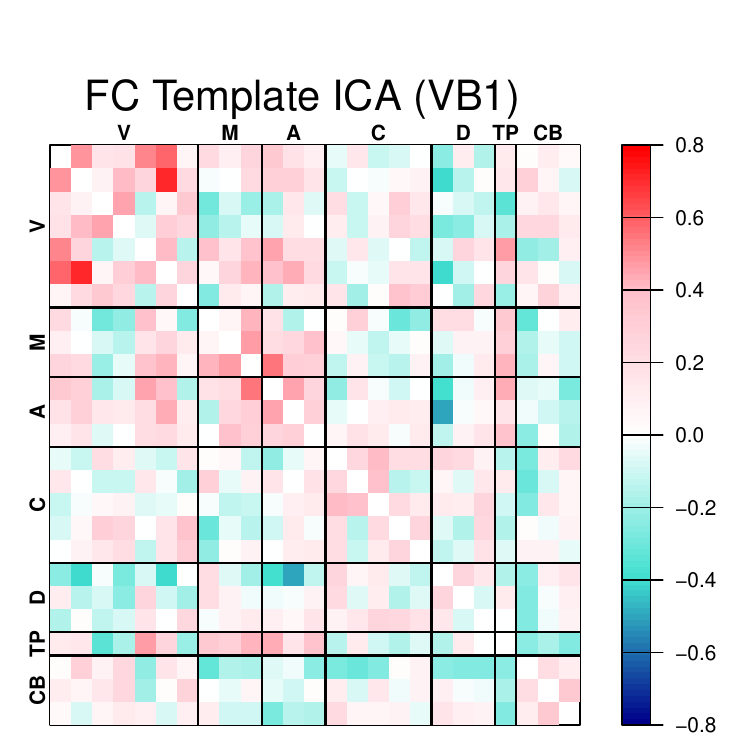} &
\includegraphics[height=35mm, page=2, trim = 5mm 0 25mm 21mm, clip]{plots/GICA25/FC_subj1_sess1.pdf} &
\includegraphics[height=35mm, page=3, trim = 5mm 0 0 21mm, clip]{plots/GICA25/FC_subj1_sess1.pdf} \\
\begin{picture}(10,90)\put(5,45){\rotatebox[origin=c]{90}{Subject 2}}\end{picture} & 
\includegraphics[height=35mm, page=1, trim = 5mm 0 25mm 21mm, clip]{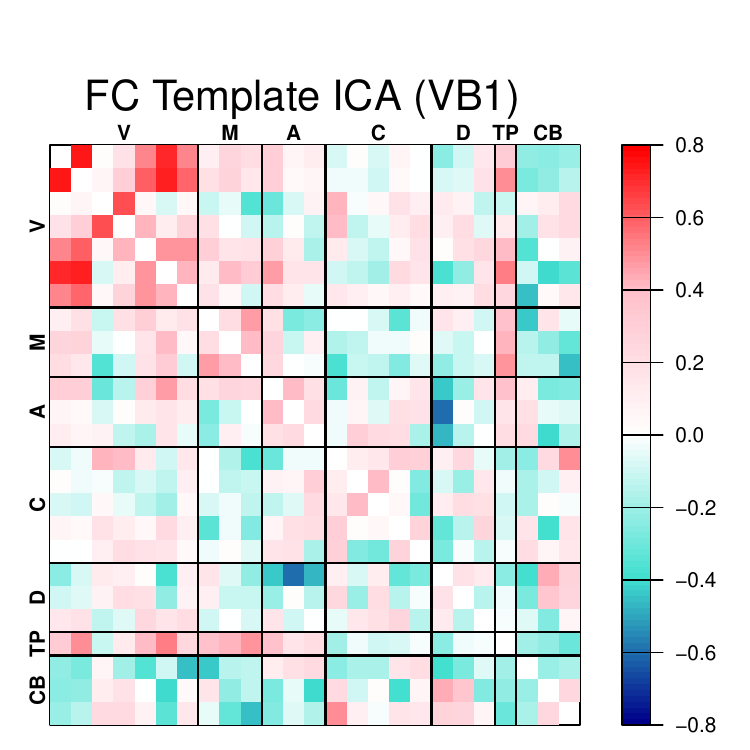} &
\includegraphics[height=35mm, page=2, trim = 5mm 0 25mm 21mm, clip]{plots/GICA25/FC_subj2_sess1.pdf} &
\includegraphics[height=35mm, page=3, trim = 5mm 0 0 21mm, clip]{plots/GICA25/FC_subj2_sess1.pdf} \\
\begin{picture}(10,90)\put(5,45){\rotatebox[origin=c]{90}{Subject 3}}\end{picture} & 
\includegraphics[height=35mm, page=1, trim = 5mm 0 25mm 21mm, clip]{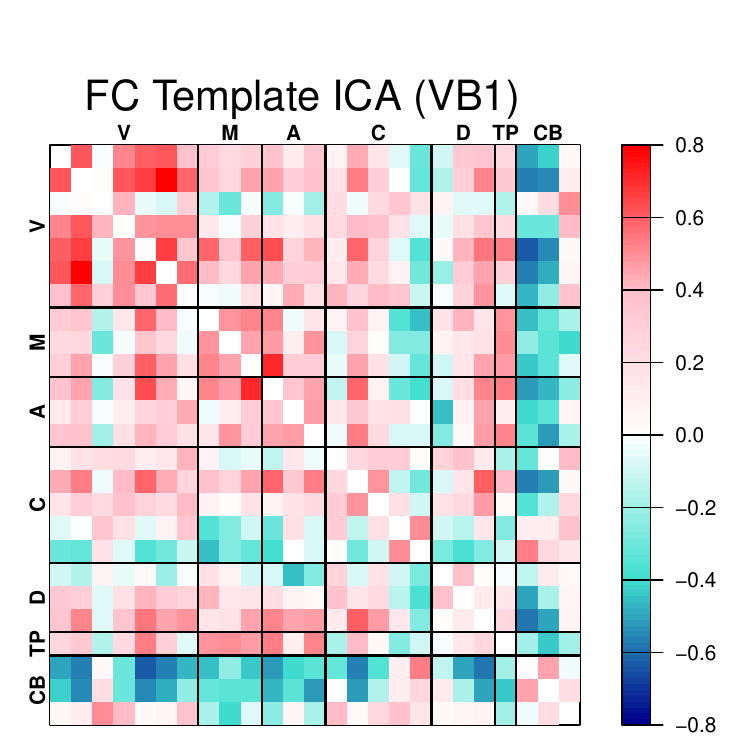} &
\includegraphics[height=35mm, page=2, trim = 5mm 0 25mm 21mm, clip]{plots/GICA25/FC_subj3_sess1.pdf} &
\includegraphics[height=35mm, page=3, trim = 5mm 0 0 21mm, clip]{plots/GICA25/FC_subj3_sess1.pdf} \\
\begin{picture}(10,90)\put(5,45){\rotatebox[origin=c]{90}{Population SD}}\end{picture} & 
\includegraphics[height=35mm, page=1, trim = 5mm 0 25mm 21mm, clip]{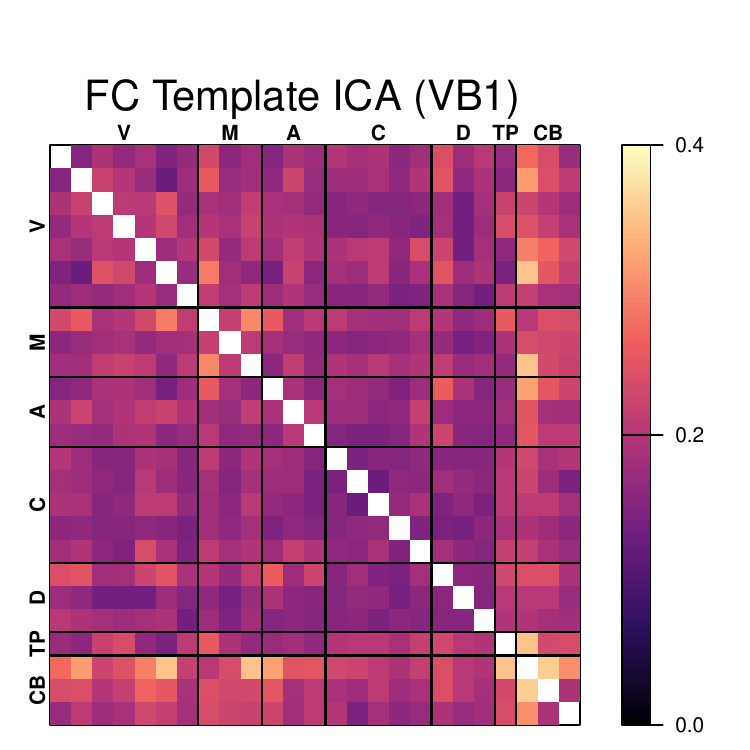} &
\includegraphics[height=35mm, page=2, trim = 5mm 0 25mm 21mm, clip]{plots/GICA25/FC_dist.pdf} &
\includegraphics[height=35mm, page=3, trim = 5mm 0 0 21mm, clip]{plots/GICA25/FC_dist.pdf} \\
\end{tabular}
    \caption{\textit{Example FC estimates and population variability.} For three randomly chosen subjects, FC matrices estimated via FC template ICA (FC-tICA) and standard template ICA (tICA) are shown using visit 1 data. The last row shows the standard deviation (SD) of subject-level estimates with respect to the training set population mean, the basis of the FC priors. The SD is lower with VB2, indicating greater shrinkage toward the population mean with our novel pChol prior.}
    \label{fig:DA:FC_estimates}
\end{figure}

Figure \ref{fig:DA:FC_estimates} shows estimated FC matrices for three example subjects. Unique connectivity features can be seen across subjects and appear quite similar across methods. The last row shows the standard deviation of subject-level estimates around the population mean, an indication of the degree of shrinkage of FC estimates in FC-tICA.  The population variance is smaller with FC-tICA compared with tICA, and is smaller for VB2 compared with VB1. This illustrates that the pChol prior for the FC results in more shrinkage toward the population mean, compared with no prior (tICA) or with the less informative IW prior (VB2).


\begin{figure}
\centering
\begin{tabular}{cccc}
& FC-tICA (VB1) & FC-tICA (VB2) & tICA  \\[10pt]
\begin{picture}(10,120)\put(5,60){\rotatebox[origin=c]{90}{ICC}}\end{picture} & 
\includegraphics[height=44mm, page=1, trim = 5mm 0 25mm 21mm, clip]{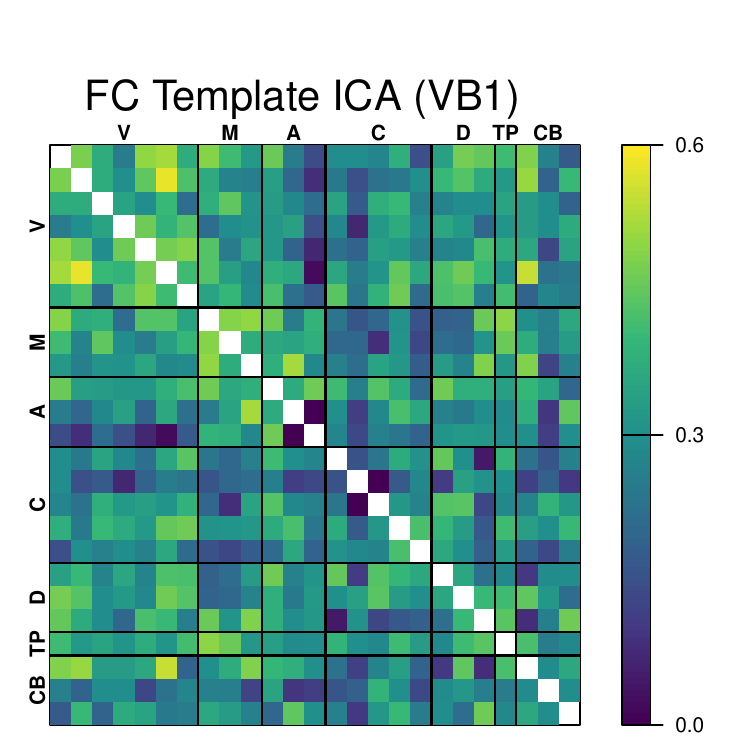} &
\includegraphics[height=44mm, page=2, trim = 5mm 0 25mm 21mm, clip]{plots/GICA25/FC_ICC.pdf} &
\includegraphics[height=44mm, page=3, trim = 5mm 0 25mm 21mm, clip]{plots/GICA25/FC_ICC.pdf} \\
& \multicolumn{3}{c}{\includegraphics[width=5cm]{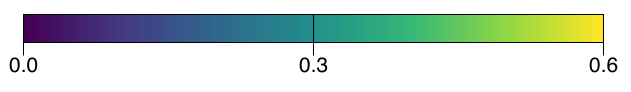}} \\
\begin{picture}(0,120)\put(-5,60){\rotatebox[origin=c]{90}{Change vs. tICA}}\end{picture} & 
\includegraphics[height=44mm, page=1, trim = 5mm 0 25mm 21mm, clip]{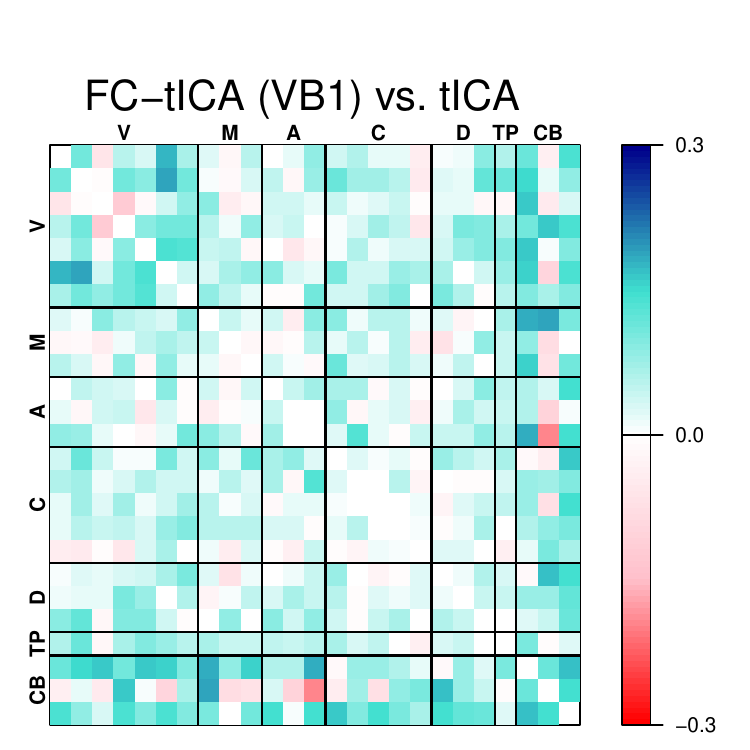} &
\includegraphics[height=44mm, page=2, trim = 5mm 0 25mm 21mm, clip]{plots/GICA25/FC_ICC_change.pdf} \\
& \multicolumn{2}{c}{\includegraphics[width=5cm]{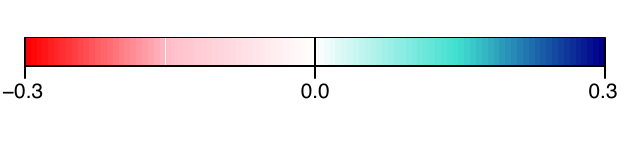}} \\
\end{tabular}
    \caption{\textit{Reliability of FC matrices.} Values shown are the intra-class correlation coefficient (ICC), i.e. the proportion of variability in the estimates that represents unique subject-level information. The difference between FC template ICA (FC-tICA) and standard template ICA (tICA) is shown on the second row. Nearly all FC pairs exhibit improved reliability using FC-tICA. The improvement is greater with VB2 versus VB1, suggesting an advantage of our novel pChol prior over the IW at reducing noise levels relative to signal variance.}
    \label{fig:DA:FC_ICC}
\end{figure}

Figure \ref{fig:DA:FC_ICC} summarizes the reliability of FC in terms of intra-class correlation coefficient (ICC). Without access to the ground truth, test-retest error measures like MSE or MAE can be spuriously improved due to Bayesian shrinkage towards a group mean. By contrast, ICC represents the proportion of variance attributable to unique individual features. Thus, higher ICC indicates improved estimation of unique subject-level FC features. The difference images in the second row show that FC-tICA produces more individually reliable FC estimates compared with standard tICA. Comparing the two FC-tICA VB algorithms, VB2 produces more dramatic improvements for some FC pairs, suggesting an advantage of our pChol prior compared to the IW.


Turning finally to the spatial IC maps representing functional brain topography, Figure \ref{fig:DA:IC_estimates} shows example spatial IC estimates and posterior standard deviations for two subjects. Visually, FC-tICA and tICA produce very similar estimates---unsurprising, given that they utilize the same population-derived priors on the spatial IC maps. The individuals' spatial patterns show some clear differences in the precise locations of engagement, including in the temporal lobe and lateral frontal lobe. To quantify the individuality and reliability of these spatial maps, Figure \ref{fig:DA:S_ICC} displays the their vertex-wise intra-class correlation coefficient (ICC). As expected, ICC is higher within areas of engagement for each IC, since ``background'' regions exhibit little true between-subject variability. Figure \ref{fig:DA:S_ICC_vs_tICA} compares the ICC of FC-tICA and tICA and reveals a subtle improvement with FC-tICA. This suggests that the inclusion of informative priors on the ICA mixing matrix in FC-tICA may have downstream benefits for estimation of the spatial ICs.

\section{Discussion}
\label{sec:discussion}

In this paper, we develop a computationally efficient Bayesian ICA framework for subject-level analysis of functional MRI data leveraging population information on spatial topography and functional connectivity (FC), an important measure of brain function. We employ population-derived priors on both the latent spatial source signals and their associated temporal activity, which can improve single-subject analysis and reduce the effects of high noise levels in fMRI data. We propose a new informative prior for correlation matrices, which can be used to draw samples that exhibit a desired mean and variance structure. To our knowledge, no similarly informative prior for correlation matrices has been proposed previously in the literature. This novel pChol prior has many potential applications within and beyond neuroimaging in settings where the within-subject correlation between variables is of interest and prior information is available.

Through extensive simulation studies and analysis of a large fMRI dataset, we show that the proposed techniques outperform two existing methods: our earlier Bayesian ICA model with priors on spatial topography alone, and dual regression, an ad-hoc method popular in practice. We chose these two benchmark methods because they can also be used to analyze single-subject data to generate subject-level versions of IC maps matched to an established group ICA atlas and the FC between them.  Other methods, such as hierarchical ICA and single-subject ICA, are not considered here because they either cannot be applied to single-subject data or do not produce ICs matched to established ICA atlases. Both of these requirements are essential for a method with clinical applicability, which is our motivation here. The proposed methods are computationally efficient, presenting a realistic alternative to ad-hoc methods.

Compared with existing ICA methods, our Bayesian ICA model has several advantages.  First, it allows for a rectangular mixing matrix, avoiding dimension reduction and reducing the risk of eliminating relevant signals \citep{risk2019linear}. Retaining the full richness of the temporal signals is necessary for precise estimation of the FC. Second, inference is possible through the posteriors, unlike in ad-hoc approaches or machine learning algorithms. Identifying significantly engaged brain locations, or significant functional connectivity pairs, provides a convenient way to summarize ICA results or to identify promising features for biomarker development.

Bayesian ICA also has a scientific advantage over the popular ad-hoc method dual regression. Dual regression obtains subject-level IC time courses, and hence FC, based on \textit{group-average} IC maps.  Misalignment between the individual and group averages may induce systematic bias into FC estimates. Recent work has suggested that functional topography is highly individualized (see Figure \ref{fig:DA:S_ICC}) and that observed differences in FC may be driven by unaccounted-for differences in spatial topography \citep{bijsterbosch2018relationship}. Along with FC, topography is biologically relevant \citep{kong2019spatial}. It is therefore important to disentangle subject-level functional topography from true FC differences, which is possible using Bayesian ICA.

 A limitation of our variational Bayes algorithms is the poor coverage of the FC credible intervals observed in our simulation studies. While this is not surprising given the known issue of posterior variance underestimation with VB, we do not recommend the use of these credible intervals for inference, particularly if using the IW prior as in VB1. The use of our novel pChol prior results in much
better coverage, although there is still room for improvement. One possible reason for the poor coverage of the credible intervals is failure to account for spatial dependencies in the IC maps $\bS$. Previously we developed a Bayesian ICA model with spatial priors on $\bS$ \citep{mejia2023template}. Future work should focus on incorporating spatial priors into this model, a computationally burdensome extension, but one which may be worthwhile to facilitate valid inference at the individual level. For now, we recommend combining ICA results from multiple subjects to perform inference on group averages, group differences, and covariate effects of FC. 

While here we consider FC to be fixed throughout the session, there is growing interest in ``dynamic'' or time-varying FC \citep{lindquist2014evaluating}.  Our Bayesian ICA framework produces IC time courses that can be used for downstream dynamic FC analyses.  Furthermore, we can obtain posterior samples of $\bA$, which can be used to perform Bayesian inference on dynamic FC metrics like dwell times,  particularly if future work produces algorithms that better quantify posterior uncertainty.  Alternatively, it may be possible to explicitly model dynamic FC within the Bayesian ICA model by introducing a time-varying latent FC state. 

One potential application of the proposed framework is for the analysis of small studies. If the focal study population is similar to that of a large repository like the HCP, it may be reasonable to apply an externally-derived prior even if the acquisition and processing protocols differ, since the population-derived priors are designed to encode the population distribution of latent brain features, not noise properties. A thorough investigation of the performance of externally-derived priors is warranted in future work. In other contexts, it may be more appropriate estimate study-specific priors, as we have done previously \citep{gaddis2022psilocybin, derman2023individual}. An interesting alternative is a hierarchical ICA framework, wherein externally-derived priors would be updated based on the data to better represent the specific population or groups being analyzed. This would also facilitate comparing groups and accounting for covariates.

This study has several limitations that should be addressed in future work. First, we have not investigated the biological relevance of the functional brain features we obtain, an important next step. Second, our model assumes temporal residual independence. In our experience, this assumption is fairly reasonable in ICA due to the high number of neural and artifactual signals being accounted for, but this assumption should be checked. Prewhitening, common in task fMRI models, could be a simple solution. Third, our framework does not account for covariates such as age and sex. Doing so at the stage of prior estimation could improve model performance by accounting for some of the between-subject variability, thus reducing prior variance. We are working to account for covariates in future extensions of the model. Finally, here we have analyzed a single-site study of healthy young adults.  More complex datasets, including multi-site datasets and those representing heteogeneous populations, may require additional nuance and care.

As a final point, here we have not performed a thorough investigation of the asymptotic properties of our estimators. Prior theoretical work in ICA mostly considers the effects of $V\to\infty$ on convergence of the square unmixing matrix \citep{risk2019linear, eloyan2013semiparametric}. The effects of $T\to\infty$ are rarely considered, since temporal dimension reduction via PCA is almost always performed as a pre-processing step, in contrast to our context where the temporal dimension is retained. If $T\to\infty$ or $V\to\infty$, this increases not only the sample size, but the number of temporal (or spatial) features that need to be estimated, complicating the asymptotics. If we instead focus on estimation of the FC, the dimensionality is fixed, which may provide a more promising avenue for asymptotic investigation. While this is out of the scope of the current manuscript, it is an important future direction.

\section*{Data Availability}

Code reproducing the simulation and experimental data analysis is available at the following Github repository: \url{https://github.com/mandymejia/FC-TemplateICA-paper/}.  The R package templateICAr is available on Github (\url{https://github.com/mandymejia/templateICAr}) and CRAN (\url{https://cran.r-project.org/web/packages/templateICAr/index.html}).

\bibliographystyle{biorefs}
\bibliography{mybib}

\newpage

\section*{Appendix}

\renewcommand\thesection{\Alph{section}}
\renewcommand\thesubsection{\thesection.\arabic{subsection}}
\renewcommand\thetable{\thesection.\arabic{table}}
\renewcommand\thefigure{\thesection.\arabic{figure}}
\setcounter{figure}{0}
\setcounter{section}{0}
\setcounter{table}{0}

{\black
\section{Population-derived prior on $\bS$}\label{app:priorS}

For the prior on the elements of $\mathbf{S}$, we estimate the prior parameters (mean and variance) based on a training dataset, representing the population from which the focal subject comes.  This process has been described previously  \citep{mejia2020template} and is implemented in the {templateICAr} R package on CRAN. In short, we utilize test-retest fMRI data from a (preferably large) set of subjects or, in the absence of multiple sessions, we create pseudo-test-retest data by splitting the time series into two halves. We assume access to a set of group-average ICA maps, which can be either provided by established studies like the HCP \citep{van2013wu} or estimated from the training data using standard software like MELODIC \citep{beckmann2004probabilistic} or GIFT \citep{calhoun2001method}.  We perform dual regression, a popular ad-hoc method, to obtain noisy estimates of the subject- and session-specific IC maps. We then perform a variance decomposition to obtain a non-negative estimate of the between-subject variance of each IC at each cortical vertex. This non-negative estimate contains a positive bias that converges to zero as the amount of fMRI data for each subject goes to infinity.  Note that dual regression also produces estimates of $\bA$, so we also obtain a set of noisy test-retest FC matrix estimates, i.e. $Cor(\bA)$, which will be used to produce a population-derived prior for $\bG$, as described elsewhere.

\section{Permuted Cholesky prior}\label{app:CP_prior}
Let $\bX_i$ be a FC matrix for one training session $i$, and let $\bX_i=\bL_i\bL_i^\top$ be its Cholesky factorization, where $\bL_i$ is lower-triangular. Note that for a correlation matrix, the sum of squares of each row of $\mathbf{L}_i$ must equal 1, and the diagonal values must be positive.  The sum-of-squares constraint also implies that no entry can have magnitude greater than $1$. To satisfy the element-wise constraints in the sampling procedure below, we first apply appropriate transformations to map the values to the real line: logit transformation for the diagonal values and Fisher z-transformation for the off-diagonal values.

It may be tempting to build individual univariate priors on the elements of $\bL$, but these would not capture complex dependencies between the elements of $\bL$. Instead, we first perform principal component analysis (PCA) and build univariate priors on the PCA scores. Let $\bm_i\in\mathbb{R}^p$, $p=Q(Q+1)/2$ be a vector of the transformed lower triangular elements of $\bL_i$. Form the matrix $\bM\in\mathbb{R}^{N\times p}$ with $i$th row $\bm_i^\top$, where $N$ is the number of training sessions. With some slight abuse of notation, let $\bM$ be centered. Through singular value decomposition, we have $\bM = \bU\bD\bV^\top$, where $\bV\in\mathbb{R}^{p\times k}$ contains $k\leq p$ principal components and $\bU\in\mathbb{R}^{N\times k}$ contain session-specific scores associated with those principal components. We retain all principal components ($k=p$, assuming $\bM$ is full rank) to avoid eliminating subtle individual differences, since our goal is not dimension reduction but to capture multivariate dependencies between the Cholesky elements while facilitating univariate sampling.

Conveniently, the columns of $\bU$ are orthonormal, so that we can reasonably assume them to arise from independent random variables with mean zero and a common variance.  They also tend to exhibit Gaussianity in practice. Thus, we assume a univariate Gaussian prior on each of the columns of $\bU$, from which we draw samples $\bu_*$. We obtain corresponding samples $\bm_* = \bu_*^\top\bD\bV^\top$. To produce a sample Cholesky factor $\bL_*$, we add back the mean lost in centering $\bM$ to $\bm_*$, reverse-transform the values to the original scale, and rescale each row to satisfy the sum-of-squares constraint. We finally produce a sample correlation matrix as $\bX_* = \bL_*\bL_*^\top$.

This procedure satisfies most of our requirements: it produces samples $\bX_*$ that are correlation matrices, and those samples approximate the element-wise mean and variance in the training sample.  However, it produces biased element-wise variance, with certain elements having too-low variance, and other elements having too-high variance. Specifically, correlation elements that are a function of more Cholesky elements have lower variance due to averaging.  To mitigate this, we randomly permute the rows and columns of the training samples $\bX_i$ prior to Cholesky factorization. Let $\bP$ be a $Q\times Q$ permutation matrix, and let $\bX_P = \bP\bX\bP^\top$ be a permuted version of a given training sample $\bX$. Let $\bL_P$ be the Cholesky factor associated with $\bX_P$, such that $\bm_*=\bL_P\bL_P^\top$, and note that $\bX = \bP^\top\bL_P\bL_P^\top\bP$. We can thus apply the PCA-based procedure described above to generate samples of $\bL_P$ for a set of randomly generated permutation matrices to generate samples of $\bX$.  Letting $\bL_{P*}$ be a sample permuted Cholesky factor produced as described above based on the permuted correlations $\bX_P=\bP\bX$ from the training sample.  We produce a sample correlation matrix as $\bX_* = \bP^\top\bL_*\bL_*^\top\bP$.  A sufficient number of random permutations should be generated to effectively mitigate the bias patterns in the element-wise variance. In our experiments, we find $100$ permutations to be effective, but the required number will likely vary with the characteristics of the training sample and the dimensionality of the correlation matrices.

\section{VB Derivation}\label{app:VB}

{\black In our VB estimation strategy, we assume that the joint posterior factorizes over the spatial IC maps $\bS$, the IC mixing matrix $\bA$, and the noise variance $\tau^2$.  Here we derive the approximate posteriors for those four groups of variables.  For $\bA$, the approximate posterior depends on the choice of hyperprior for $\bG$: the Inverse-Wishart, which is conjugate (VB1), or our proposed informative prior for correlation matrices, which requires sampling from $p(\bG)$ (VB2).

Recall that the FC template ICA model is given by
\begin{align*}
    \by_v &= \bA \bs_v + \be_v, \mbox{ where } \be_v \sim N(0, \tau^2\bI_v), \\
    \bs_v &= \bs^0_v + \bdelta_v, \mbox{ where } \bdelta_v \sim N(0, \bD_v), \\
    \ba_t &\sim MVN(\mathbf{0}, \bG), \mbox{ where } \bG \sim p(\bG), \\
    \tau^2 &\sim InverseGamma(\alpha_0, \beta_0).
\end{align*}
}

\subsection{Approximate posterior of $\bS$}

{\black The density $q(\bs|\bY)$ factorizes over locations $v$, with the approximate posterior for $\bs_v$ given by
\begin{align*}
\log \ q(\bs_v|\bY) 
&= \EX_{\bA, \tau^2} \left[ \log p(\by_v|\bA, \bs, \tau^2) + \log p(\bs) \right] + const \\
&\propto -\frac{1}{2} \EX_{\bA, \tau^2} \left[ \frac{1}{\tau^2}\big(\by_v - \bA\bs_v\big)^\top\big(\by_v - \bA\bs_v\big) + (\bs_v - \bs_v^0)^\top \bD_v^{-1} (\bs_v - \bs_v^0) \right] \\
&\propto -\frac{1}{2}  \EX_{\bA, \tau^2} \left[ 
    \bs_v^\top\big(\frac{1}{\tau^2}\bA^\top\bA + \bD_v^{-1}\big)\bs_v 
    -2\bs_v^\top\big(\frac{1}{\tau^2}\bA^\top\by_v + \bD_v^{-1}\bs_v^0\big)
    \right].
\end{align*}

\noindent Hence, $q(\bs_v|\bY)$ is Normal with mean and covariance given by 
$$
\hat\bs_{v} = \VX(\bs_{v}) \left(\frac{1}{\hat\tau^2}\hat\bA^\top\by_v + \bD_v^{-1}\bs_v^0 \right)\quad\text{and}\quad
\VX(\bs_{v}) = \left( \frac{1}{\hat\tau^2}\EX[\bA^\top\bA] +\bD_v^{-1}  \right)^{-1},
$$
\noindent where $\hat\tau^2$ is the mean of the approximate posterior of $\tau^2$, the $t$th row of $\hat\bA$ is the mean of the approximate posterior of $\ba_t$, and
$
\EX[\bA^\top\bA] =  \sumt \EX\big[\ba_t\ba_t^\top \big] 
= \sumt\big[\VX(\ba_t) + \hat\ba_t\hat\ba_t^\top\big].
$ 

\subsection{Approximate posterior of $\bA$}

\subsubsection{Inverse-Wishart prior on $\bG$.}

For VB1 where $p(\bG)\sim IW(\bPsi_0,\nu_0)$, we first derive the marginal prior for $\ba_t$, 
$
p(\ba_t) = \int_{\bG} p(\ba_t|\bG) p(\bG) d\bG,
$
which is a multivariate $t$ distribution:
\begin{align*}
p(\ba_t) &= \int_{\bG}  p(\ba_t|\bG) p(\bG) d\bG \\
&\propto \int_{\bG} 
|\bG|^{-\frac{1}{2}}\text{exp}\left(-\tfrac{1}{2}\ba_t^\top\bG^{-1}\ba_t\right) 
|\bG|^{-\frac{1}{2}(\nu_0+Q+1)}\text{exp}\left(-\tfrac{1}{2}\text{Tr}\{\bPsi_0\bG^{-1}\}\right) d\bG \\
&=  \int_{\bG} 
|\bG|^{-\frac{1}{2}(\nu_0+Q+2)}\text{exp}\left(-\tfrac{1}{2}\text{Tr}\{(\bPsi_0 + \ba_t\ba_t^\top)\bG^{-1}\}\right) d\bG \\
&\propto \big|\bPsi_0 + \ba_t\ba_t^\top\big|^{-\frac{1}{2}(\nu_0+1)} 
\quad\text{(since the integrand is proportional to an IW density)}\\
&\propto \big|\bI + \bPsi_0^{-1}\ba_t\ba_t^\top\big|^{-\frac{1}{2}(\nu_0+1)}  
= \big(1+\ba_t^\top\bPsi_0^{-1}\ba_t\big)^{-\frac{1}{2}(\nu_0+1)} 
\propto t_{\nu_a}(\bzero, \bSigma_a),
\end{align*}
\noindent with $\nu_a = \nu_0+1-Q$ and $\bSigma_a = \nu_a^{-1}\bPsi_0$, which can be represented as a scale-mixture of Normals:  if $u\sim Gamma(\nu_a/2,\nu_a/2)$ (shape-rate parameterization), then $p(\ba_t|u) \sim N(\bzero, u^{-1}\bSigma_a)$. Hence, conditional on $u$, the approximate posterior of $\ba$ is Gaussian. First, we re-write the likelihood in terms of $\ba = (\ba_1^\top,\dots,\ba_T^\top)^\top$, defining $\by = (\by_1^\top,\dots,\by_T^\top)^\top$, $\bS_\otimes = \bI_T \otimes \bS$, and $\be = (\be_1^\top,\dots,\be_T^\top)^\top$:
$$
\by = \bS_\otimes^\top\ba + \be,\quad \be\sim MVN(\bzero, \tau^2\bI_{TV}) 
$$
\noindent Conditionally on $u$, the approximate log posterior for $\ba$ is given by
\begin{align*}
\log q(\ba|\bY,u) &\propto \EX_{\bS,\tau^2}\big[\log p(\ba|\bY,\bS,\tau^2)\big] + \sumt \log p(\ba_t|u) \\
&\propto -\frac{1}{2}\EX_{\bS,\tau^2}\left[\frac{1}{\tau^2}(\by-\bS_\otimes^\top\ba)^\top(\by-\bS_\otimes^\top\ba)\right] -\frac{1}{2}\sumt \ba_t^\top(u^{-1}\bSigma_a)^{-1}\ba_t \\
&\propto -\frac{1}{2}\sumt\EX_{\bS,\tau^2}\left[\frac{1}{\tau^2}(\ba_t^\top\bS\bS^\top\ba_t - 2\ba_t^\top\bS\by_t)\right] -\frac{1}{2}\sumt \ba_t^\top(u\nu_a\bPsi_0^{-1})\ba_t \\
&= -\frac{1}{2}\sumt \left\{ \ba_t^\top\left( \frac{1}{\hat\tau^2}\EX[\bS\bS^\top] + u\nu_a\bPsi_0^{-1}\right)\ba_t - 2\ba_t^\top \left(\frac{1}{\hat\tau^2}\hat\bS\by_t\right) \right\}.
\end{align*}

\noindent Thus, $q(\ba|\bY,u)$ is Normal with mean and covariance given by
$$
\EX[\ba_t|u] = \VX(\ba_t|u)\left(\frac{1}{\hat\tau^2}\hat\bS\by_t\right)
\quad\text{and}\quad
\VX(\ba_t|u) = \Big( \frac{1}{\hat\tau^2}\EX[\bS\bS^\top] + u\nu_a\bPsi_0^{-1}\Big)^{-1},
$$
\noindent where $\hat\tau^2$, $\hat\bS$, and $\EX[\bS\bS^\top]$ are approximate posterior moments of $\tau^2$ and $\bS$, respectively. To update those approximate posteriors, we require $\EX[\ba_t]$ and $\VX(\ba_t)$, which we can obtain through law of total expectation/covariance:
\begin{align*}
    \hat\ba_t = \EX[\ba_t] &= \EX_u\Big[\EX[\ba_t|u]\Big] 
    = \EX_u\left[ \Big( \frac{1}{\hat\tau^2}\EX[\bS\bS^\top] + u\nu_a\bPsi_0^{-1}\Big)^{-1}\right]
    \left(\frac{1}{\hat\tau^2}\hat\bS\by_t\right), \\
    \VX(\ba_t) &= \EX_u \big[ Cov(\ba_t|u) \big] + \VX_u\big(\EX[\ba_t|u]\big) \\
    &= \EX_u\left[ \Big( \frac{1}{\hat\tau^2}\EX[\bS\bS^\top] + u\nu_a\bPsi_0^{-1}\Big)^{-1}\right] 
    + \VX_u\left( \Big( \frac{1}{\hat\tau^2}\EX[\bS\bS^\top] + u\nu_a\bPsi_0^{-1}\Big)^{-1} \left(\frac{1}{\hat\tau^2}\hat\bS\by_t\right)\right),
\end{align*}
which are estimated via Monte Carlo using samples from $u \sim Gamma(\nu_a/2, \nu_a/2)$.

Recall that the FC matrix $Cov(\bA)$ is a $Q\times Q$ matrix representing the temporal synchrony between the columns of the $T\times Q$ matrix $\bA$.  We estimate $Cov(\bA)$ as the empirical correlation using the posterior mean of $\bA$. For each Gamma sample $u$, 
$
Cov(\hat{\bA}_u) = \bB_u Cov(\hat{\bA}_0)\bB_u,
$
where the $t$th row of $\hat\bA_0$ is $\hat\tau^{-2}\hat\bS\by_t$, and $\bB_u = \Big( \frac{1}{\hat\tau^2}\EX[\bS\bS^\top] + u\nu_a\bPsi_0^{-1}\Big)^{-1}$. These $Cov(\hat\bA_u)$ form a set of samples for $Cov(\bA)$. Based on these samples, we can construct posterior credible intervals for individual elements of $Cov(\bA)$ to identify significant positive and negative connections.

\subsubsection{Permuted Cholesky Prior on $\bG$.} For our novel CP prior for correlation matrices, we have samples from $p(\bG)$. Conditional on $\bG$, the approximate log posterior for $\ba$ factorizes over $t$, and
\begin{align*}
\log q(\ba_t|\bY,\bG) &\propto \EX_{\bS,\tau^2}\big[\log p(\ba_t|\by_t,\bS,\tau^2)\big] + \log p(\ba_t|\bG) \\
&\propto -\frac{1}{2}\EX_{\bS,\tau^2}\left[\frac{1}{\tau^2}(\by_t-\bS^\top\ba_t)^\top(\by_t-\bS^\top\ba_t)\right] -\frac{1}{2}\ba_t^\top\bG^{-1}\ba_t \\
&\propto -\frac{1}{2} \EX_{\bS,\tau^2}\left[\frac{1}{\tau^2}(\ba_t^\top\bS\bS^\top\ba_t - 2\ba_t^\top\bS\by_t)\right] -\frac{1}{2} \ba_t^\top\bG^{-1}\ba_t \\
&= -\frac{1}{2} \left\{ \ba_t^\top\left( \frac{1}{\hat\tau^2}\EX[\bS\bS^\top] + \bG^{-1}\right)\ba_t - 2\ba_t^\top \left(\frac{1}{\hat\tau^2}\hat\bS\by_t\right) \right\},
\end{align*}
\noindent so $q(\ba_t|\bY,\bG)$ is Normal with mean and covariance given by
$$
\EX[\ba_t|\bG] = \VX(\ba_t|\bG)\left(\frac{1}{\hat\tau^2}\hat\bS\by_t\right)
\quad\text{and}\quad
\VX(\ba_t|\bG) = \Big( \frac{1}{\hat\tau^2}\EX[\bS\bS^\top] + \bG^{-1}\Big)^{-1}.
$$
To compute $\EX[\ba_t]$ and $\VX(\ba_t)$, we use the law of total expectation and law of total covariance. Defining $\bV_k = \Big( \frac{1}{\hat\tau^2}\EX[\bS\bS^\top] + \bG_k^{-1}\Big)^{-1}$:
\begin{align*}
    \hat\ba_t = \EX[\ba_t] &= \EX_{\bG}\Big[\EX[\ba_t|\bG]\Big] 
    =\frac{1}{K} \sum_{k=1}^K \bV_k\left(\frac{1}{\hat\tau^2}\hat\bS\by_t\right) 
    = \left(\frac{1}{K} \sum_{k=1}^K \bV_k\right)\left(\frac{1}{\hat\tau^2}\hat\bS\by_t\right)\\
    \VX(\ba_t) &= \EX_{\bG} \Big[ \VX(\ba_t|\bG) \Big] + \VX_{\bG}\Big(\EX[\ba_t|\bG]\Big) 
    = \frac{1}{K}\sum_{k=1}^K \bV_k + Cov_k\left( \frac{1}{\hat\tau^2}\bV_k\hat\bS\by_t\right).
\end{align*}
Here, $\bV_k$ can be rewritten, using the Cholesky factorization $\bE := \frac{1}{\hat\tau^2}\EX[\bS\bS^\top] = \bL_E\bL_E^\top$, as
$$
\bV_k = \Big( \frac{1}{\hat\tau^2}\EX[\bS\bS^\top] + \bG_k^{-1}\Big)^{-1}
= \Big( \bL_E\bL_E^\top + \bG_k^{-1}\Big)^{-1} 
= \bL_E^{-\top}\Big( \bI_Q + \bL_E^{-1}\bG_k^{-1}\bL_E^{-\top}\Big)^{-1} \bL_E^{-1}.
$$

\noindent The only difficult part is computing $\Big( \bI_Q + \bL_E^{-1}\bG_k^{-1}\bL_E^{-\top}\Big)^{-1}$ for many $k$, since $\bL_E \in \mathbb{R}^{Q\times Q}$ and its inverse are easy to compute, and the $\bG_k^{-1}$ are pre-computed for each sample $k$. For this purpose, we will use the following approximation:
$$
\left(\bI + \bA \right)^{-1} \approx
\bI - \bA + \bA^2,
$$

\noindent which holds if the eigenvalues of $\bA$ all have magnitude less than $1$. Note the following facts: 
(1) $eig(\bL_E^{-1}\bG_k^{-1}\bL_E^{-\top}) = eig(\bL_E^{-\top}\bL_E^{-1}\bG_k^{-1}) = eig(\bE^{-1}\bG_k^{-1})$; 
(2) the maximum eigenvalue of a product of two matrices satisfies $\lambda_{max}(\bA\bB) \leq \lambda_{max}(\bA)\lambda_{max}(\bB)$;
(3) the minimum eigenvalue satisfies ${\lambda_{min}(\bA\bB) \leq \lambda_{min}(\bA)\lambda_{min}(\bB)}$; and
(4) $\bG_k$ and $\bE$ are positive definite so all eigenvalues are positive. 
Therefore, the above approximation therefore holds as long as
$
\lambda_{max}(\bE^{-1})\lambda_{max}(\bG_k^{-1}) < 1.
$

While this is somewhat conservative relative to the condition $\lambda_{max}(\bE^{-1}\bG_k^{-1}) < 1$, it is much more efficient since $\lambda_{max}(\bG_k^{-1})$ can be computed during template estimation, so that only $\lambda_{max}(\bE^{-1})$ needs to be computed at each VB iteration. Hence, for each sample $k$, after checking that the above condition holds, we can compute
$$
\bV_k = \Big( \bE + \bG_k^{-1}\Big)^{-1} 
\approx \bL_E^{-\top}\Big( \bI_Q - \bL_E^{-1}\bG_k^{-1}\bL_E^{-\top} + \bL_E^{-1}\bG_k^{-1}\bL_E^{-\top}\bL_E^{-1}\bG_k^{-1}\bL_E^{-\top}\Big)\bL_E^{-1},
$$

\noindent which makes computing the sums involved in $\EX[\ba_t]$ and $\VX(\ba_t)$ straightforward.  At each VB iteration, we exclude any samples $\bG_k$ for which the condition above does not hold.  In practice, we find that the condition holds in the vast majority (over $99.5\%$) of samples. Finally, to improve accuracy, at the final iteration we compute $\bV_k$ exactly using all samples.

\subsection{Approximate posterior of $\tau^2$}

For $\tau^2$, $\log q(\tau^2|\bY)$ is given by
\begin{align*}
\log q(\tau^2|\bY) &= \EX_{\bA, \bS} \left[ \sumv\sumt \log p(y_{vt}|\ba_t, \bs_v, \tau^2) + \log p(\tau^2) \right] + const \\
&= \EX_{\bA, \bS} \Bigg[ \sumv\sumt \log g(y_{tv}:\ba_t^\top \bs_v,\tau^2) + \log p(\tau^2) \Bigg] \\
&\propto \EX_{\bA, \bS} \Bigg[ -\frac{TV}{2}\log(\tau^2) - \frac{1}{2\tau^2} \sumv\sumt (y_{tv} - \ba_t^\top \bs_v)^2  -(\alpha_0+1)\log(\tau^2)-\frac{\beta_0}{\tau^2} \Bigg] \\
&= -\left( \alpha_0 + \frac{TV}{2} + 1 \right)\log(\tau^2) 
-\frac{1}{\tau^2}\Bigg\{\beta_0 + \frac{1}{2}\sumv\sumt y_{tv}^2 - \sumv\Big(\sumt y_{tv}\hat\ba_t^\top\Big)\hat\bs_v \\
&\qquad + \frac{1}{2}\EX\Bigg[ \sumv\bs_v^\top\Big(\sumt \ba_t\ba_t^\top\Big)\bs_v \Bigg]\Bigg\}.
\end{align*}
Using the trace-expectation trick, the last term can be written as 
$$
    \sumv Tr\left\{\EX\Big[ \Big(\sumt \ba_t\ba_t^\top\Big)\bs_v\bs_v^\top \Big]\right\}
    = Tr\left\{ \sumt \EX\Big[\ba_t\ba_t^\top\Big] \sumv \EX \Big[\bs_v\bs_v^\top\Big] \right\}
    = Tr\left\{ \EX\Big[\bA^\top\bA\Big] \EX \Big[\bS\bS^\top\Big] \right\},
$$
\noindent which only involves the second moments of $\ba_t$ and $\bs_v$, which are given above. Therefore, $q(\tau^2|\bY) \sim IG(\alpha,\hat\beta)$ with mean $\hat\tau^2 = \frac{\hat\beta}{\alpha - 1}$, where $\alpha = \alpha_0 + \frac{TV}{2}$ and
$$
    \hat\beta = \beta_0 + \frac{1}{2}\sumv\sumt y_{tv}^2 - \sumv\Big(\sumt y_{tv}\hat\ba_t^\top\Big)\hat\bs_v + \frac{1}{2}Tr\left\{ \EX\Big[\bA^\top\bA\Big] \EX \Big[\bS\bS^\top\Big] \right\}.
$$
}

\subsection{Effective sample size adjustment}

We estimate the ESS of $\bA$ using the initial estimate of $\bA$.  For each column of $\bA$, we use the Yule-Walker equations \citep{brockwell1991time} to fit an autoregressive model of order 10. This is considered a high model order in fMRI analysis and has been shown to be more than sufficient to capture its temporal autocorrelation \citep{parlak2023sources}. We then construct the temporal covariance matrix $\bSigma$ based on the estimated AR coefficients and estimate the temporal effective sample size as $T_\text{eff}=\text{Tr}(\bSigma)^2/\text{Tr}(\bSigma^2)$ \citep{bretherton1999effective}. $T_\text{eff}$ can be used to adjust posterior moments of $\bA$ involving sums over $t$, i.e. $\EX[\bA^\top\bA]=\sumt \EX\Big[\ba_t\ba_t^\top\Big]$ appearing in $V(\bs_{v})$ is replaced with $({T_\text{eff}}/{T})\sumt \EX\Big[\ba_t\ba_t^\top\Big]$.
}

\newpage
\section{Additional Simulation Figures}
\label{app:sim_results}

\subsection{Data Generation}
\label{app:sim_data_generation}

\begin{figure}[H]
    \centering
    \begin{tabular}{ccccccl}
    & IC1  & IC2 & IC3 & IC4 & IC5 & \\
    \hline \\[-10pt]
    \begin{picture}(10,95)\put(5,45){\rotatebox[origin=c]{90}{Generating Mean}}\end{picture} &
    \includegraphics[width=0.9in, trim=0 1in 0 1in, clip]{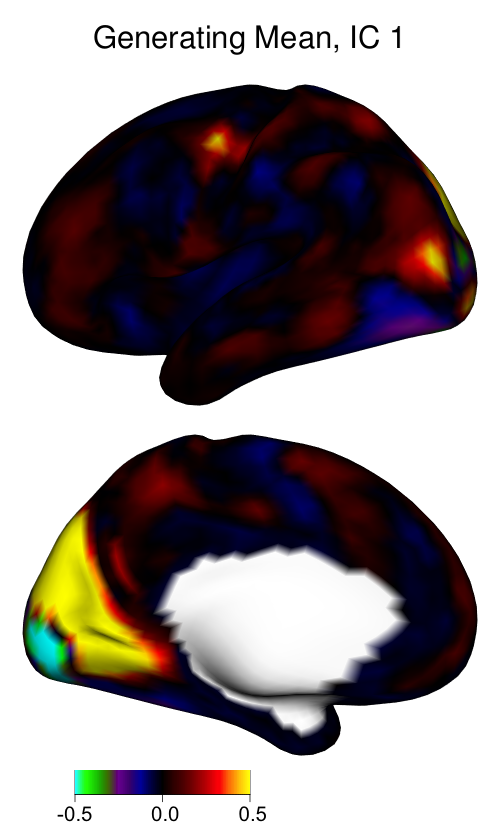} &
    \includegraphics[width=0.9in, trim=0 1in 0 1in, clip]{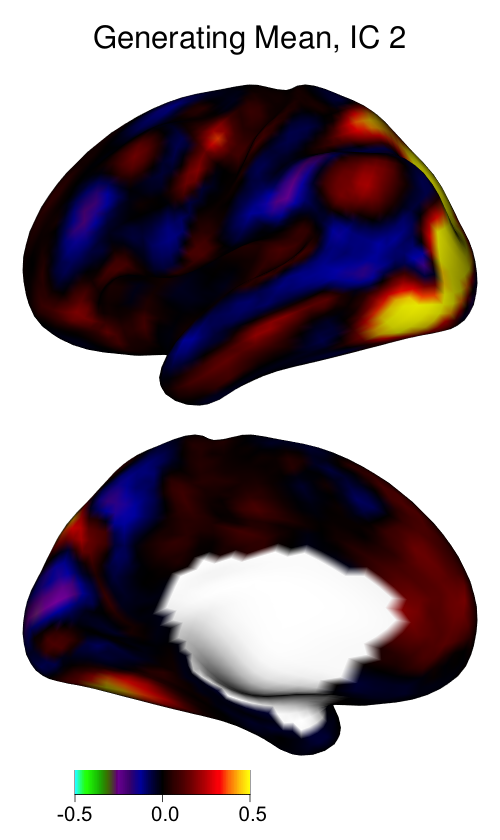} &
    \includegraphics[width=0.9in, trim=0 1in 0 1in, clip]{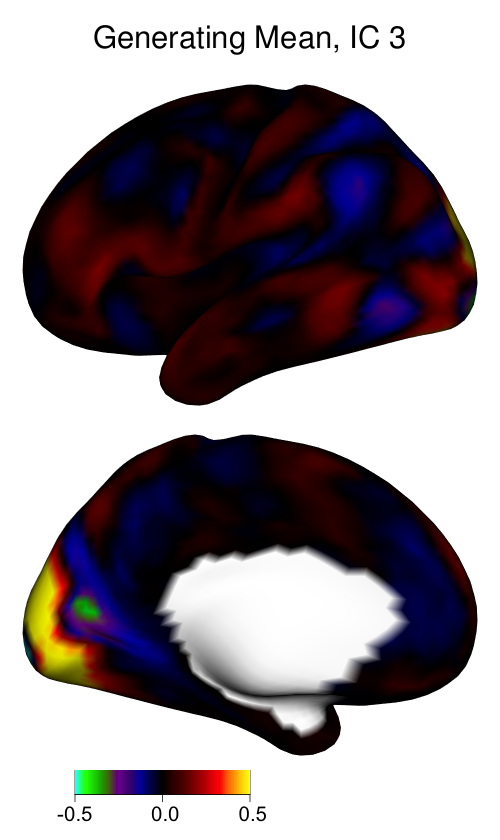} &
    \includegraphics[width=0.9in, trim=0 1in 0 1in, clip]{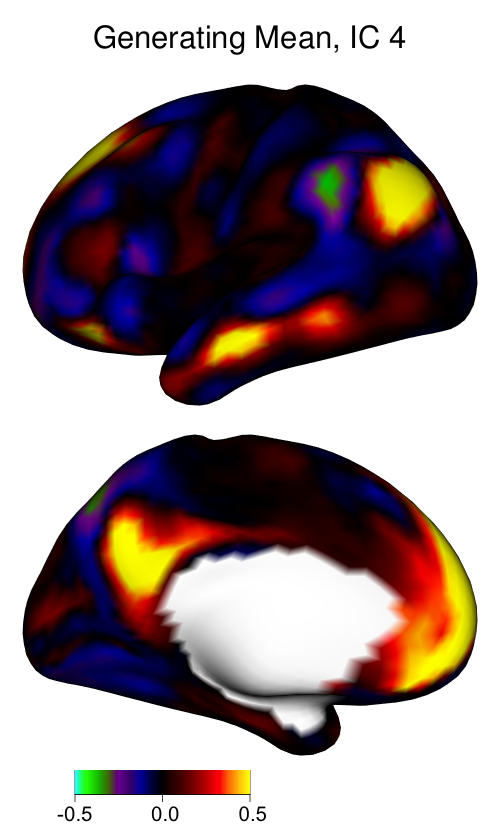} &
    \includegraphics[width=0.9in, trim=0 1in 0 1in, clip]{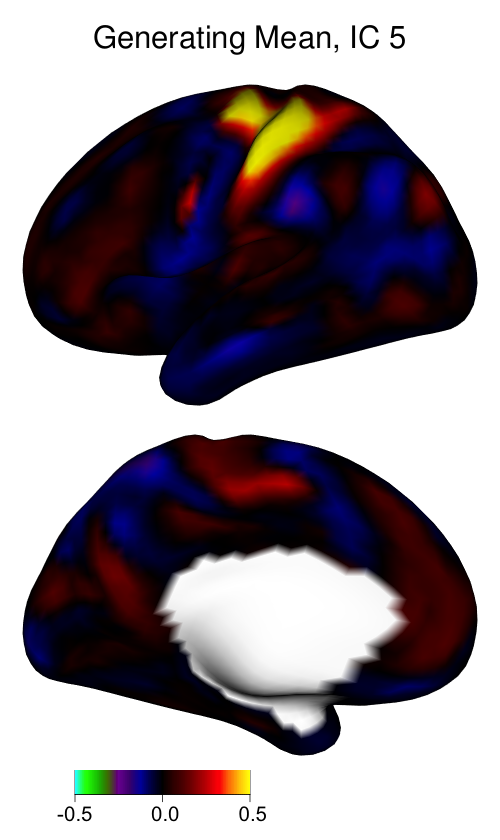} &
    \includegraphics[width=0.9in, angle=90, trim=-1cm 5mm 3in 10.5in, clip]{simulation/templates/generating_mean5.png} \\
    \hline \\[-12pt]
    \begin{picture}(10,95)\put(5,45){\rotatebox[origin=c]{90}{Generating Variance}}\end{picture} &
    \includegraphics[width=0.9in, trim=0 1in 0 1in, clip]{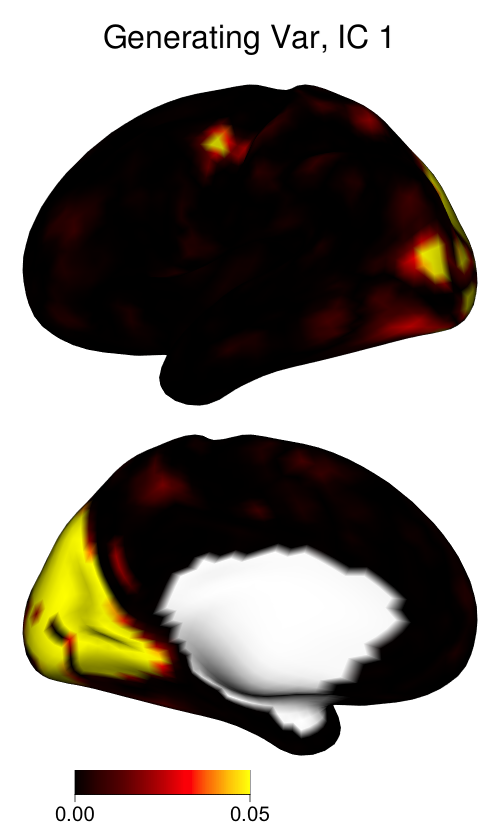} &
    \includegraphics[width=0.9in, trim=0 1in 0 1in, clip]{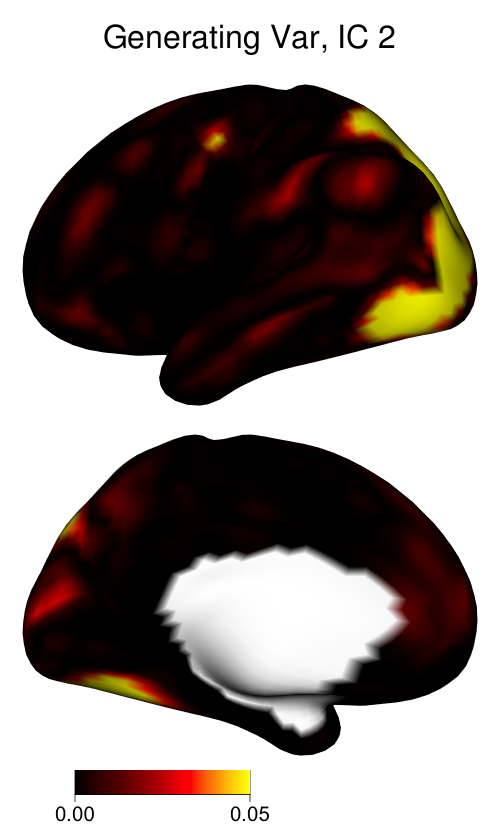} &
    \includegraphics[width=0.9in, trim=0 1in 0 1in, clip]{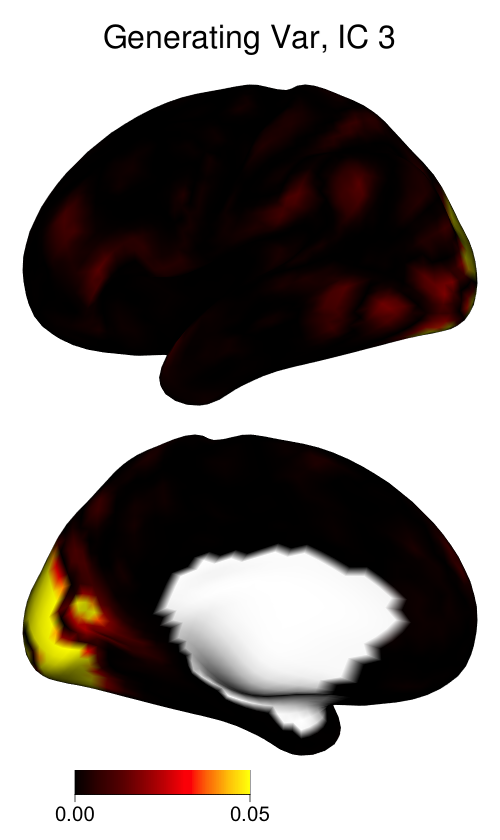} &
    \includegraphics[width=0.9in, trim=0 1in 0 1in, clip]{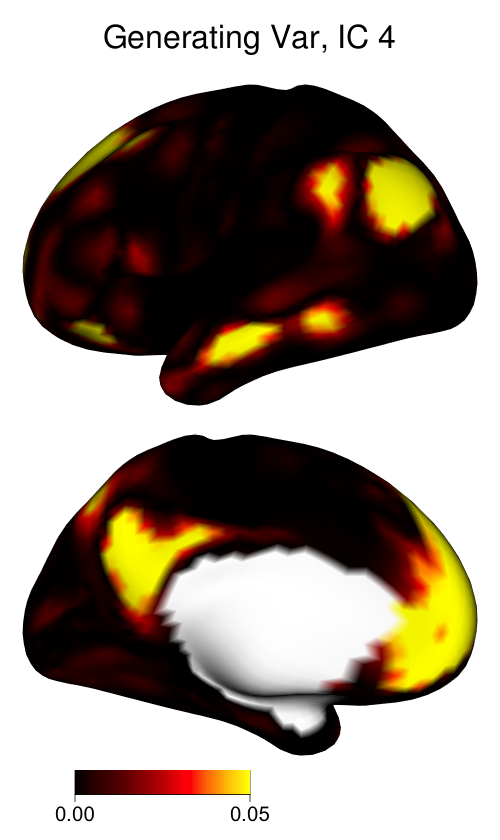} &
    \includegraphics[width=0.9in, trim=0 1in 0 1in, clip]{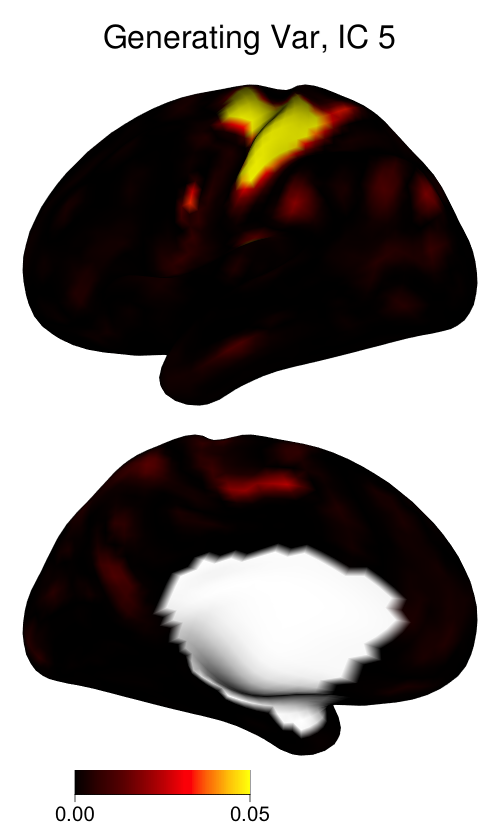} &
    \includegraphics[width=0.9in, angle=90, trim=-1cm 5mm 3in 10.5in, clip]{simulation/templates/generating_var5.png} \\
    \hline \\[-12pt]
    \begin{picture}(10,95)\put(5,45){\rotatebox[origin=c]{90}{Example Deviations}}\end{picture} &
    \includegraphics[width=0.9in, trim=0 1in 0 1in, clip]{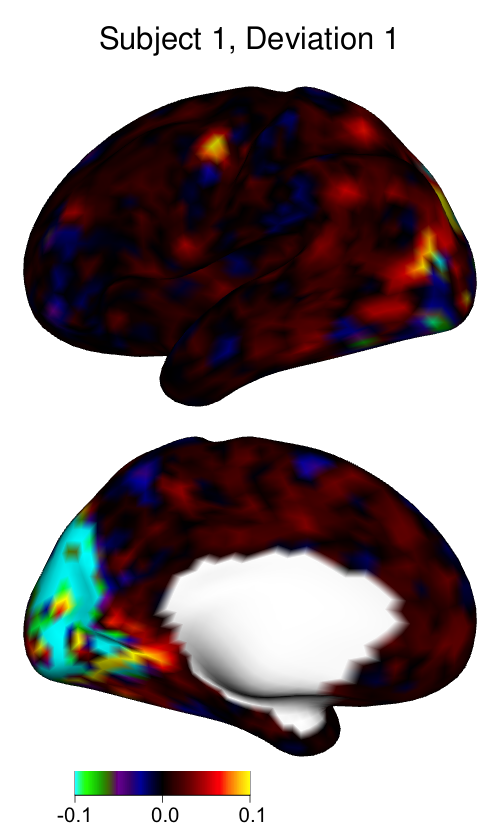} &
    \includegraphics[width=0.9in, trim=0 1in 0 1in, clip]{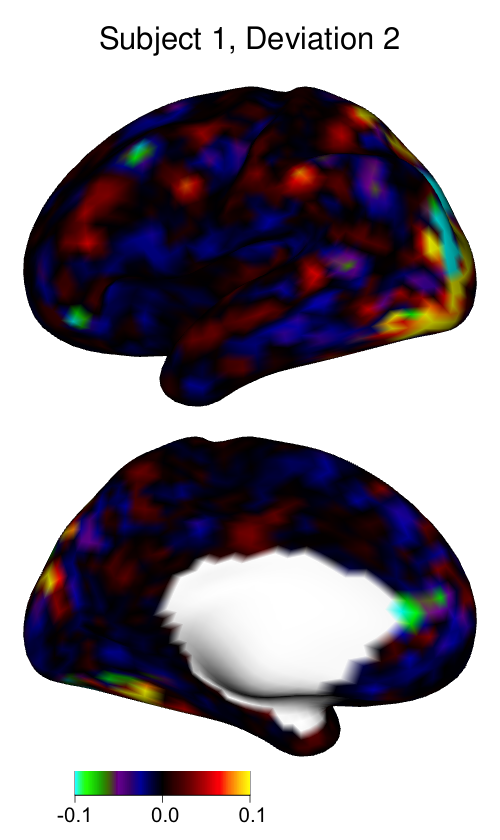} &
    \includegraphics[width=0.9in, trim=0 1in 0 1in, clip]{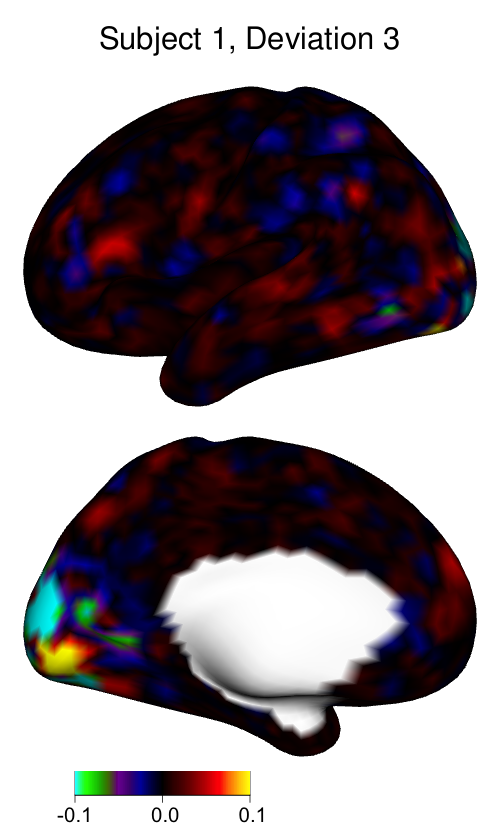} &
    \includegraphics[width=0.9in, trim=0 1in 0 1in, clip]{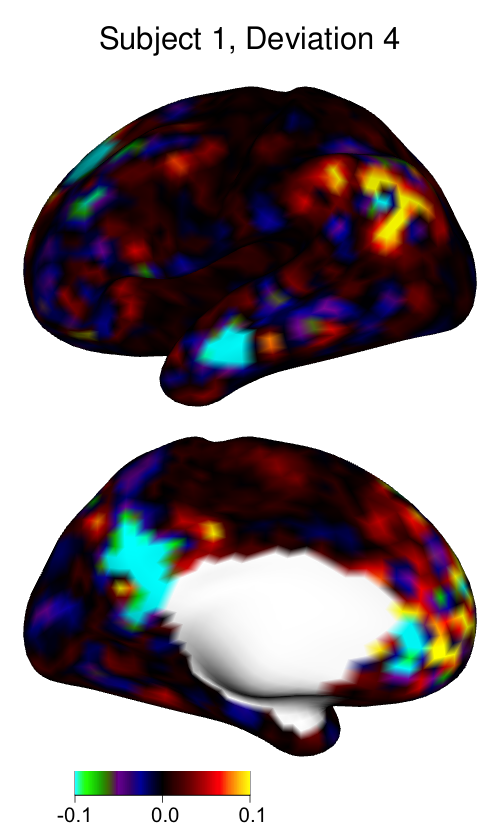} &
    \includegraphics[width=0.9in, trim=0 1in 0 1in, clip]{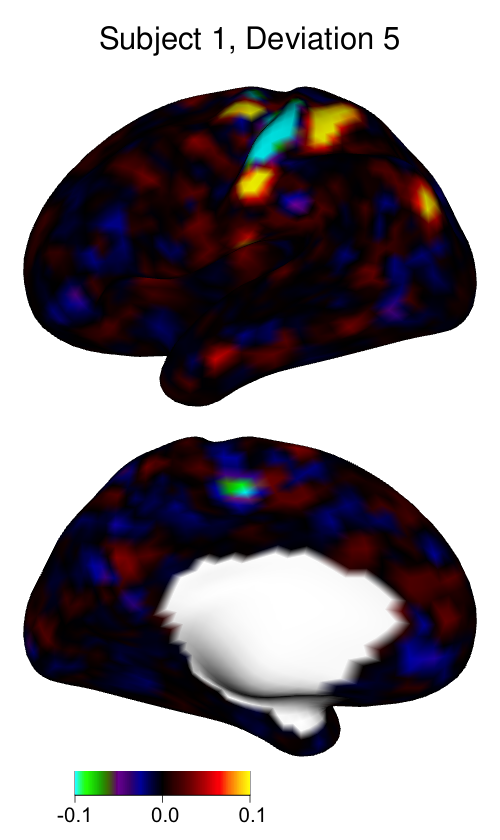} &
    \includegraphics[width=0.9in, angle=90, trim=-1cm 5mm 3in 10.5in, clip]{simulation/exampleICs/subject1_dev5.png} \\
    \hline \\[-12pt]
    \begin{picture}(10,95)\put(5,45){\rotatebox[origin=c]{90}{Example ICs}}\end{picture} &
    \includegraphics[width=0.9in, trim=0 1in 0 1in, clip]{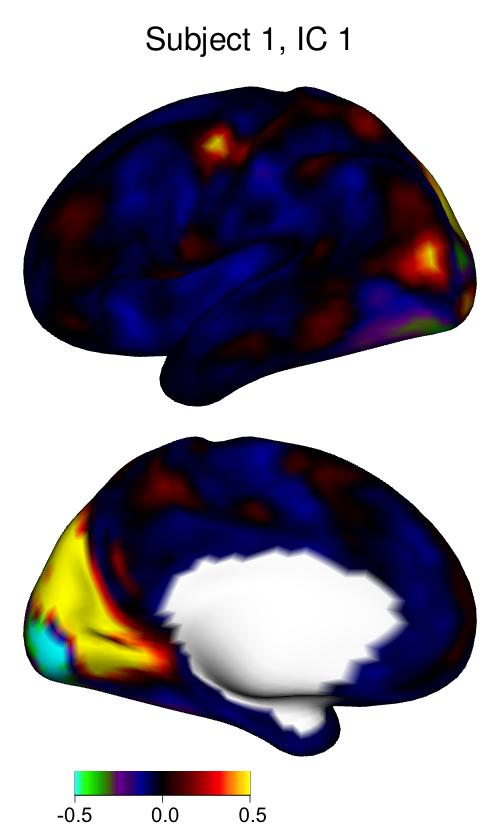} &
    \includegraphics[width=0.9in, trim=0 1in 0 1in, clip]{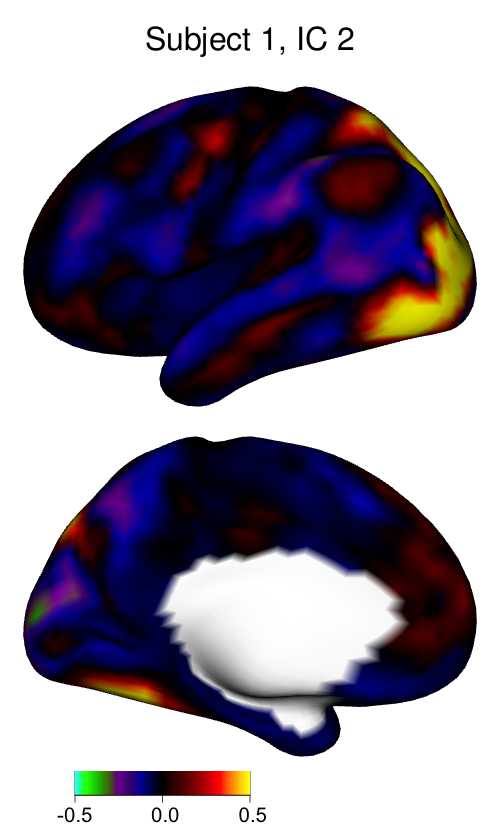} &
    \includegraphics[width=0.9in, trim=0 1in 0 1in, clip]{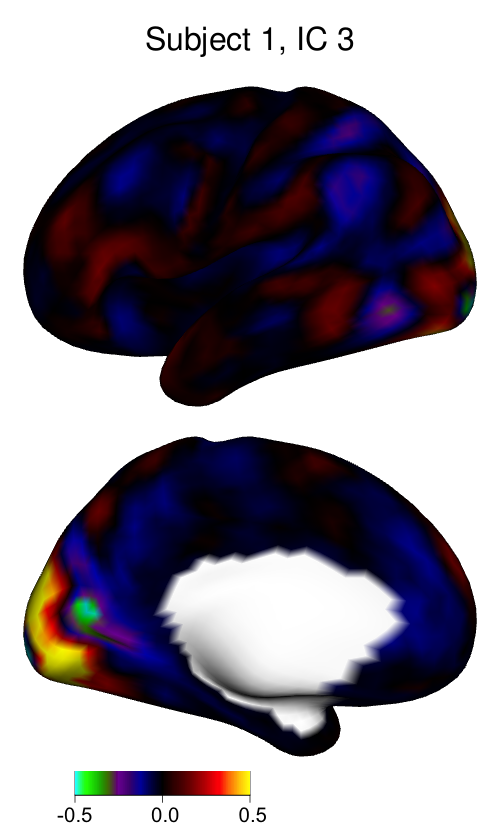} &
    \includegraphics[width=0.9in, trim=0 1in 0 1in, clip]{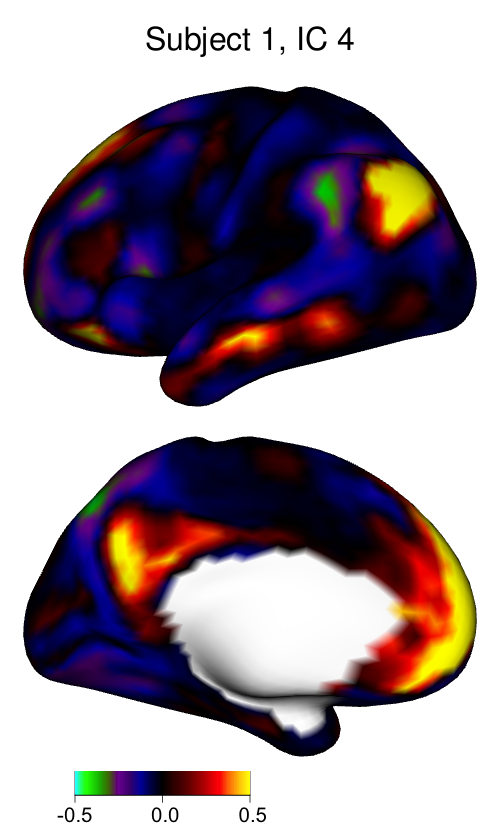} &
    \includegraphics[width=0.9in, trim=0 1in 0 1in, clip]{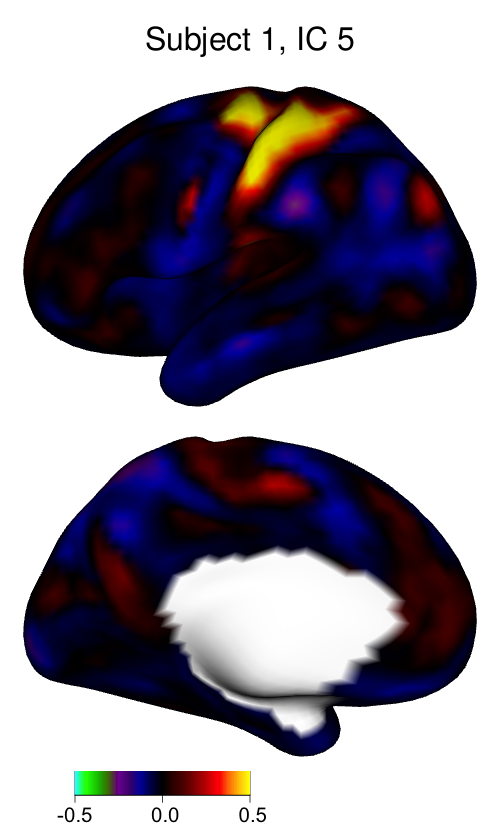} &
    \includegraphics[width=0.9in, angle=90, trim=-1cm 5mm 3in 10.5in, clip]{simulation/exampleICs/subject1_IC5.png} \\
    \hline \\[-20pt]
    \end{tabular}
    \caption{\textit{Generation of simulated subject-level IC maps.} Unique subject-level deviation maps (third row) are created by generating mean-zero Normal draws with variance equal to the generating variance, then spatially smoothing the resulting maps along the surface. Unique subject-level IC maps (fourth row) are created by adding those deviation maps to the generating mean maps. The result is subject-specific ICs that are similar to the group-average ICs but with subtle differences.}
    \label{fig:sim:subjICs}
\end{figure}

Figure \ref{fig:sim:subjICs} illustrates the data generating process for the simulated spatial IC maps.  We first select five real group-average ICs from the Human Connectome Project (HCP) 25-component group ICA \citep{van2013wu} representing three visual network components (IC1, IC2 and IC3), one default mode network (DMN) component (IC4), and one motor network component (IC5). We use these as the basis for the generating mean and set the generating standard deviation (SD) proportional to the mean. For computational convenience, we focus on the left hemisphere and resample (interpolate) to approximately 3000 cortical vertices. To generate a unique subject-level version of each IC, we draw Normal mean-zero samples with the generating standard deviation at each cortical vertex. We then spatially smooth each deviation map using a surface-based Gaussian kernel with 8mm full width at half maximum (FWHM). Smoothing and resampling were performed using the Connectome Workbench \citep{marcus2011informatics} via the {ciftiTools} R package \citep{pham2022ciftitools}.  We then add the resulting map, representing the subject-specific deviation, to the generating mean to produce the subject-specific IC maps.  

\begin{figure}[H]
\centering
\includegraphics[width=1\linewidth]{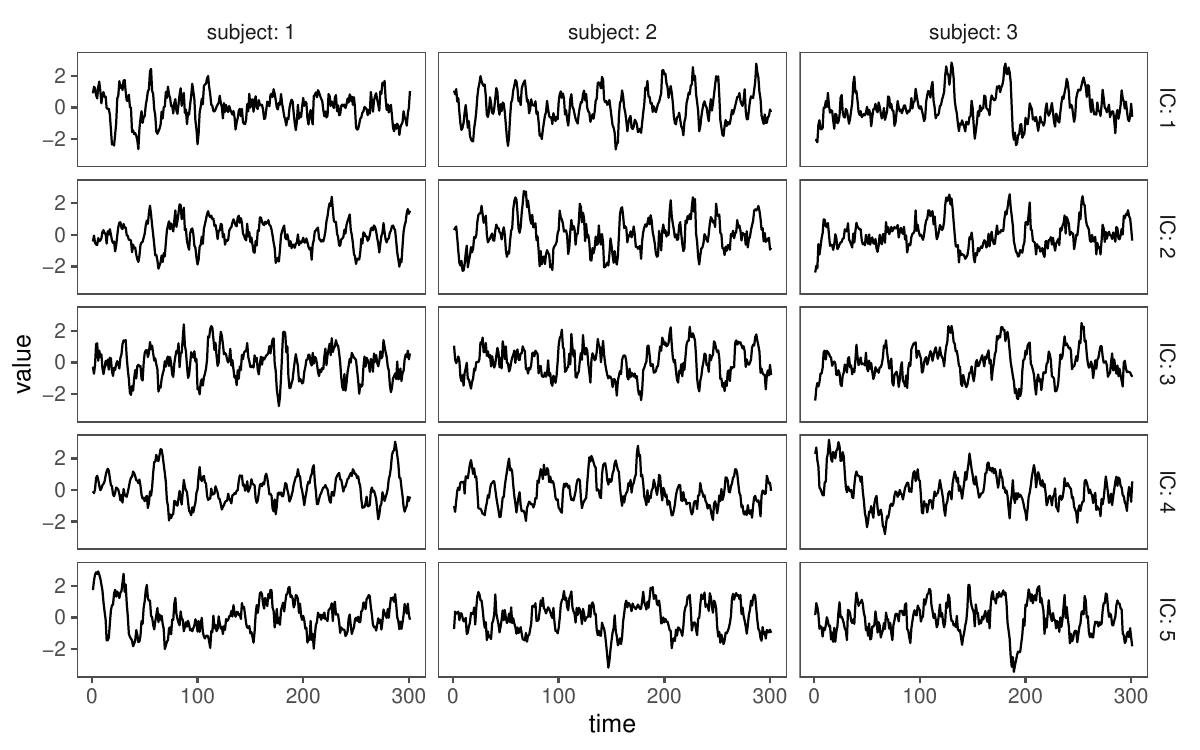}\\
\includegraphics[width=0.32\linewidth, page=1]{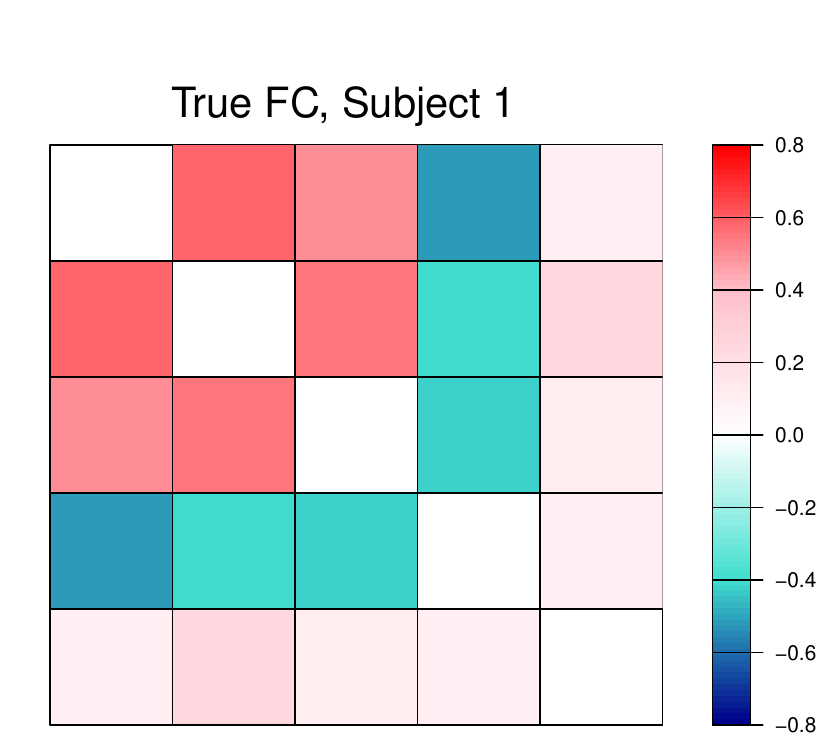}
\includegraphics[width=0.32\linewidth, page=2]{simulation/TCs_real/TCs_real_FC.pdf}
\includegraphics[width=0.32\linewidth, page=3]{simulation/TCs_real/TCs_real_FC.pdf}
\caption{{True IC time courses and functional connectivity (FC) for three example subjects in simulation study.} The time courses (arbitrary units) are based on real fMRI data and exhibit realistic features like autocorrelation. Only 300 of the total 1200 time points are shown. The FC matrices (based on all 1200 time points) show similar but unique patterns across the subjects.}
\label{fig:sim:TC_FC_real}
\end{figure}

\subsection{Simulation Results}

\begin{figure}[H]
    \centering
    \begin{tabular}{ccccccc}
    & IC1  & IC2 & IC3 & IC4 & IC5 &  \\
    \hline \\[-12pt]
    \begin{picture}(10,90)\put(5,45){\rotatebox[origin=c]{90}{Oracle Mean}}\end{picture} &
    \includegraphics[width = 0.8in, trim=0 1in 0 1in, clip]{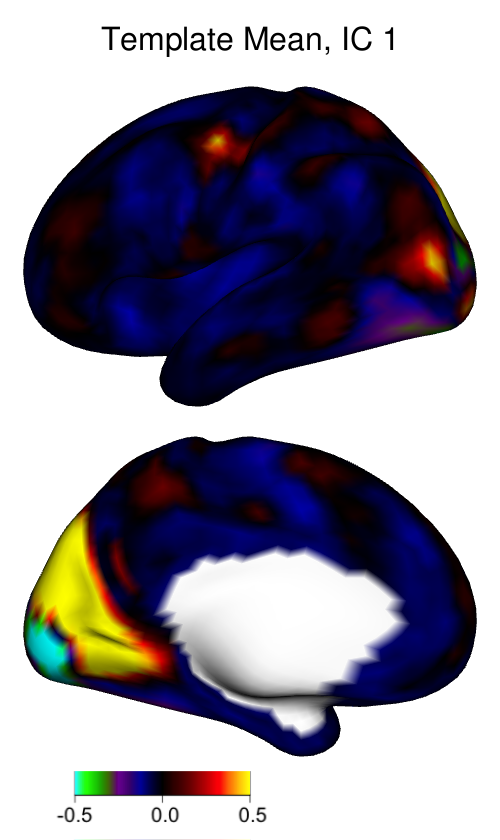} &
    \includegraphics[width = 0.8in, trim=0 1in 0 1in, clip]{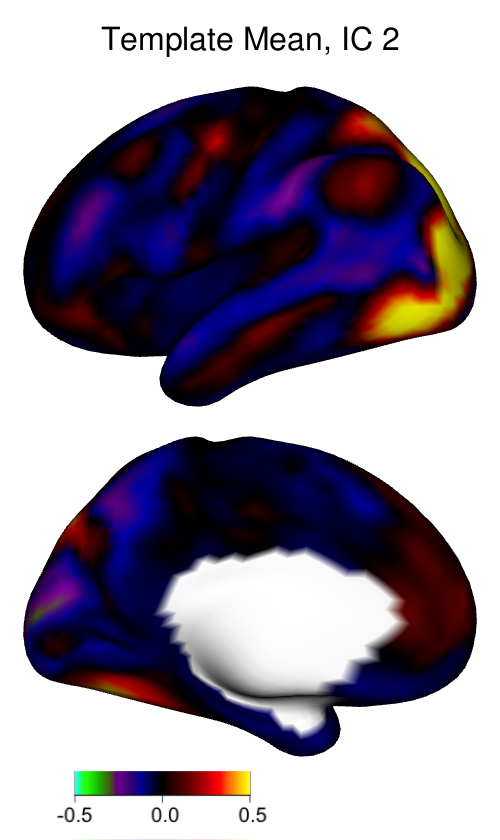} &
    \includegraphics[width = 0.8in, trim=0 1in 0 1in, clip]{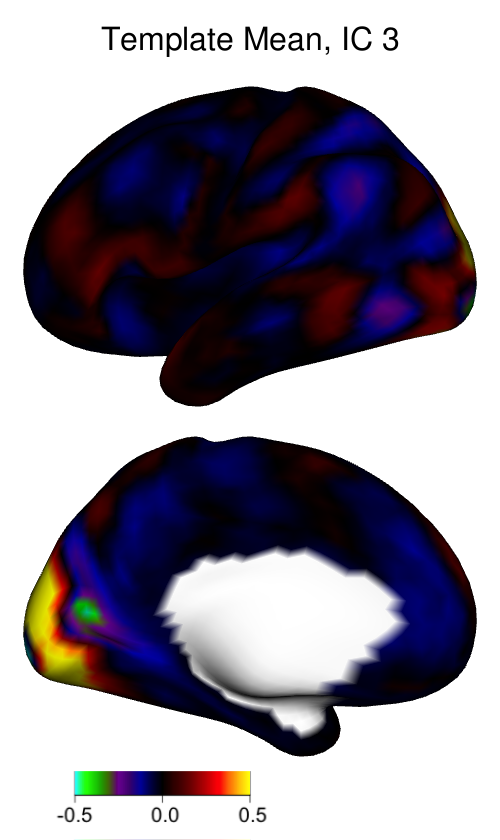} &
    \includegraphics[width = 0.8in, trim=0 1in 0 1in, clip]{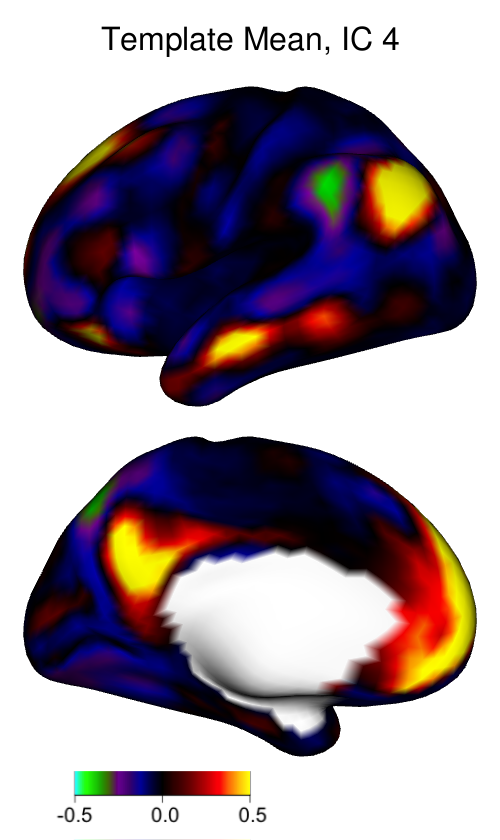} &
    \includegraphics[width = 0.8in, trim=0 1in 0 1in, clip]{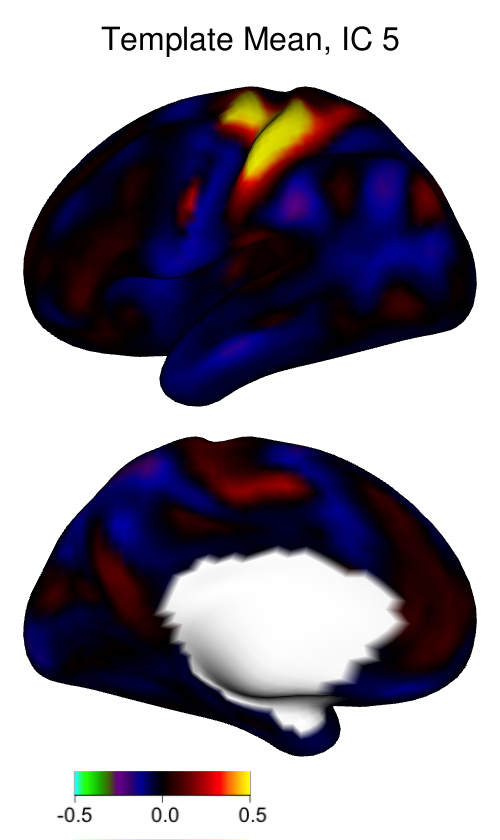} &
    \includegraphics[width = 0.8in, angle=90, trim=-1cm 5mm 3in 10.5in, clip]{simulation/templates/template_mean5.png} \\
    \hline \\[-12pt]
    \begin{picture}(10,90)\put(5,45){\rotatebox[origin=c]{90}{Template Mean}}\end{picture} &
    \includegraphics[width = 0.8in, trim=0 1in 0 1in, clip]{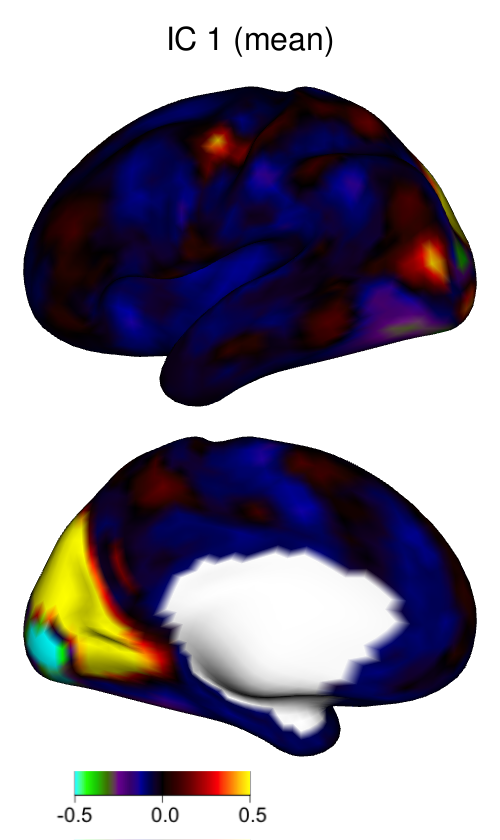} &
    \includegraphics[width = 0.8in, trim=0 1in 0 1in, clip]{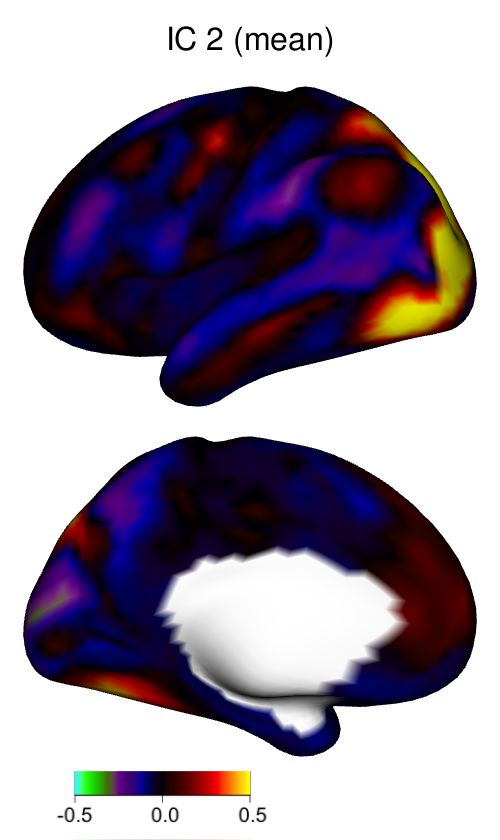} &
    \includegraphics[width = 0.8in, trim=0 1in 0 1in, clip]{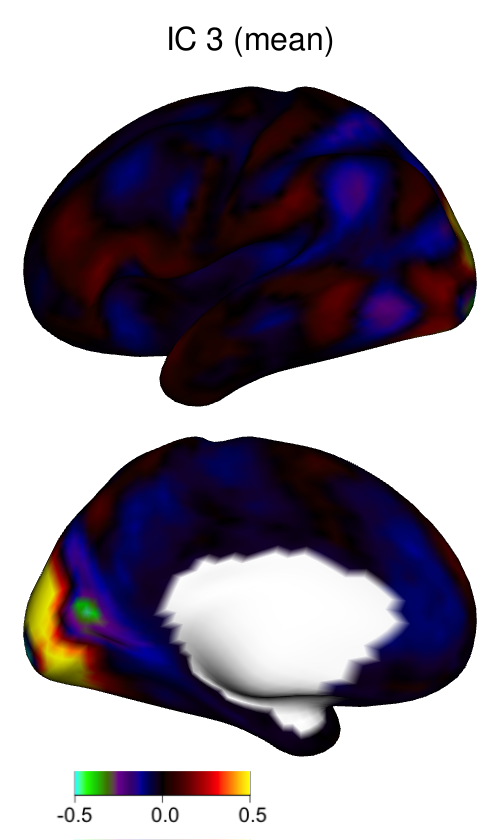} &
    \includegraphics[width = 0.8in, trim=0 1in 0 1in, clip]{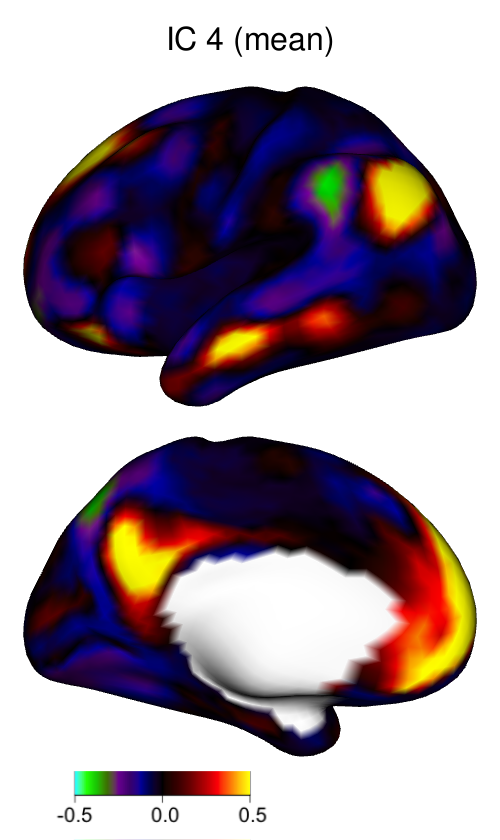} &
    \includegraphics[width = 0.8in, trim=0 1in 0 1in, clip]{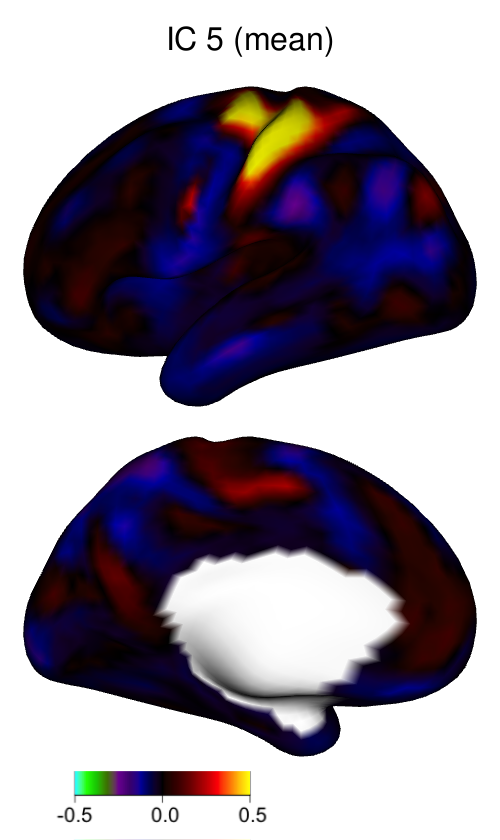} &
    \includegraphics[width = 0.8in, angle=90, trim=-1cm 5mm 3in 10.5in, clip]{simulation/templates/template_est_mean_IC_5.png}\\
    \hline \\[-12pt]
    \begin{picture}(10,90)\put(5,45){\rotatebox[origin=c]{90}{Oracle Var}}\end{picture} &
    \includegraphics[width = 0.8in, trim=0 1in 0 1in, clip]{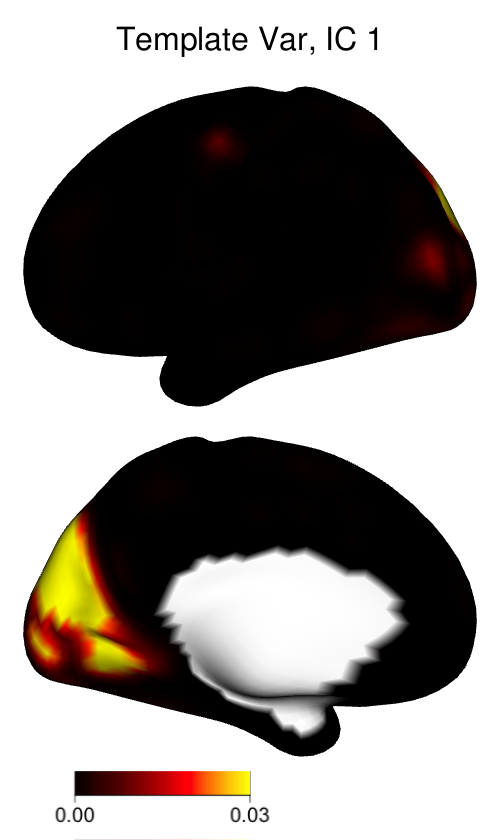} &
    \includegraphics[width = 0.8in, trim=0 1in 0 1in, clip]{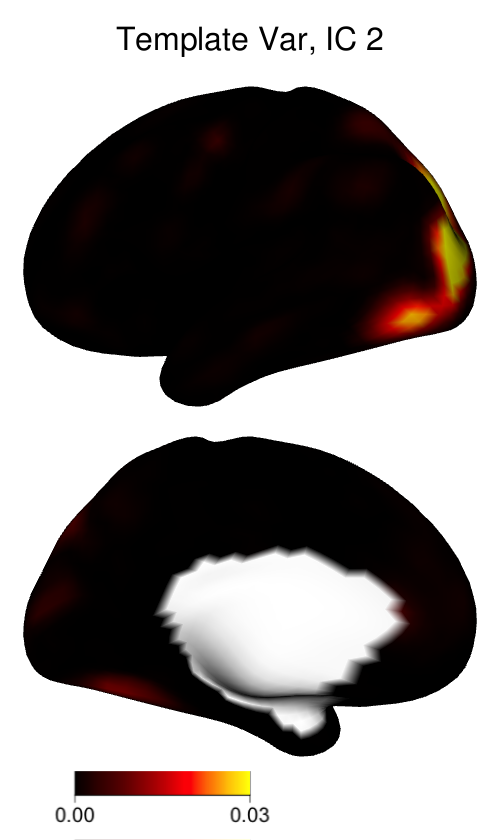} &
    \includegraphics[width = 0.8in, trim=0 1in 0 1in, clip]{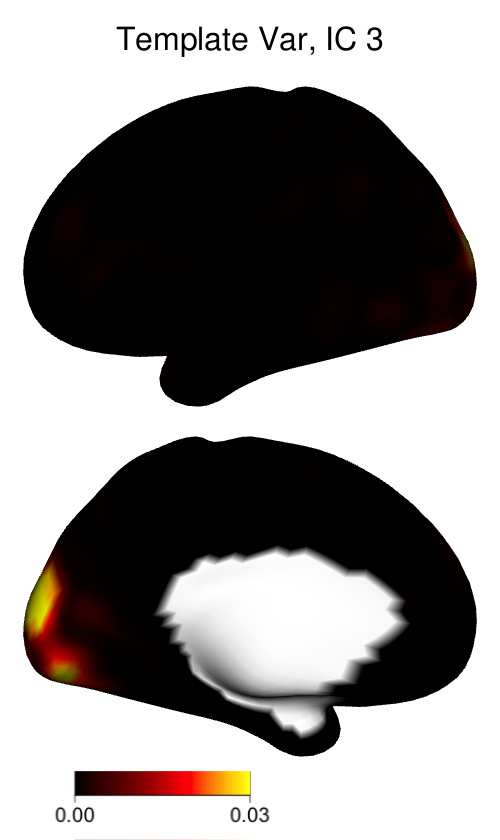} &
    \includegraphics[width = 0.8in, trim=0 1in 0 1in, clip]{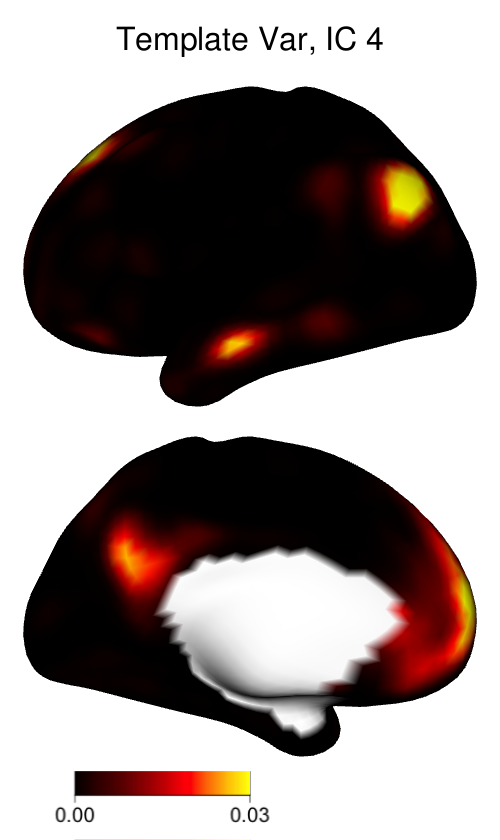} &
    \includegraphics[width = 0.8in, trim=0 1in 0 1in, clip]{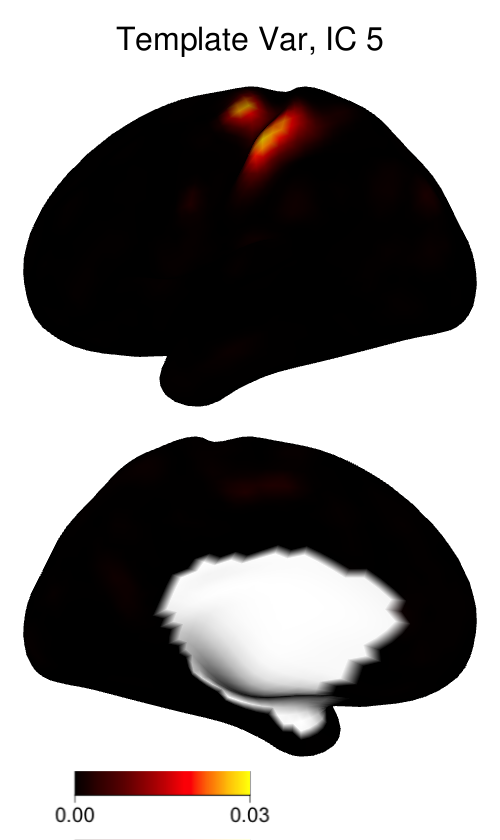} &
    \includegraphics[width = 0.8in, angle=90, trim=-1cm 5mm 3in 10.5in, clip]{simulation/templates/template_var5.png} \\
    \hline \\[-12pt]
    \begin{picture}(10,90)\put(5,45){\rotatebox[origin=c]{90}{Template Var}}\end{picture} &
    \includegraphics[width = 0.8in, trim=0 1in 0 1in, clip]{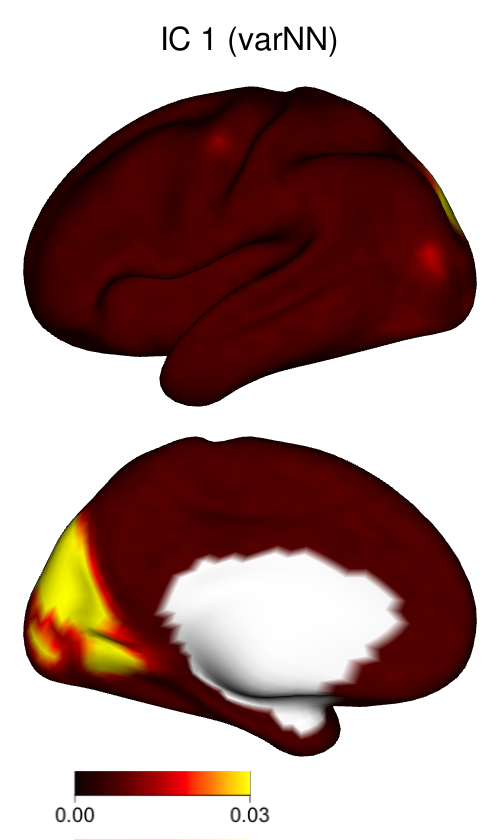} &
    \includegraphics[width = 0.8in, trim=0 1in 0 1in, clip]{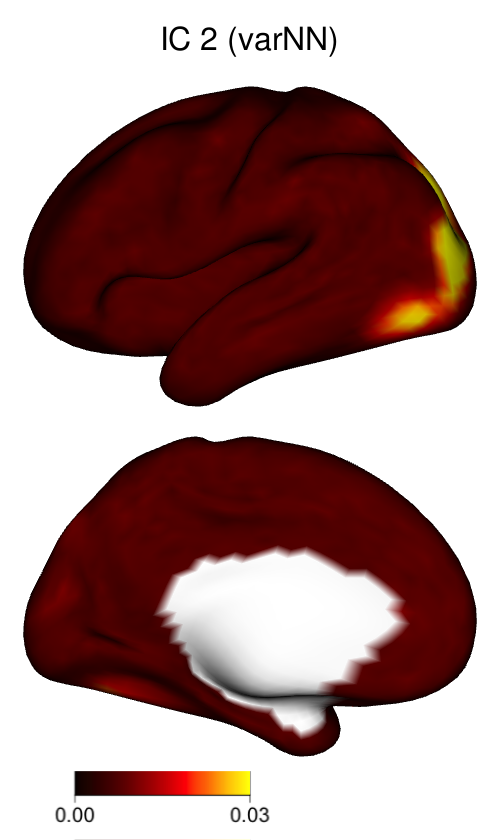} &
    \includegraphics[width = 0.8in, trim=0 1in 0 1in, clip]{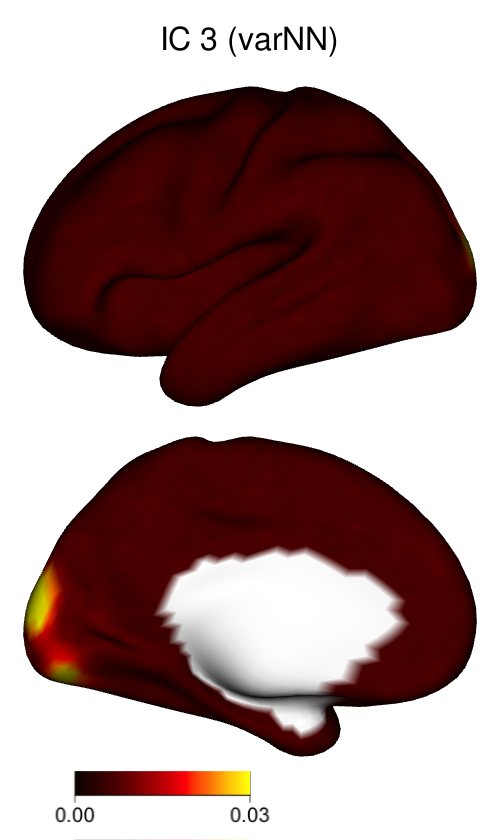} &
    \includegraphics[width = 0.8in, trim=0 1in 0 1in, clip]{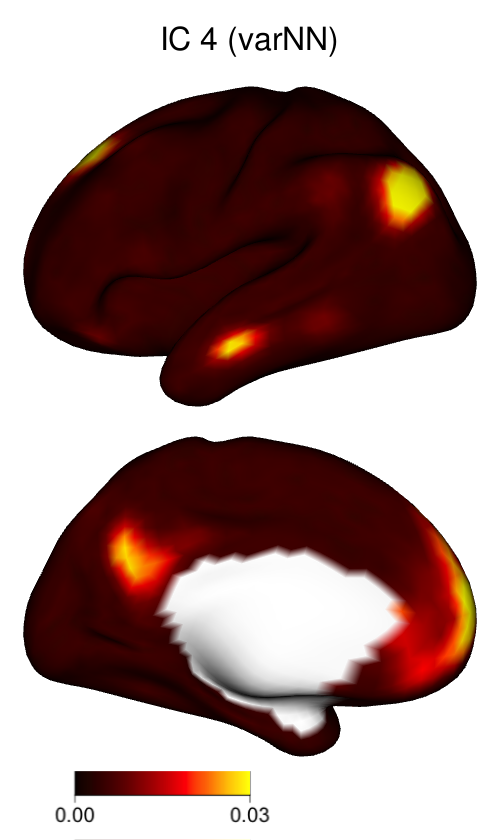} &
    \includegraphics[width = 0.8in, trim=0 1in 0 1in, clip]{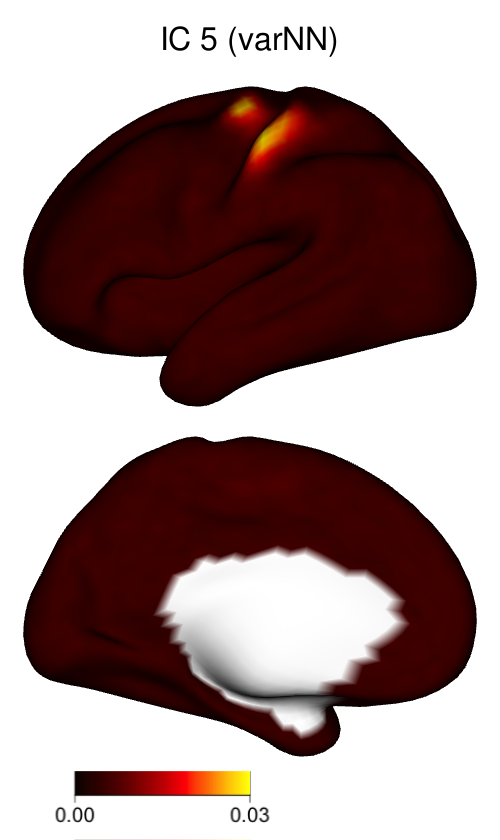} &
    \includegraphics[width = 0.8in, angle=90, trim=-1cm 5mm 3in 10.5in, clip]{simulation/templates/template_est_var_IC_5.png}\\
        \hline \\[-20pt]
    \end{tabular}
    \caption{\textit{Prior mean and variance for spatial IC maps in simulation study.} The oracle values are based on the true spatial IC maps, while the template values are based on extraction of the latent signals from the fMRI timeseries, as described in Section 2.1. The estimated template means are very close to the oracle means. The estimated template variances are very similar to the oracle variances in higher-variance regions, but they are notably higher than the oracle variance in background (low-variance) areas. This is an expected result of the biased non-negative variance estimation approach, which avoids under-estimation of the variance for a less informative prior.}
    \label{fig:sim:template_vs_oracle}
\end{figure}

\begin{figure}[H]
\centering
\begin{tabular}{cccc}
& Population & IW Prior & Permuted Cholesky Prior \\
\begin{picture}(10,100)\put(5,50){\rotatebox[origin=c]{90}{Mean}}\end{picture} &
    \includegraphics[height=38mm, page=1, trim = 5mm 0 25mm 22mm, clip]{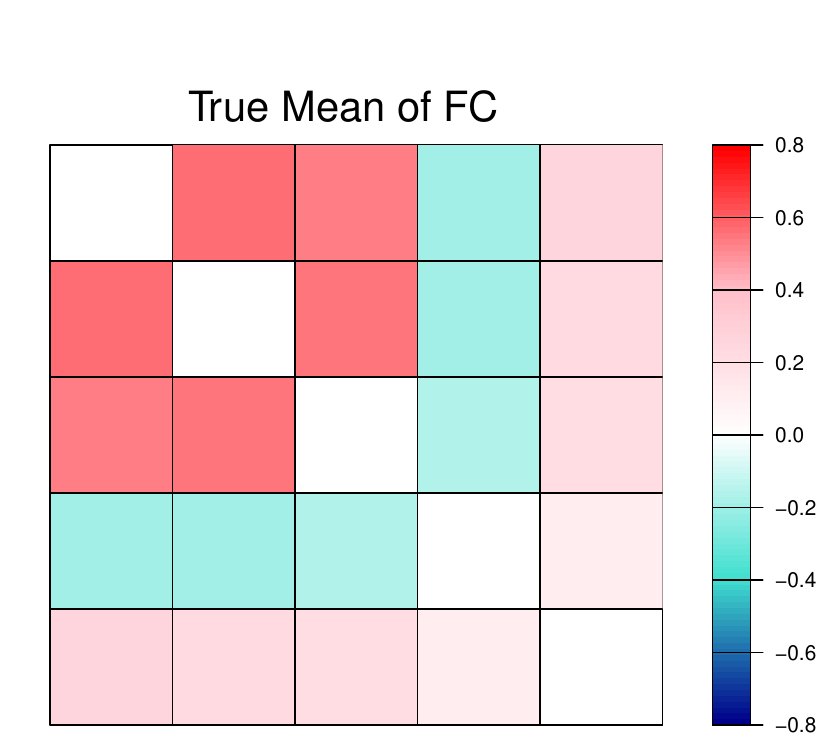} &
    \includegraphics[height=38mm, page=1, trim = 5mm 0 25mm 22mm, clip]{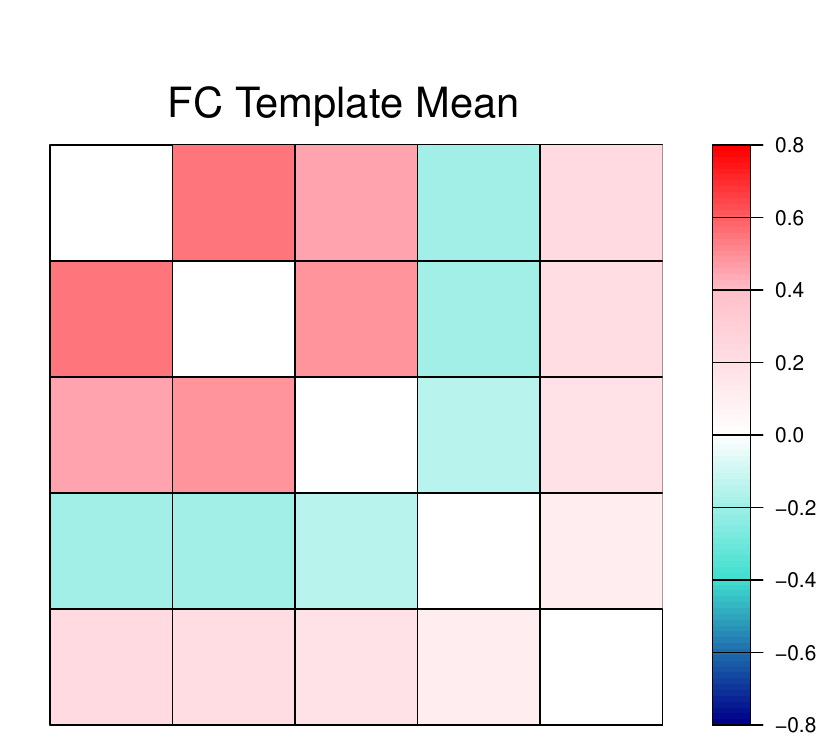} &
    \includegraphics[height=38mm, page=1, trim = 5mm 0 25mm 22mm, clip]{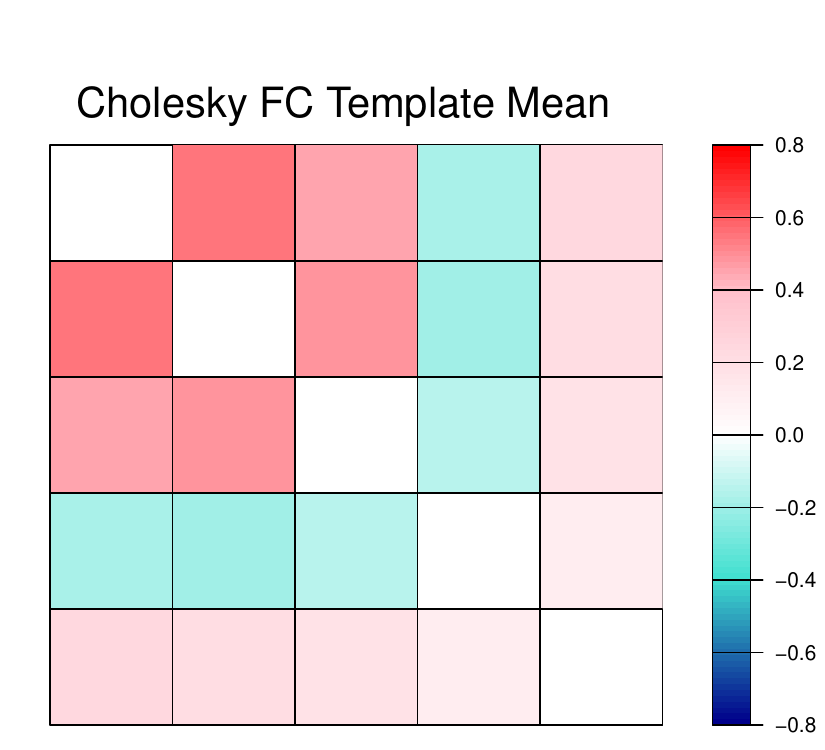} \\
    & \multicolumn{3}{c}{\includegraphics[height=5mm]{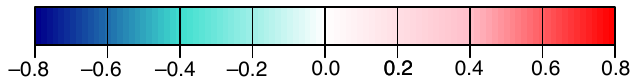}} \\
\begin{picture}(10,100)\put(5,50){\rotatebox[origin=c]{90}{SD}}\end{picture} &
    \includegraphics[height=38mm, page=2, trim = 5mm 0 25mm 22mm, clip]{simulation/plots/FC_mean_and_SD.pdf} &
    \includegraphics[height=38mm, page=2, trim = 5mm 0 25mm 22mm, clip]{simulation/templates/FCtemplate.pdf} &
    \includegraphics[height=38mm, page=2, trim = 5mm 0 25mm 22mm, clip]{simulation/templates/FCtemplate2.pdf} \\
    & \multicolumn{3}{c}{\includegraphics[height=5mm]{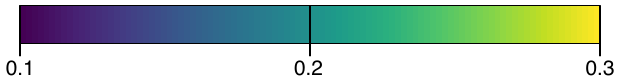}}\\
\end{tabular}
    \caption{\textit{Population and prior mean and standard deviation (SD) of functional connectivity (FC).} The prior mean and SD are shown for both priors we consider for FC: the Inverse-Wishart (IW) and our novel permuted Cholesky prior.  The prior mean is similar to the population mean for both priors.  The IW prior SD is somewhat higher than the true SD. This is by design to avoid under-estimation of the true SD for any connection, as described in Section 2.1. The permuted Cholesky prior SD closely mimics the population SD.}
    \label{fig:sim:template_FC}
\end{figure}

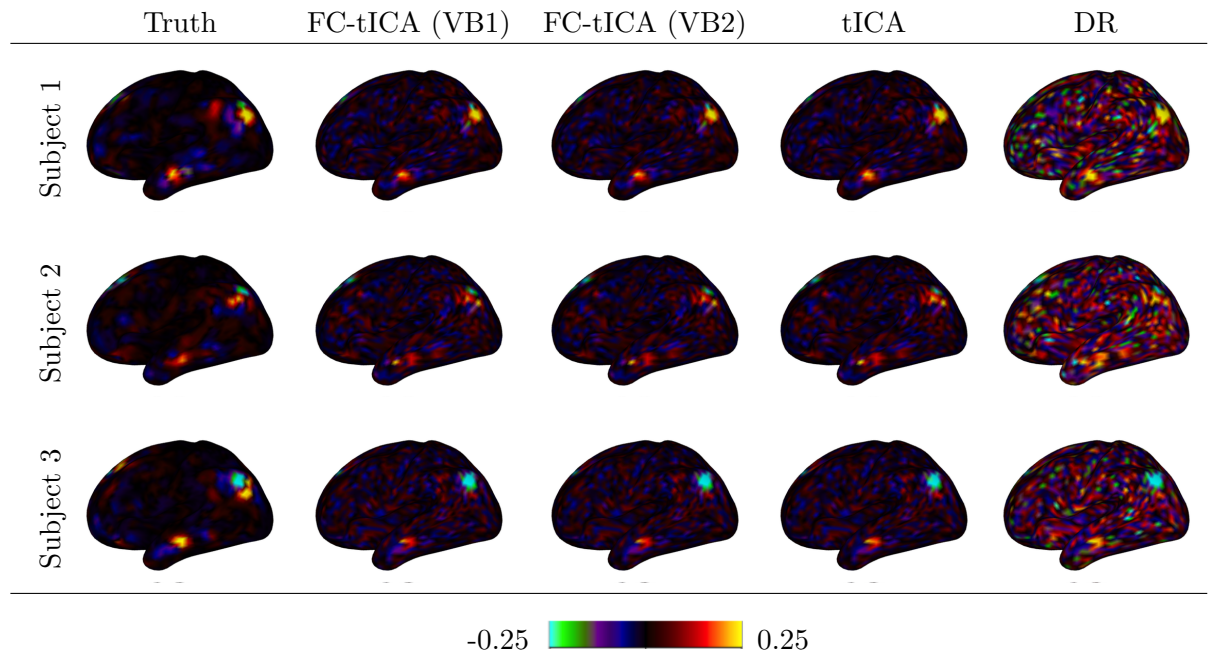
\begin{figure}[H]
    \centering
    \begin{tabular}{ccccccc}
    & Truth & FC-tICA (VB1) & FC-tICA (VB2) & tICA & DR \\
    \hline \\[-6pt]
    \begin{picture}(10,50)\put(5,25){\rotatebox[origin=c]{90}{Subject 1}}\end{picture} &
    \includegraphics[width=25mm, trim=5mm 6in 5mm 1in, clip]{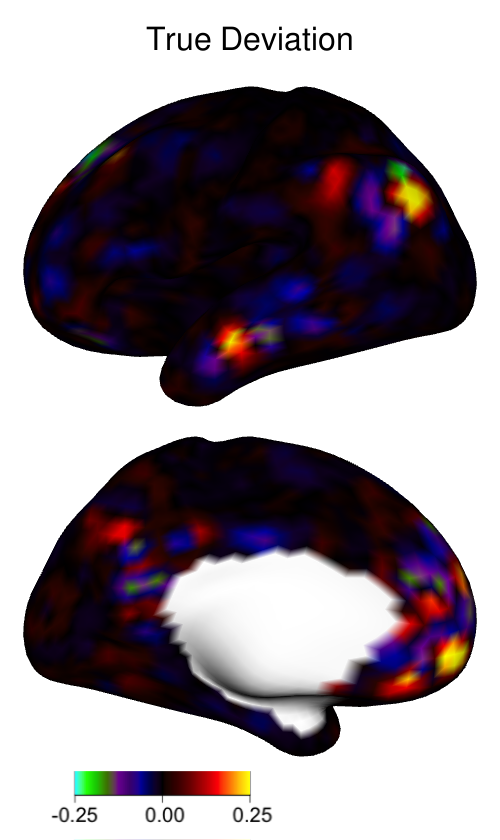} &
    \includegraphics[width=25mm, trim=5mm 6in 5mm 1in, clip]{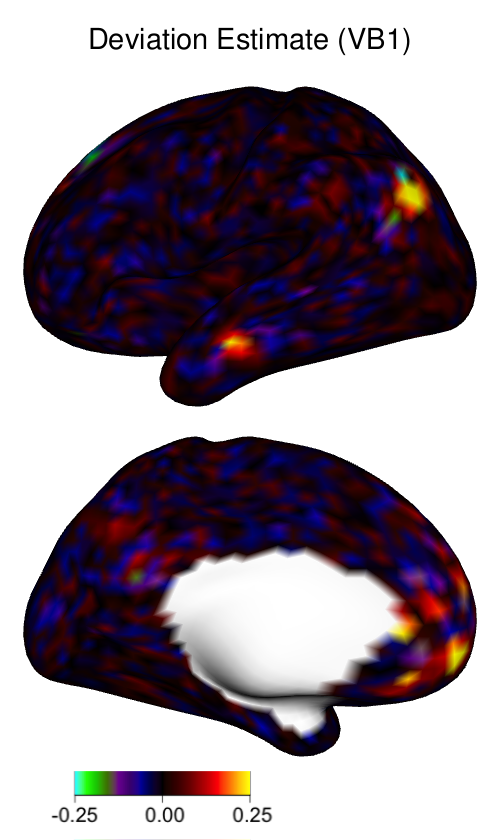} &
    \includegraphics[width=25mm, trim=5mm 6in 5mm 1in, clip]{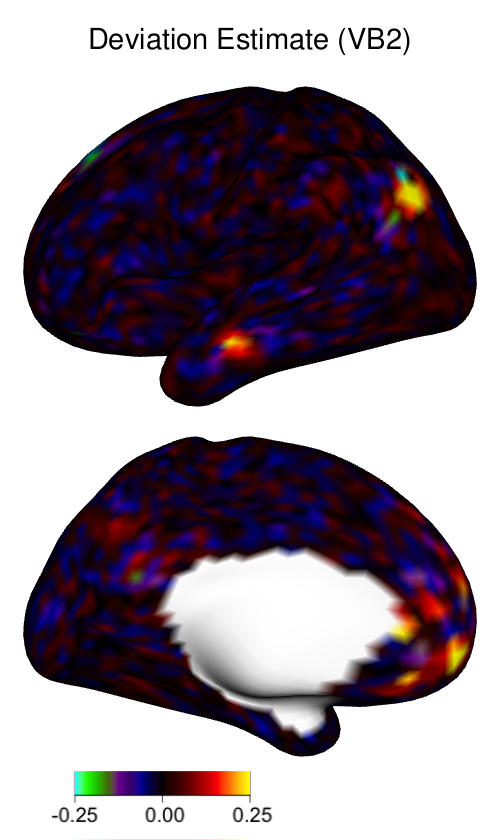} & 
    \includegraphics[width=25mm, trim=5mm 6in 5mm 1in, clip]{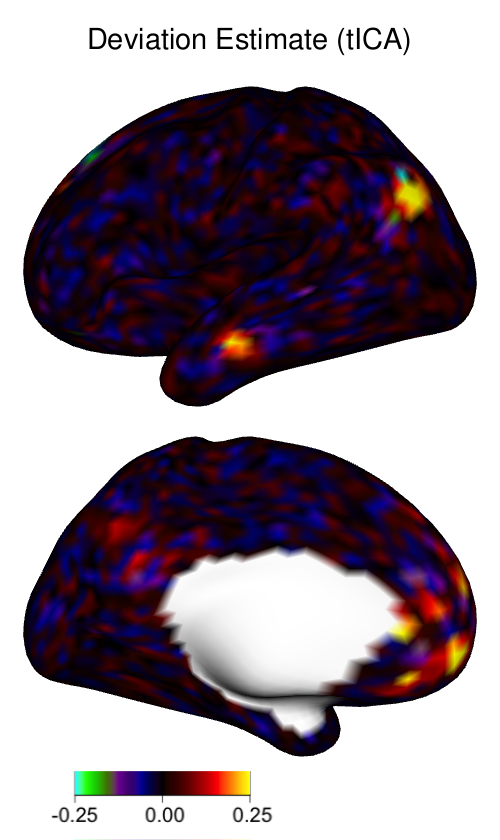} &
    \includegraphics[width=25mm, trim=5mm 6in 5mm 1in, clip]{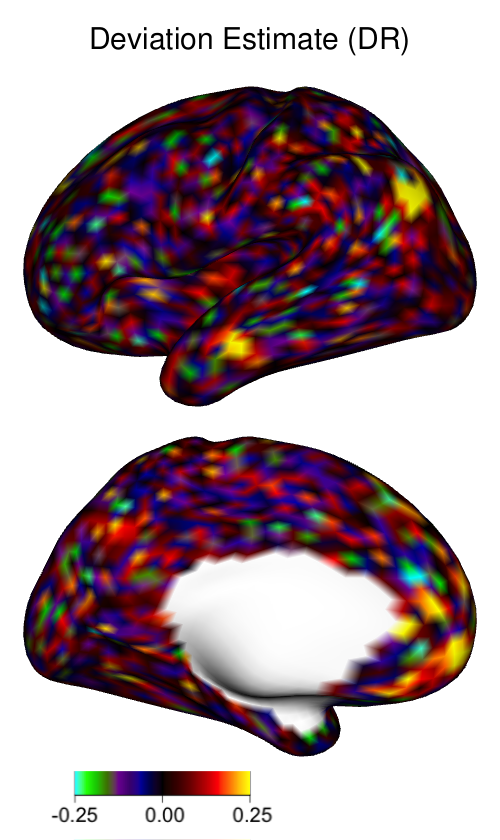} \\[10pt]
    \begin{picture}(10,50)\put(5,25){\rotatebox[origin=c]{90}{Subject 2}}\end{picture} &
    \includegraphics[width=25mm, trim=5mm 6in 5mm 1in, clip]{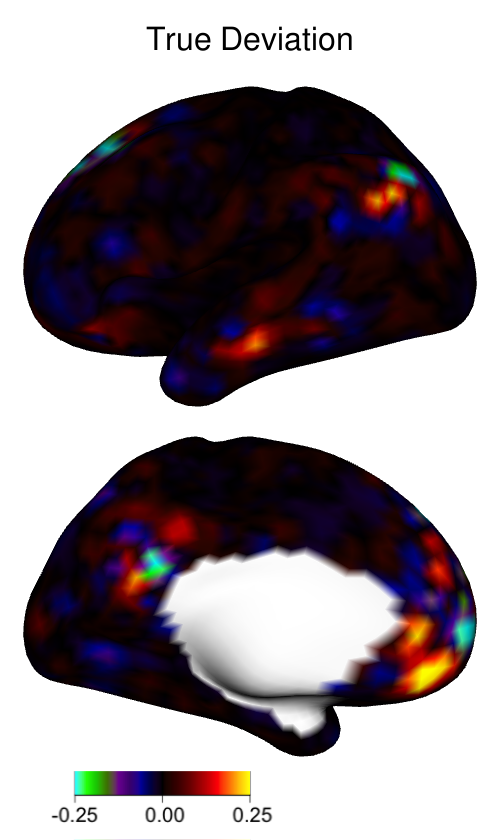} &
    \includegraphics[width=25mm, trim=5mm 6in 5mm 1in, clip]{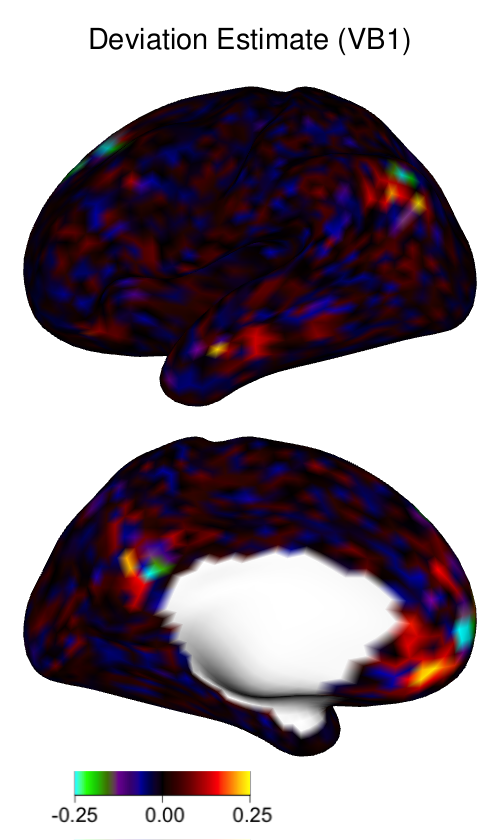} &
    \includegraphics[width=25mm, trim=5mm 6in 5mm 1in, clip]{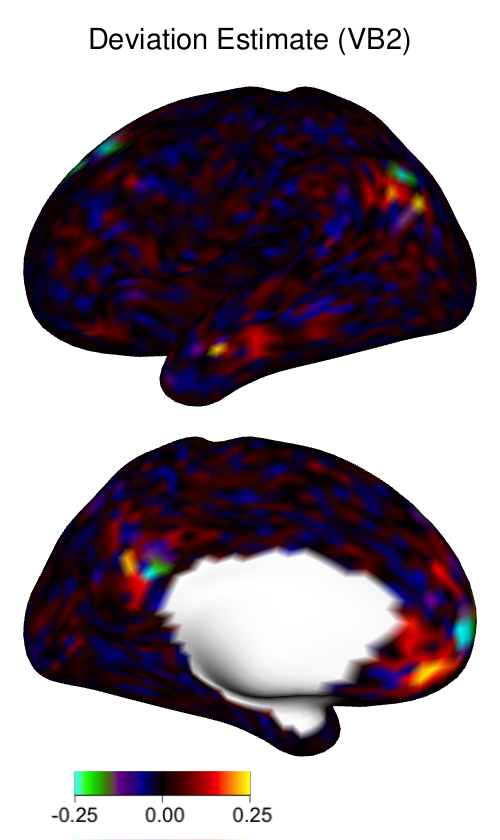} & 
    \includegraphics[width=25mm, trim=5mm 6in 5mm 1in, clip]{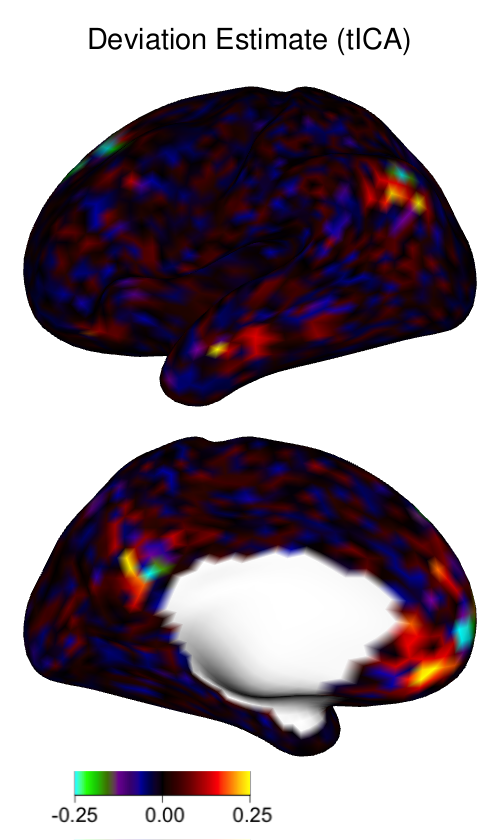} &
    \includegraphics[width=25mm, trim=5mm 6in 5mm 1in, clip]{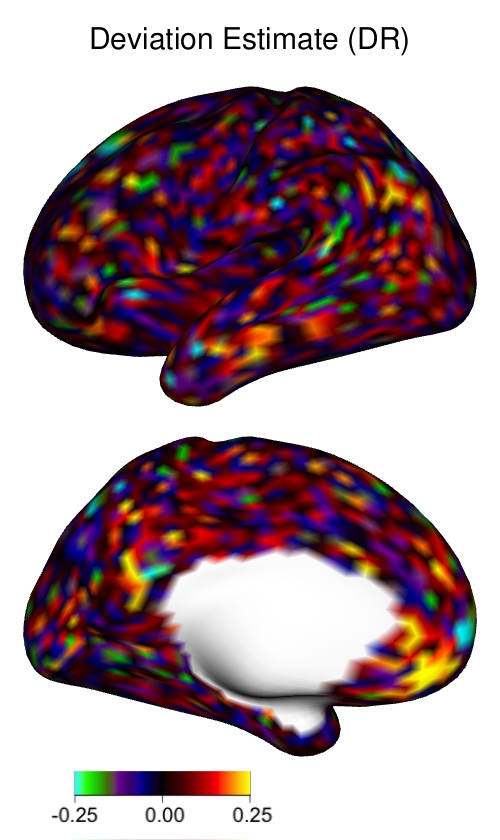} \\[10pt]
    \begin{picture}(10,50)\put(5,25){\rotatebox[origin=c]{90}{Subject 3}}\end{picture} &
    \includegraphics[width=25mm, trim=5mm 6in 5mm 1in, clip]{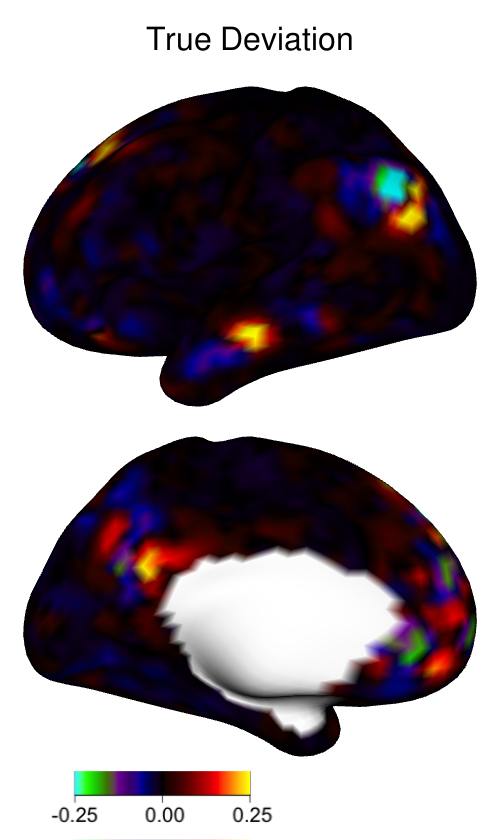} &
    \includegraphics[width=25mm, trim=5mm 6in 5mm 1in, clip]{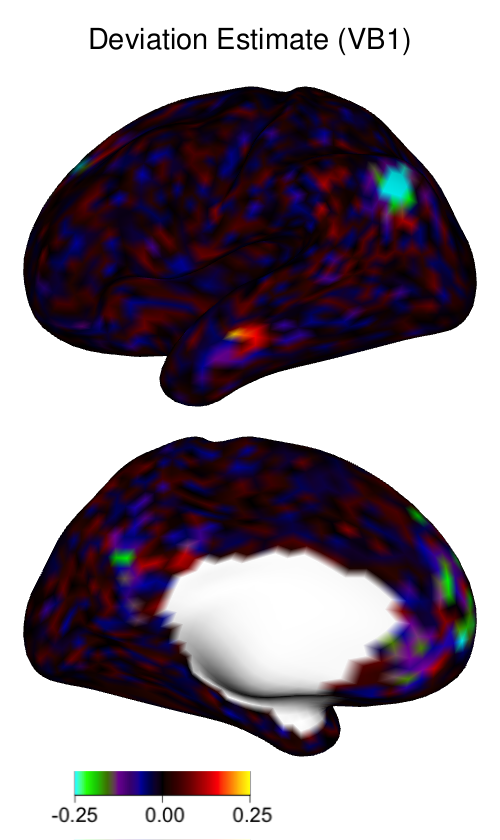} &
    \includegraphics[width=25mm, trim=5mm 6in 5mm 1in, clip]{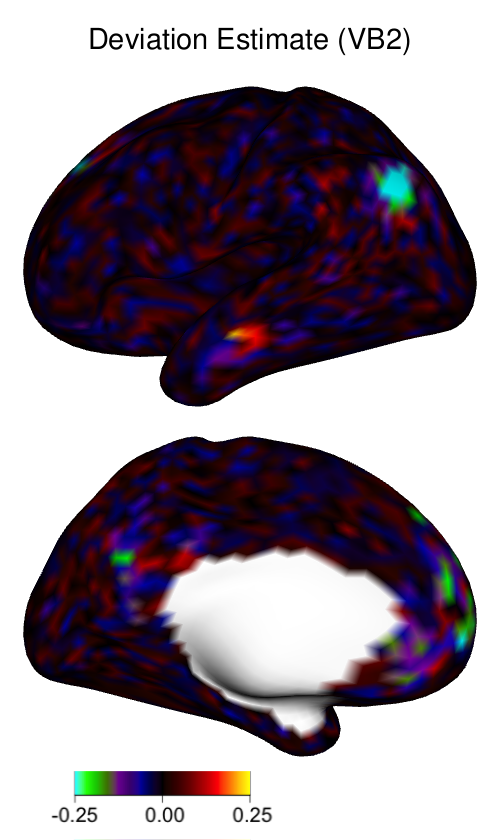} & 
    \includegraphics[width=25mm, trim=5mm 6in 5mm 1in, clip]{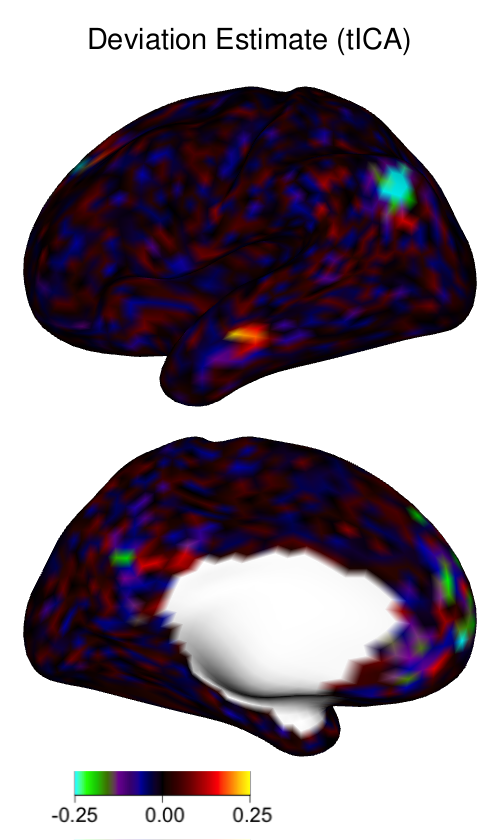} &
    \includegraphics[width=25mm, trim=5mm 6in 5mm 1in, clip]{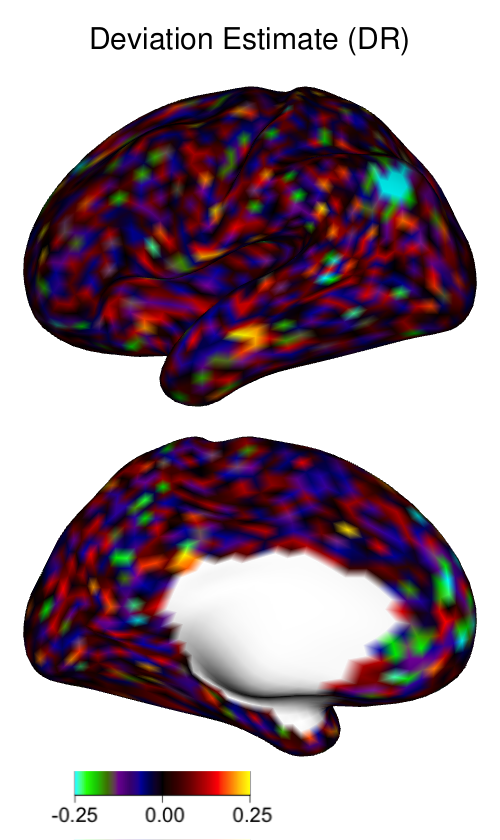} \\
    \hline \\[-10pt]
    & \multicolumn{5}{c}{-0.25\includegraphics[width=3cm, trim=3mm 13mm 3mm -3mm, clip]{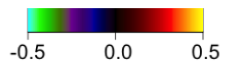}0.25}
    \end{tabular}
    \caption{\textit{Estimated subject-level deviations in simulation study.} For three randomly selected test subjects, the true and estimated deviation maps for the default mode network component (IC 4) are shown. The two proposed VB algorithms for FC template ICA (FC-tICA) and the existing methods tICA and DR all identify the unique features of each subject reasonably well. The DR estimates are noticeably noisier.} 
    \label{fig:sim:results_subjICs}
\end{figure}

\begin{figure}[H]
    \centering
    \begin{tabular}{cccccc}
    & IC1  & IC2 & IC3 & IC4 & IC5 \\
        \hline \\[-12pt]
    \begin{picture}(0,100)\put(-5,50){\rotatebox[origin=c]{90}{DR}}\end{picture} &
    \includegraphics[width=1in, trim=0 1in 0 1in, clip]{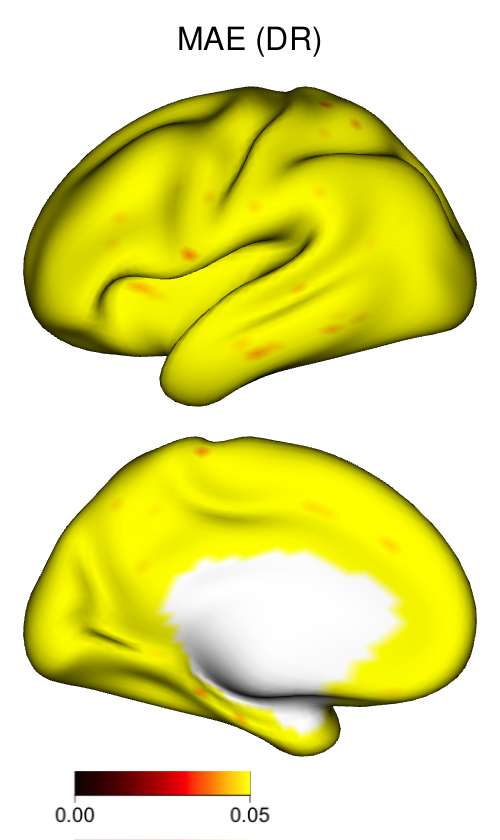} &
    \includegraphics[width=1in, trim=0 1in 0 1in, clip]{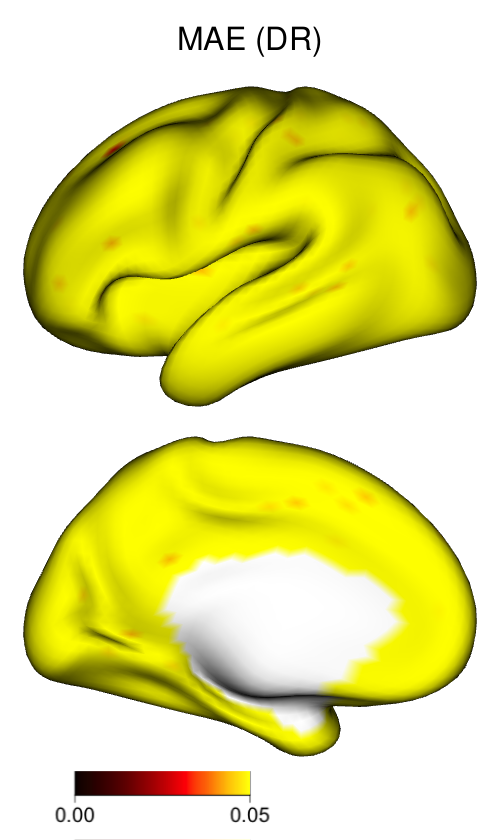} &
    \includegraphics[width=1in, trim=0 1in 0 1in, clip]{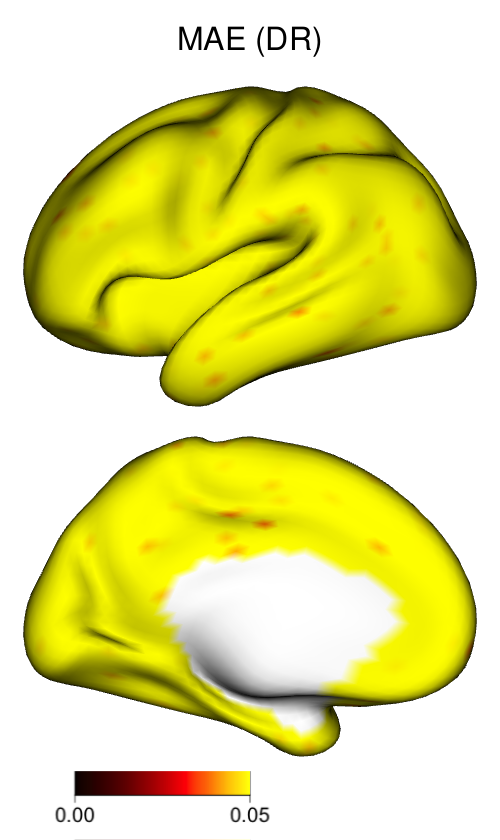} &
    \includegraphics[width=1in, trim=0 1in 0 1in, clip]{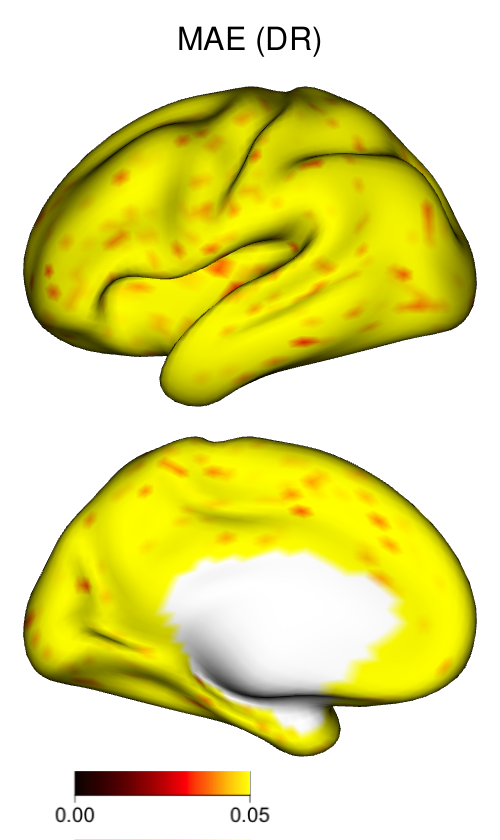} &
    \includegraphics[width=1in, trim=0 1in 0 1in, clip]{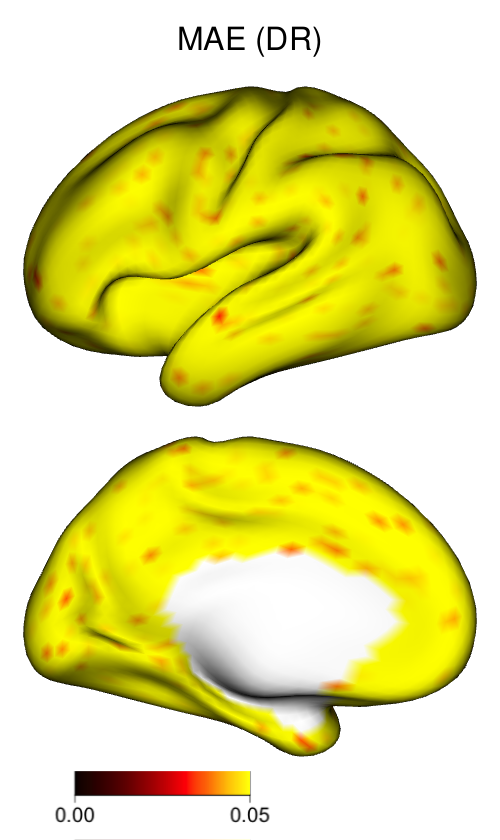} \\
        \hline \\[-12pt]    \begin{picture}(0,100)\put(-5,50){\rotatebox[origin=c]{90}{tICA}}\end{picture} &
    \includegraphics[width=1in, trim=0 1in 0 1in, clip]{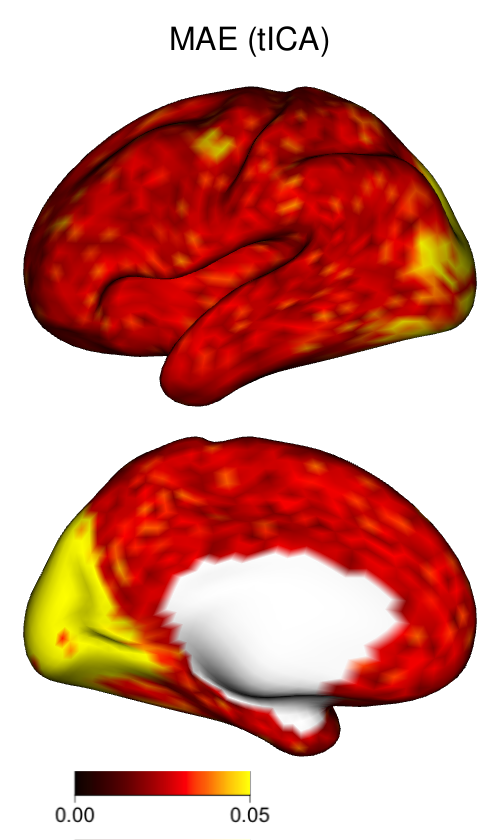} &
    \includegraphics[width=1in, trim=0 1in 0 1in, clip]{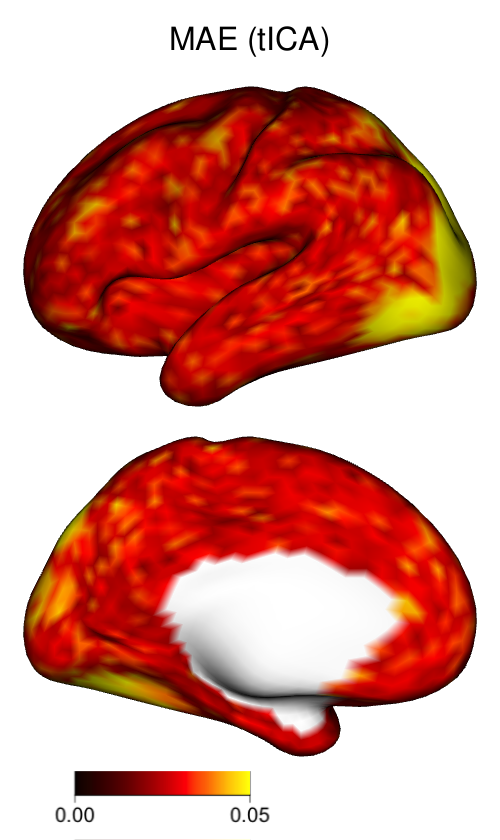} &
    \includegraphics[width=1in, trim=0 1in 0 1in, clip]{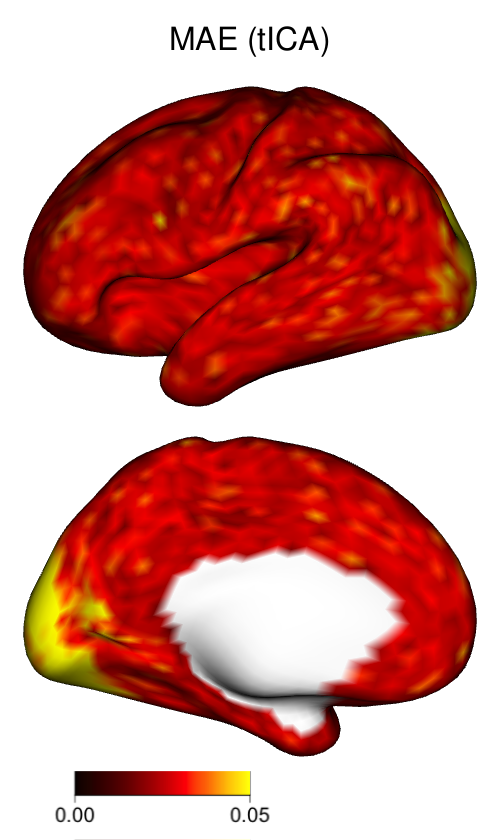} &
    \includegraphics[width=1in, trim=0 1in 0 1in, clip]{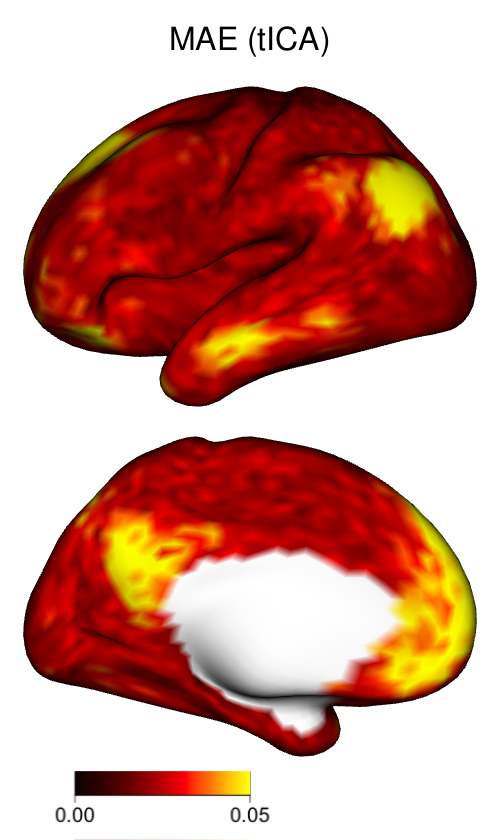} &
    \includegraphics[width=1in, trim=0 1in 0 1in, clip]{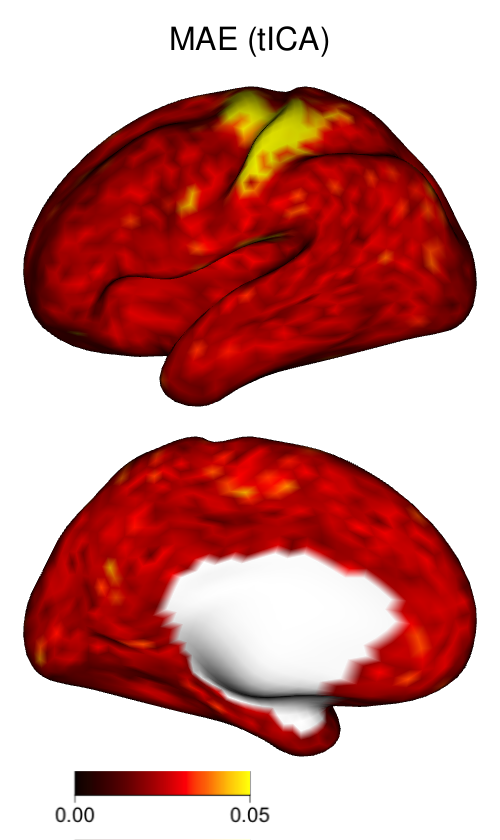} \\
        \hline \\[-12pt]
    \begin{picture}(0,100)\put(-5,50){\rotatebox[origin=c]{90}{FC-tICA (VB1)}}\end{picture} &
    \includegraphics[width=1in, trim=0 1in 0 1in, clip]{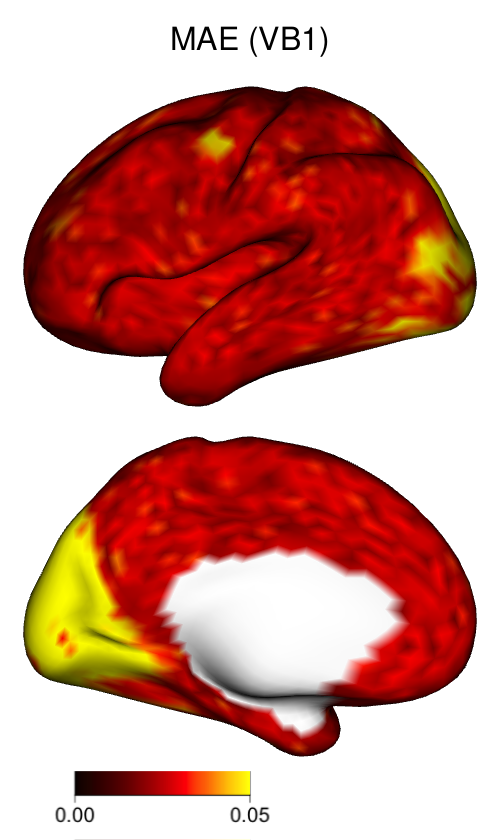} &
    \includegraphics[width=1in, trim=0 1in 0 1in, clip]{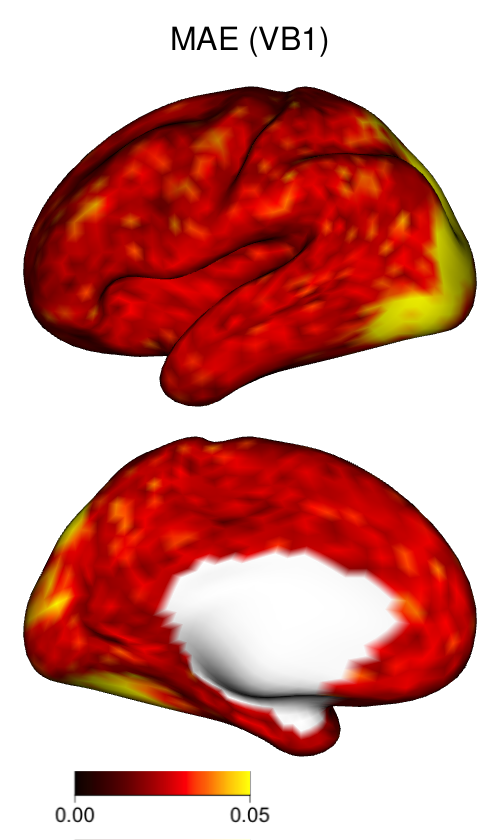} &
    \includegraphics[width=1in, trim=0 1in 0 1in, clip]{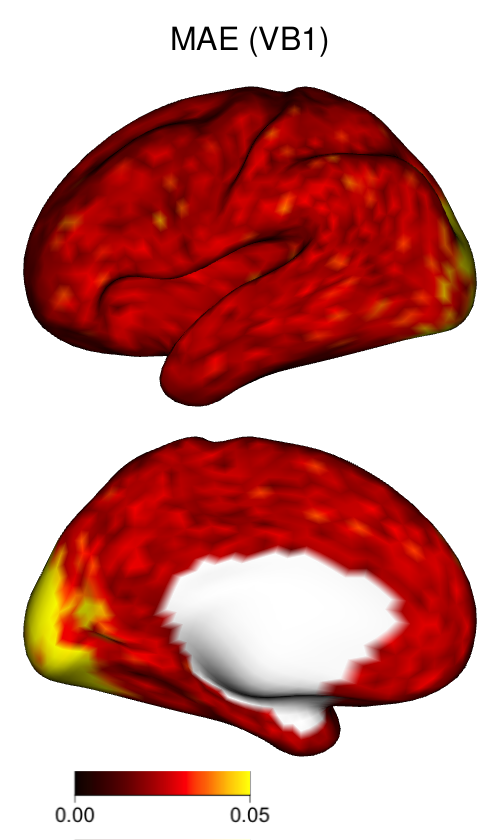} &
    \includegraphics[width=1in, trim=0 1in 0 1in, clip]{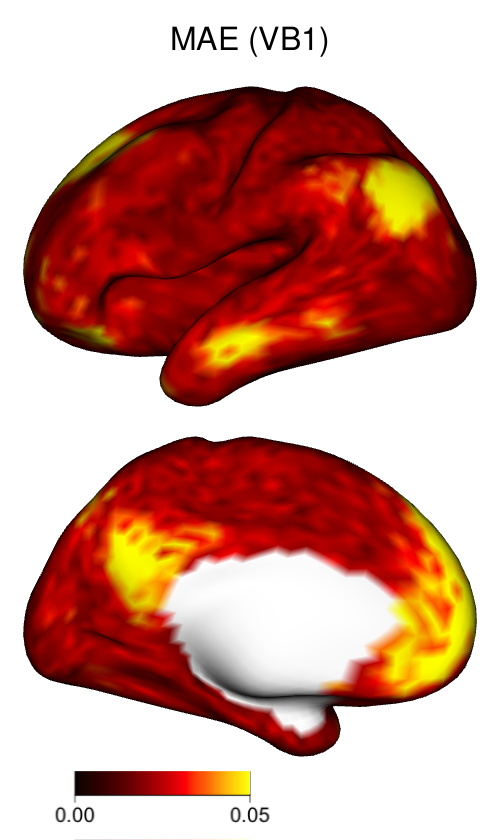} &
    \includegraphics[width=1in, trim=0 1in 0 1in, clip]{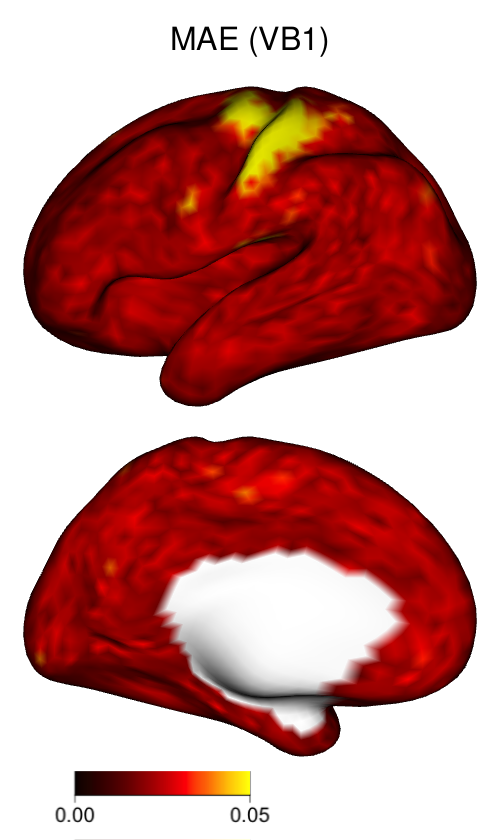} \\        
    \hline \\[-12pt]
    \begin{picture}(0,100)\put(-5,50){\rotatebox[origin=c]{90}{FC-tICA (VB2)}}\end{picture} &
    \includegraphics[width=1in, trim=0 1in 0 1in, clip]{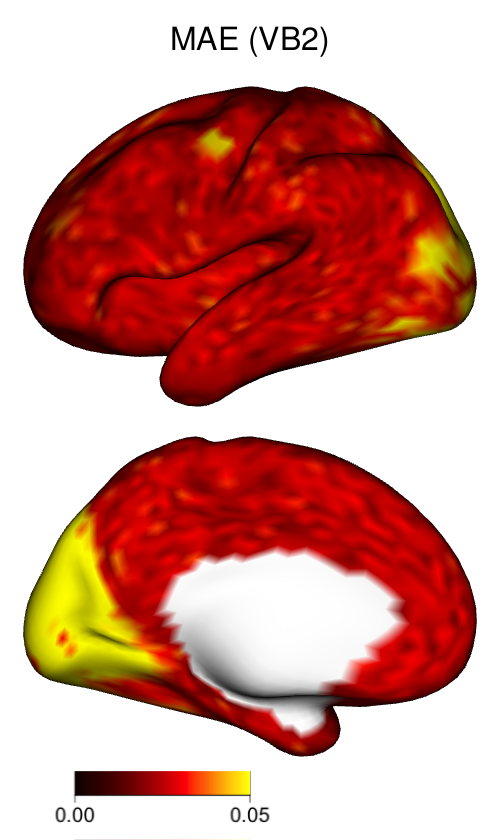} &
    \includegraphics[width=1in, trim=0 1in 0 1in, clip]{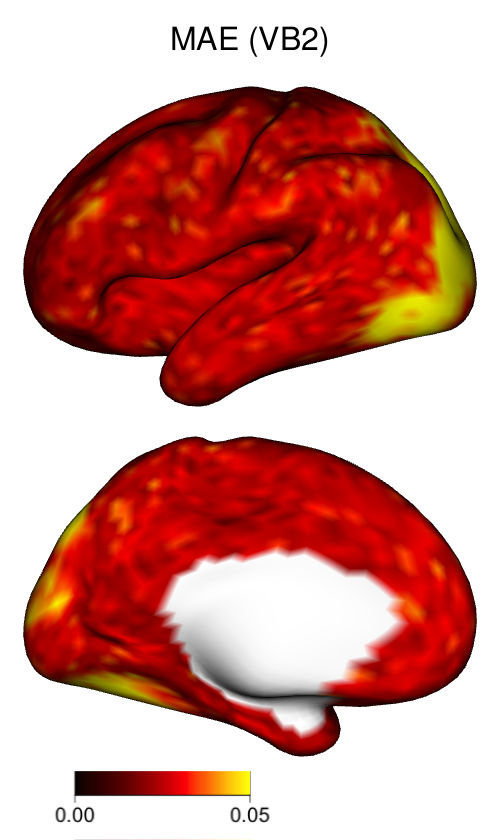} &
    \includegraphics[width=1in, trim=0 1in 0 1in, clip]{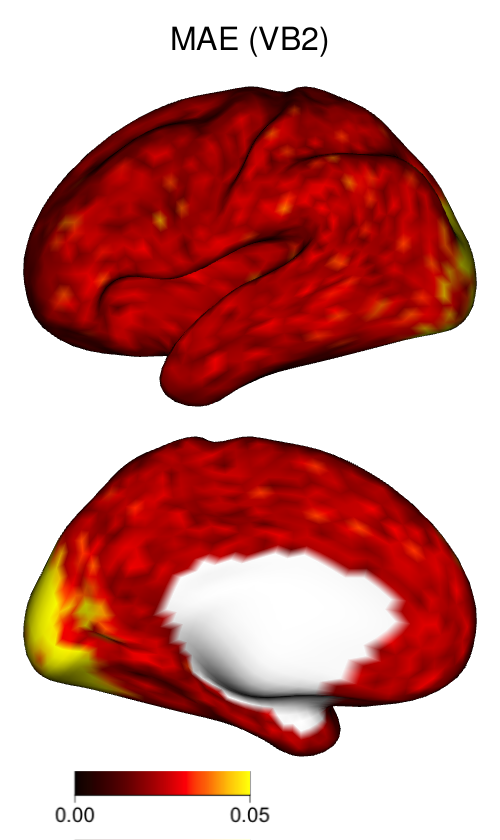} &
    \includegraphics[width=1in, trim=0 1in 0 1in, clip]{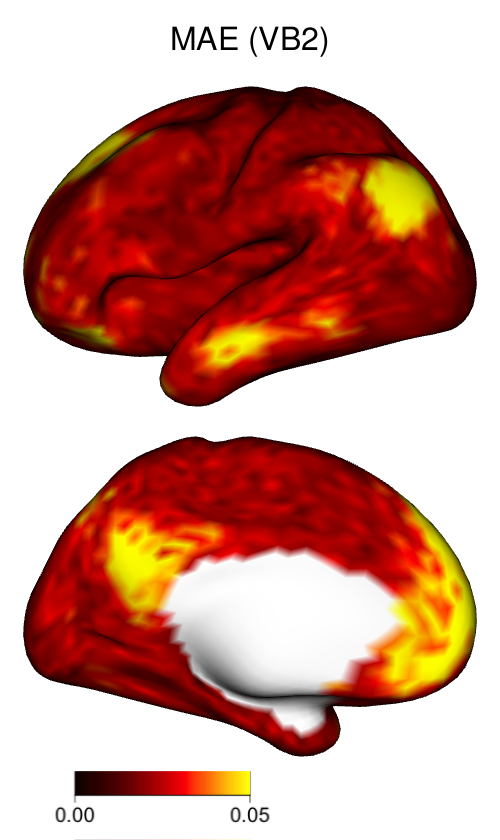} &
    \includegraphics[width=1in, trim=0 1in 0 1in, clip]{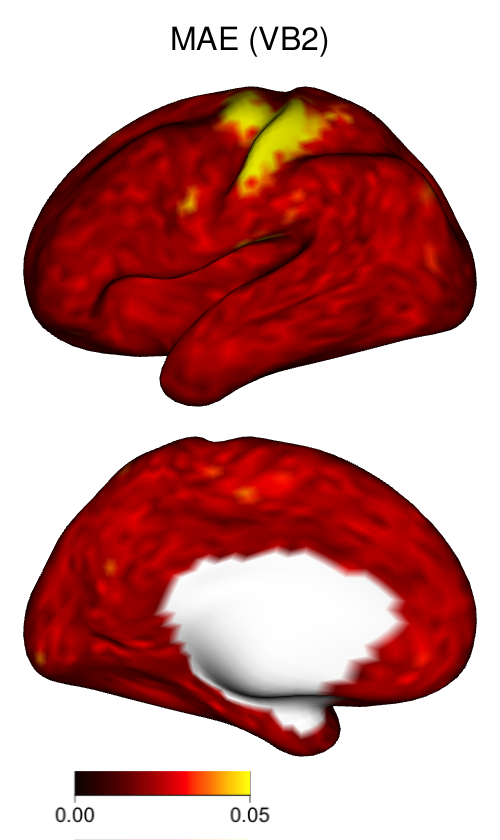} \\        
    \hline \\[-12pt]
    & \multicolumn{5}{c}{\includegraphics[width=3cm, trim=18mm 5mm 8cm 10.5in, clip]{simulation/images/MAE/MAE_VB2_IC_5.png}}
    \end{tabular}
    \caption{\textit{Median absolute error (MAE) of subject-level spatial maps over 50 simulated test subjects in the simulation study.} Similar patterns are seen across the tICA and FC-tICA algorithms, with lower error in background areas and higher error in the areas of engagement for each IC.  DR has much higher estimation error compared with tICA or FC-tICA.}
    \label{fig:sim:results_MAE_S}
\end{figure}

\begin{figure}[H]
    \centering
    \begin{tabular}{cc|cccc}
    & Retest & \multicolumn{4}{c}{Test Dataset} \\[6pt]
    \hline\\
    & Truth & FC-tICA (VB1) & FC-tICA (VB2) & tICA & DR  \\[10pt]
    \begin{picture}(10,50)\put(5,25){\rotatebox[origin=c]{90}{Subject 1}}\end{picture} &
    \includegraphics[height=25mm, page=1, trim = 5mm 3mm 25mm 22mm, clip]{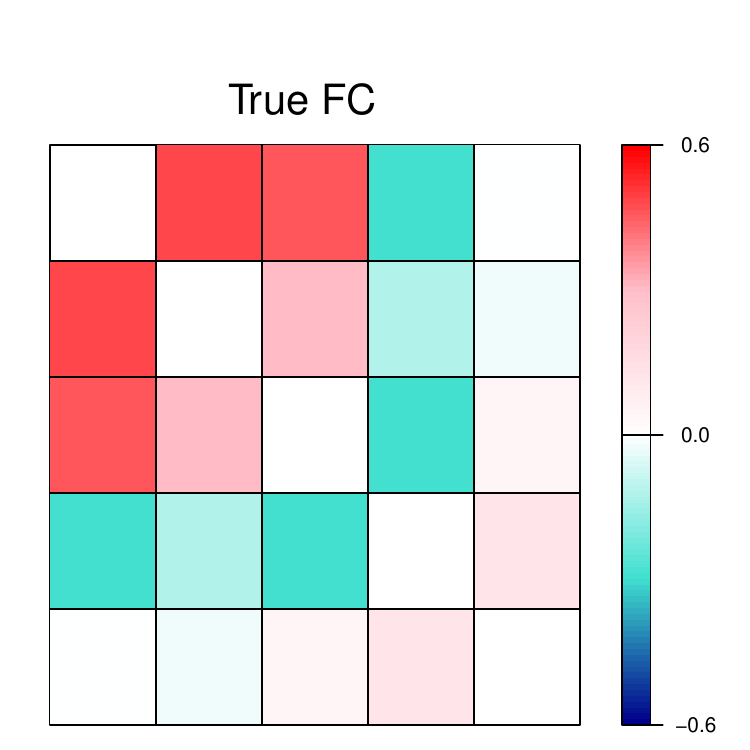} & 
    \includegraphics[height=25mm, page=3, trim = 5mm 3mm 25mm 22mm, clip]{simulation/plots/testsubj1FC.pdf} &
    \includegraphics[height=25mm, page=4, trim = 5mm 3mm 25mm 22mm, clip]{simulation/plots/testsubj1FC.pdf} &
    \includegraphics[height=25mm, page=2, trim = 5mm 3mm 25mm 22mm, clip]{simulation/plots/testsubj1FC.pdf} &
    \includegraphics[height=25mm, page=5, trim = 5mm 3mm 25mm 22mm, clip]{simulation/plots/testsubj1FC.pdf} \\
    \begin{picture}(10,50)\put(5,25){\rotatebox[origin=c]{90}{Subject 2}}\end{picture} &
    \includegraphics[height=25mm, page=1, trim = 5mm 3mm 25mm 22mm, clip]{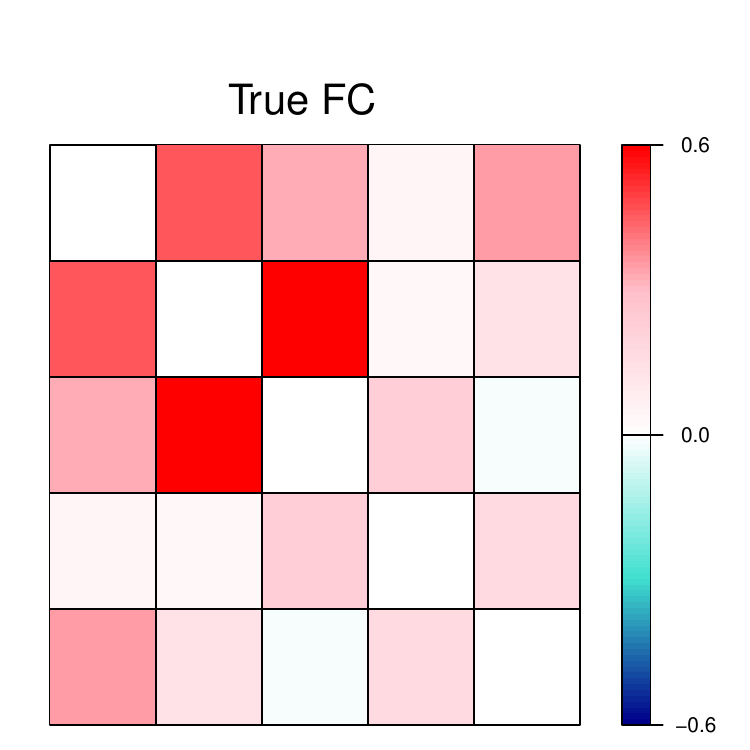} & 
    \includegraphics[height=25mm, page=3, trim = 5mm 3mm 25mm 22mm, clip]{simulation/plots/testsubj2FC.pdf} &
    \includegraphics[height=25mm, page=4, trim = 5mm 3mm 25mm 22mm, clip]{simulation/plots/testsubj2FC.pdf} &
    \includegraphics[height=25mm, page=2, trim = 5mm 3mm 25mm 22mm, clip]{simulation/plots/testsubj2FC.pdf} &
    \includegraphics[height=25mm, page=5, trim = 5mm 3mm 25mm 22mm, clip]{simulation/plots/testsubj2FC.pdf}  \\
    \begin{picture}(10,50)\put(5,25){\rotatebox[origin=c]{90}{Subject 3}}\end{picture} &
    \includegraphics[height=25mm, page=1, trim = 5mm 3mm 25mm 22mm, clip]{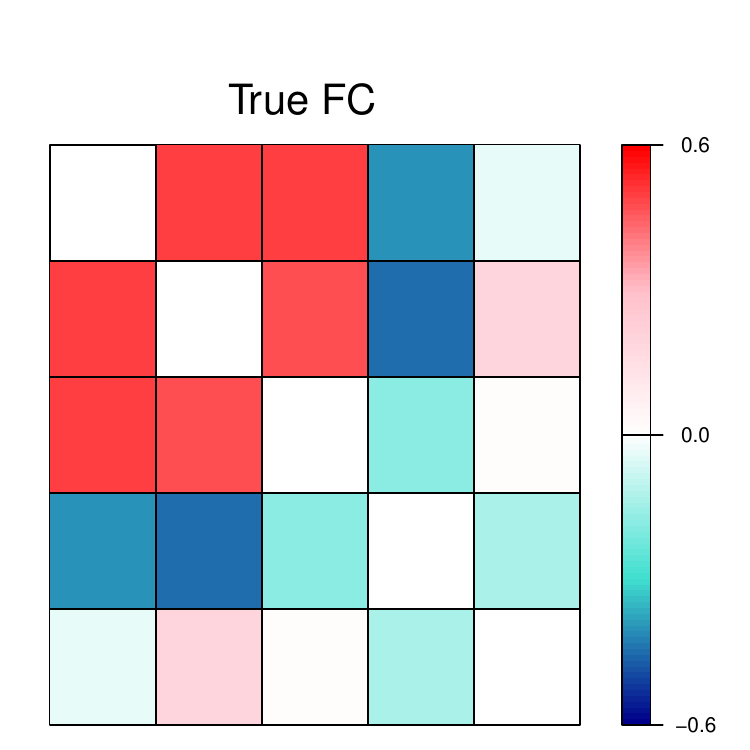} & 
    \includegraphics[height=25mm, page=3, trim = 5mm 3mm 25mm 22mm, clip]{simulation/plots/testsubj3FC.pdf} &
    \includegraphics[height=25mm, page=4, trim = 5mm 3mm 25mm 22mm, clip]{simulation/plots/testsubj3FC.pdf} &
    \includegraphics[height=25mm, page=2, trim = 5mm 3mm 25mm 22mm, clip]{simulation/plots/testsubj3FC.pdf} &
    \includegraphics[height=25mm, page=5, trim = 5mm 3mm 25mm 22mm, clip]{simulation/plots/testsubj3FC.pdf}  \\
    & \multicolumn{5}{c}{\includegraphics[height=10mm]{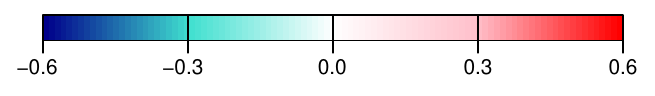}} \\
    \end{tabular}  \\
    \caption{\textit{True FC (based on a held-out ``retest'' dataset) and estimated subject-level FC matrices in the simulation study.} Template ICA (tICA), FC template ICA (FC-tICA), and the ad-hoc method dual regression (DR) are displayed for comparison. In three randomly selected test subjects, all methods produce visually similar estimates. Comparing the estimates to the retest ground truth, similar individual features can be seen, while some features differ. This is to be expected due to natural within-subject variation of FC both within and across sessions.}
    \label{fig:sim:FC_true_est}
\end{figure}

\begin{figure}[H]
    \centering
    \begin{tabular}{ccccc}
     & FC-tICA (VB1) & FC-tICA (VB2) & tICA & DR  \\[10pt]
    \begin{picture}(10,80)\put(5,40){\rotatebox[origin=c]{90}{T = 200 (2.4 min)}}\end{picture} &
    \includegraphics[height=30mm, page=2, trim = 5mm 3mm 25mm 22mm, clip]{simulation/plots/MAE_FC_byduration_VB1.pdf} & 
    \includegraphics[height=30mm, page=2, trim = 5mm 3mm 25mm 22mm, clip]{simulation/plots/MAE_FC_byduration_VB2.pdf} & 
    \includegraphics[height=30mm, page=2, trim = 5mm 3mm 25mm 22mm, clip]{simulation/plots/MAE_FC_byduration_tICA.pdf} & 
    \includegraphics[height=30mm, page=2, trim = 5mm 3mm 25mm 22mm, clip]{simulation/plots/MAE_FC_byduration_DR.pdf} \\
    \begin{picture}(10,80)\put(5,40){\rotatebox[origin=c]{90}{T = 400 (4.8 min)}}\end{picture} &
    \includegraphics[height=30mm, page=3, trim = 5mm 3mm 25mm 22mm, clip]{simulation/plots/MAE_FC_byduration_VB1.pdf} & 
    \includegraphics[height=30mm, page=3, trim = 5mm 3mm 25mm 22mm, clip]{simulation/plots/MAE_FC_byduration_VB2.pdf} & 
    \includegraphics[height=30mm, page=3, trim = 5mm 3mm 25mm 22mm, clip]{simulation/plots/MAE_FC_byduration_tICA.pdf} & 
    \includegraphics[height=30mm, page=3, trim = 5mm 3mm 25mm 22mm, clip]{simulation/plots/MAE_FC_byduration_DR.pdf} \\    
    \begin{picture}(10,80)\put(5,40){\rotatebox[origin=c]{90}{T = 600 (7.2 min)}}\end{picture} &
    \includegraphics[height=30mm, page=4, trim = 5mm 3mm 25mm 22mm, clip]{simulation/plots/MAE_FC_byduration_VB1.pdf} & 
    \includegraphics[height=30mm, page=4, trim = 5mm 3mm 25mm 22mm, clip]{simulation/plots/MAE_FC_byduration_VB2.pdf} & 
    \includegraphics[height=30mm, page=4, trim = 5mm 3mm 25mm 22mm, clip]{simulation/plots/MAE_FC_byduration_tICA.pdf} & 
    \includegraphics[height=30mm, page=4, trim = 5mm 3mm 25mm 22mm, clip]{simulation/plots/MAE_FC_byduration_DR.pdf} \\[10pt]    
    & \multicolumn{4}{c}{\includegraphics[height=8mm]{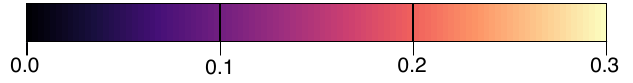}} \\
    \end{tabular}  \\
    \caption{\textit{Accuracy of FC estimates by scan duration.} The strong within-network connections, often of high scientific interest, are highlighted by the green box. The fourth row/column of each matrix corresponds to weaker DMN-visuomotor connections, while the fifth row/column of each matrix corresponds to connections with the motor IC.}
    \label{fig:sim:FC_MAE_by_duration}
\end{figure}

\begin{figure}[H]
    \centering
    \includegraphics[width=5.5in, trim=0 0 0 1cm, clip]{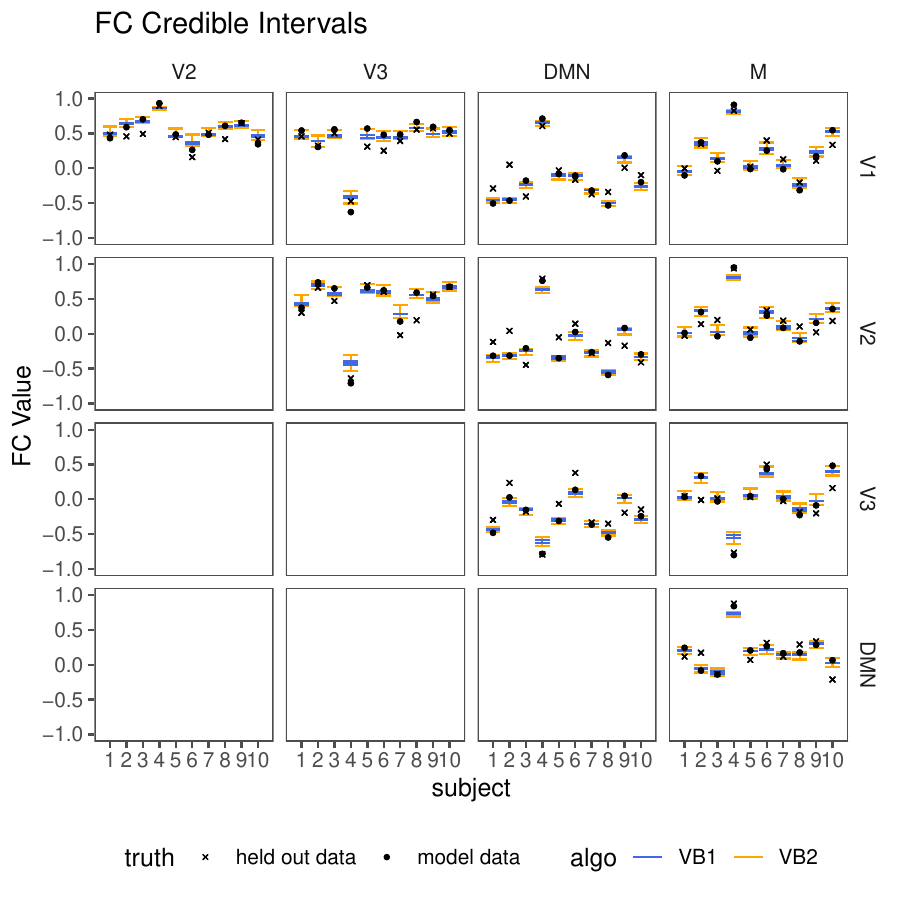}
    \caption{\textit{Credible intervals and ground truth FC values in the simulation study.} The first 10 simulated subjects are shown for illustration. The arrangement of the plots corresponds to the upper triangle of the FC matrices shown in Figure \ref{fig:sim:FC_true_est} and elsewhere. The row/column labels indicate the three visual ICs (V1-V3), the DMN IC, and the motor IC (M). Two ground truth values are shown: the true FC during the part of the session used for model estimation (``model data'') and the true FC during the part of the session held out to assess predictive accuracy (``held out data'').}
    \label{fig:sim:FC_CIs}
\end{figure}

\begin{figure}[H]
    \centering
    \includegraphics[scale=0.52, trim = 0 5mm 0 2cm, clip]{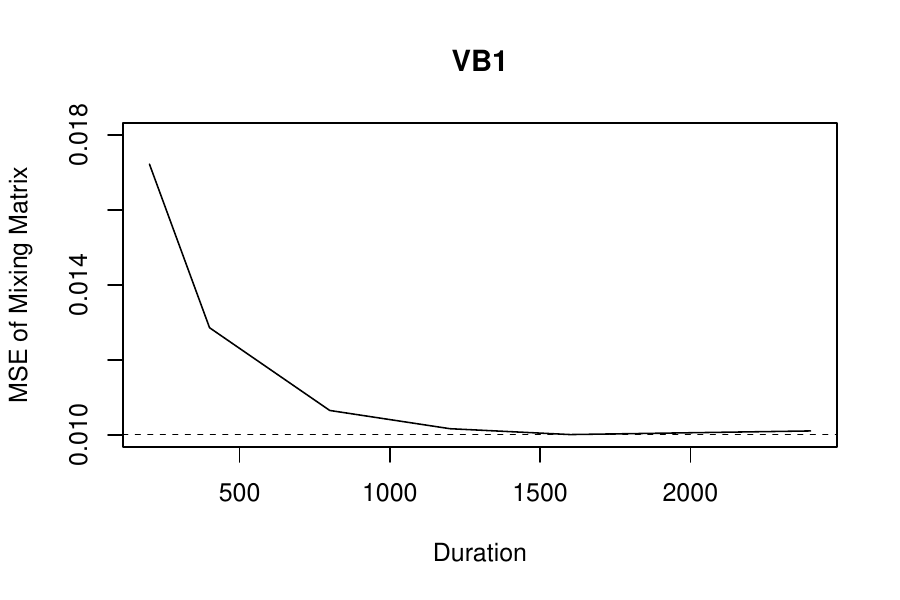}\\
    \includegraphics[scale=0.52, trim = 0 0 0 5mm, clip]{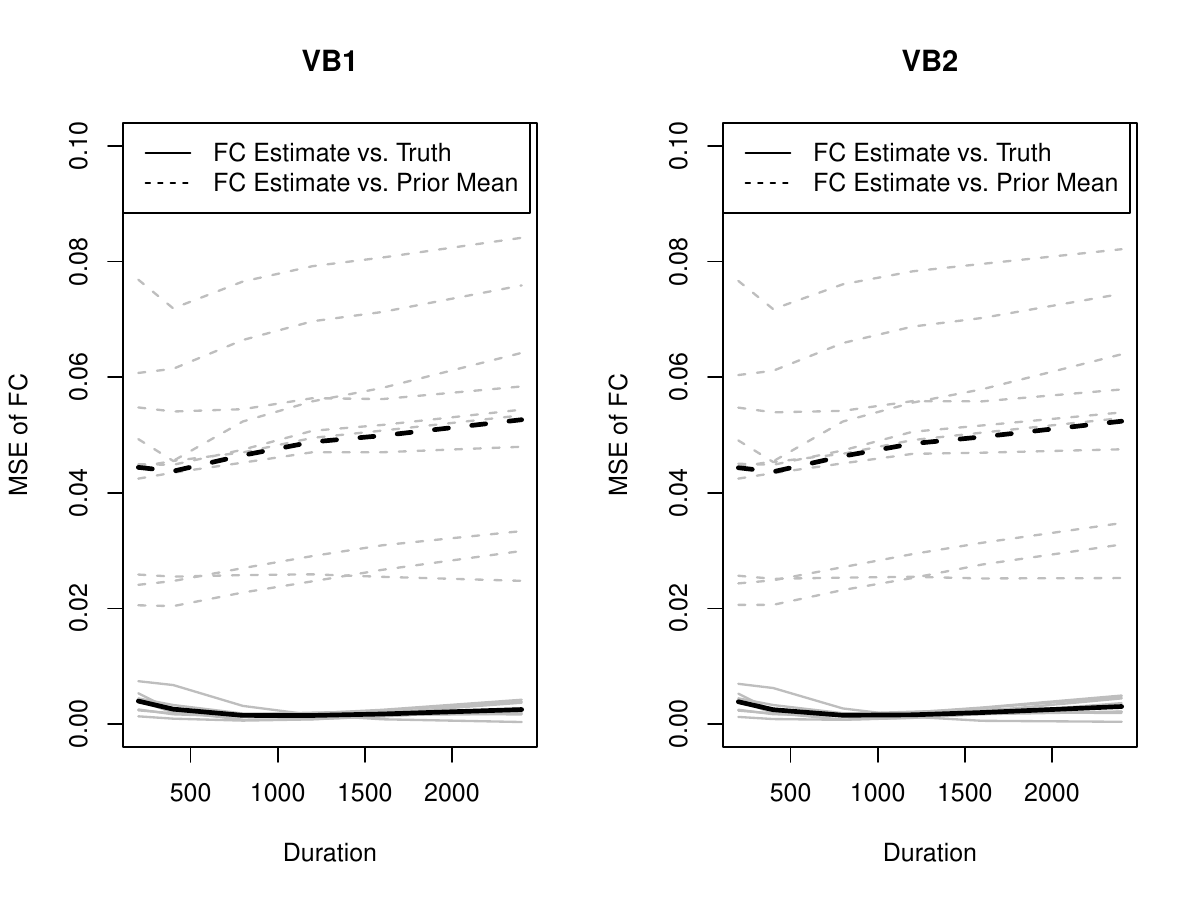}
    \caption{\black \textit{The effects of increasing $T$ in long-duration simulation study.} {Top: }MSE of the $T\times Q$ mixing matrix $\bA$. Results shown correspond to VB1 but are similar for VB2 (results not shown). {Bottom:} MSE of the FC matrix $Cor(\bA)$, relative to the true subject-level FC (solid lines) and relative to the prior mean (dashed lines). The gray lines correspond to the 10 individual FC pairs, while the black lines show their mean.}
    \label{fig:MSE_vs_T}
\end{figure}

\newpage
{\black
\subsection{Long-Duration Simulation}
\label{app:long_duration_sim}

Here, we describe an additional simulation study serving as a preliminary empirical investigation of the asymptotic properties of our model as $T\to\infty$, focusing on the estimation of FC. Here we also examine the influence of the prior when $T$ is small. We generated a set of simulated data on $n=50$ test subjects with duration varying from $T=200$ (2.5 min, very short) to $T=2400$ (30 min, very long). For each simulated subject, the true FC was set to be the empirical correlation of the time series (duration $1200$) used in the main simulation study for one HCP subject.

Time courses were generated as multivariate Normal realizations with mean zero and covariance equal to the true subject-level FC. This ensures constant FC over the time series, since in the realistic time courses used in the main simulation study, FC patterns may vary over time due to natural fluctuations in brain function. While we chose that more realistic setting for our main simulation study, such fluctuations complicate a study of convergence. To maintain the typical setting of $V \gg T$, we enlarged the spatial dimension by resampling both hemispheres to approximately 10,000 vertices, resulting in $V = 18,792$. We ran both VB algorithms to a tolerance of $1 \times 10^{-6}$ for each subject and scan duration. 

\textbf{Figure \ref{fig:MSE_vs_T}} summarizes the results of this analysis. The top panel shows the MSE over elements and subjects of the $T\times Q$ mixing matrix $\bA$. As $T$ increases to around $T = 1600$, the estimation of A becomes more accurate. This may be a downstream effect of improved estimation of $\bS$ with larger $T$, observed previously \citep{mejia2020template}. However, as $T$ becomes very large, the estimation of $\bA$ slightly worsens. This may be due to the increasingly high dimensionality of $\bA$, while the spatial resolution is fixed. For example, at $T = 2400$, $\bA$ contains $12,000$ parameters, nearing the number of spatial observations. It may also reflect diminishing ``downstream effects'' mentioned above, as the estimation of $\bS$ begins to converge. 

The bottom panel of \textbf{Figure \ref{fig:MSE_vs_T}} shows the MSE over subjects of the FC matrix, $Cor(\bA)$, relative to the true subject-level FC (solid lines) and the prior mean (dashed lines). For both VB algorithms, the FC estimates are much closer to the truth than to the prior mean, even for the shortest duration ($T = 200$). As the scan duration increases, the estimates get
further away from the prior mean, reflecting a growing influence of the data. For very large $T$, there is a slight worsening of the FC estimation accuracy mirroring the similar effect observed for $\bA$ above. Since $\bA$ determines $Cor(\bA)$, more estimation error in $\bA$ has downstream consequences for estimation of $Cor(\bA)$. Importantly, however, the data dominate the prior across all scan duration, and the role of the prior diminishes with increasing $T$.}

\newpage
\section{IC Network Assignments}
\label{app:IC_assignments}

\begin{figure}[H]
    \centering
    \includegraphics[width=0.19\textwidth]{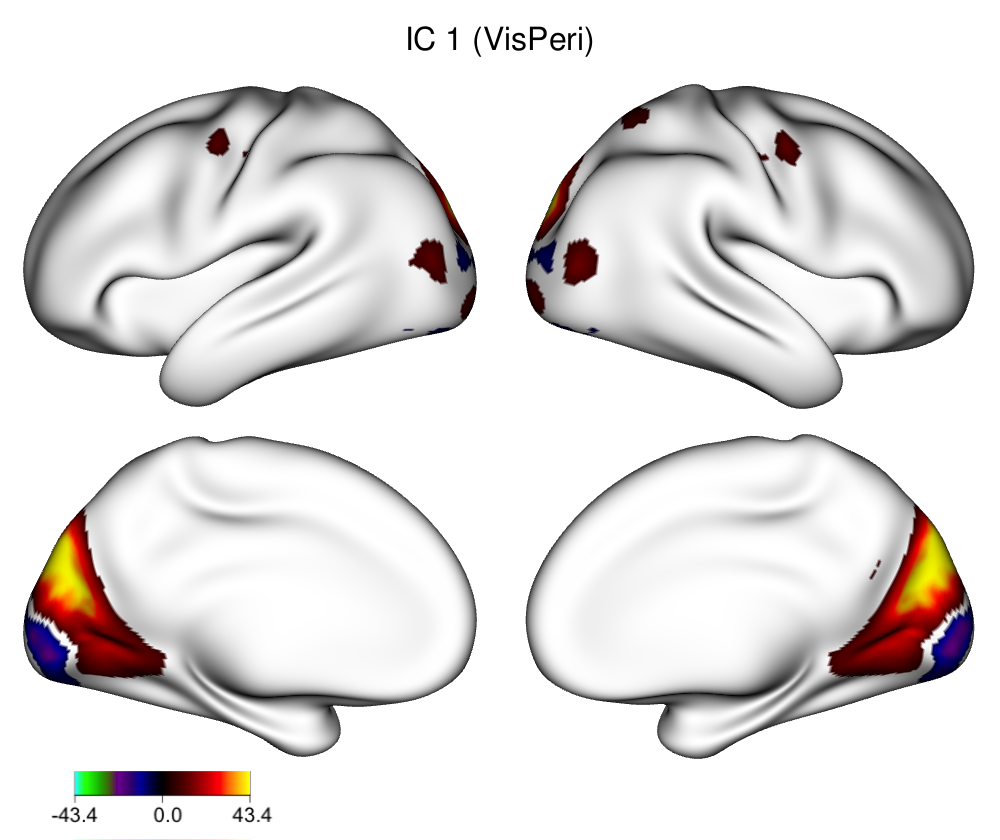}
    \includegraphics[width=0.19\textwidth]{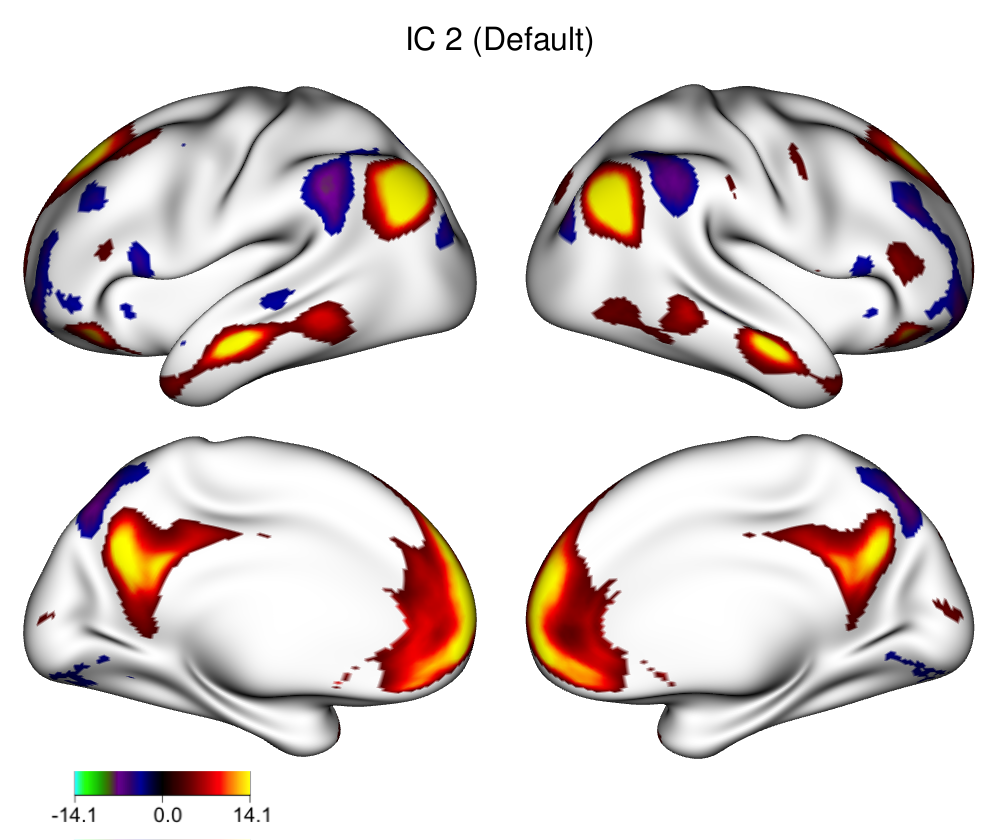}
    \includegraphics[width=0.19\textwidth]{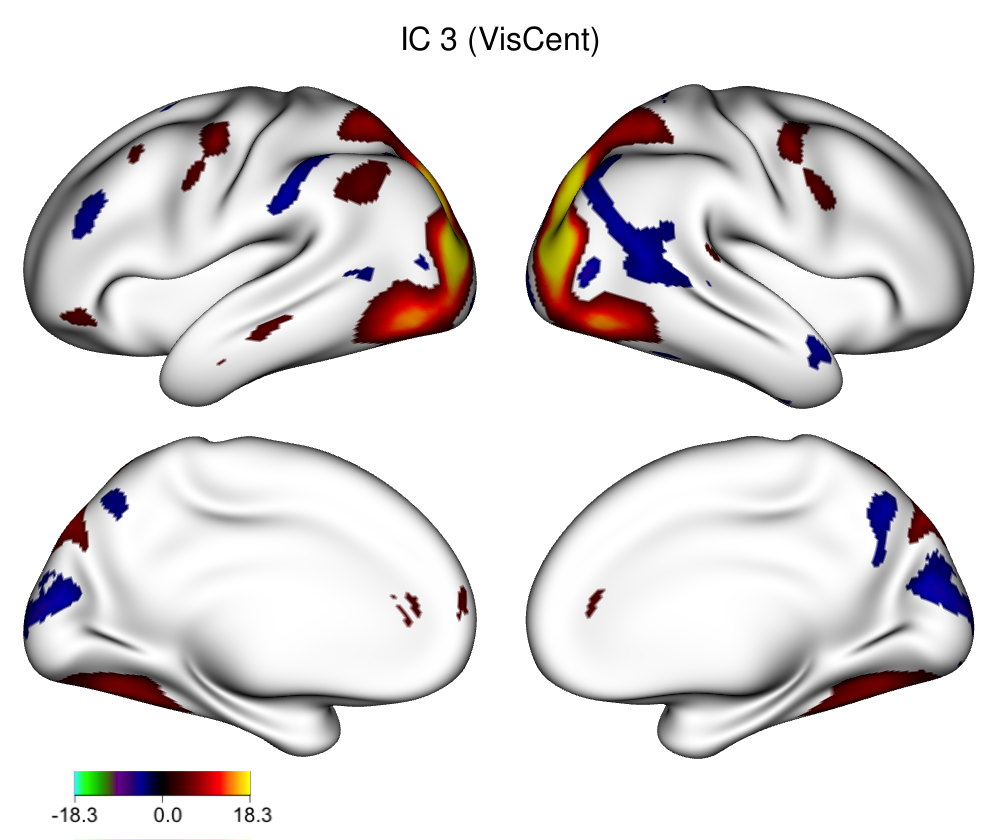}
    \includegraphics[width=0.19\textwidth]{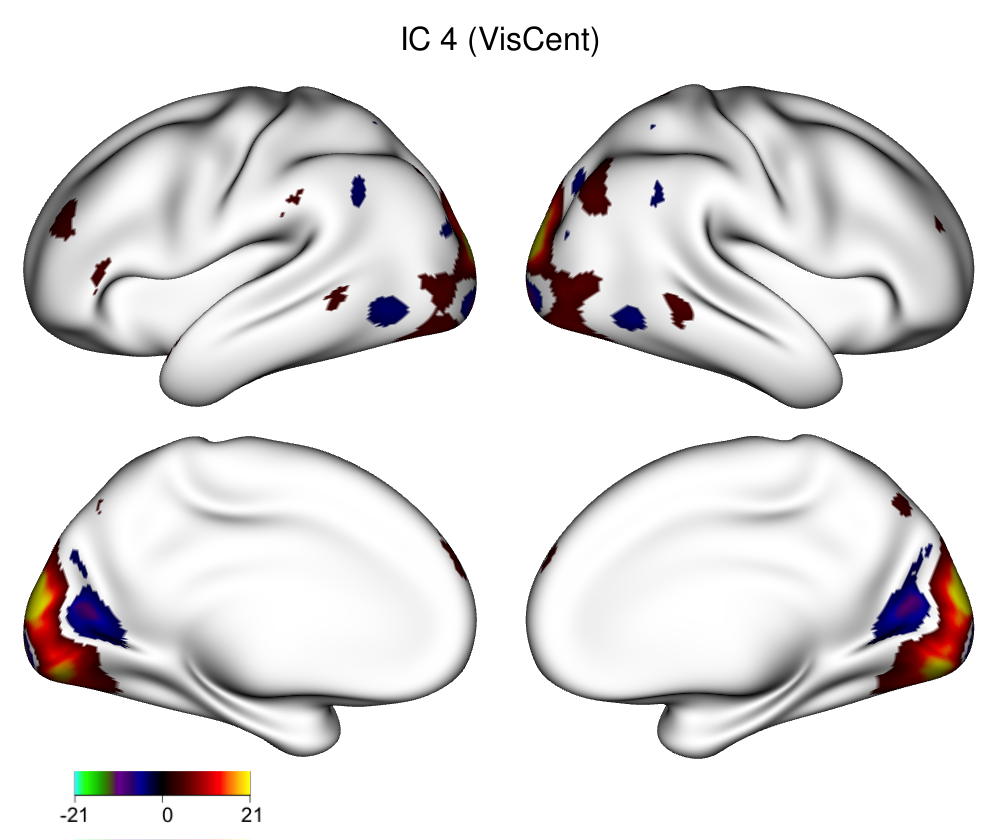}
    \includegraphics[width=0.19\textwidth]{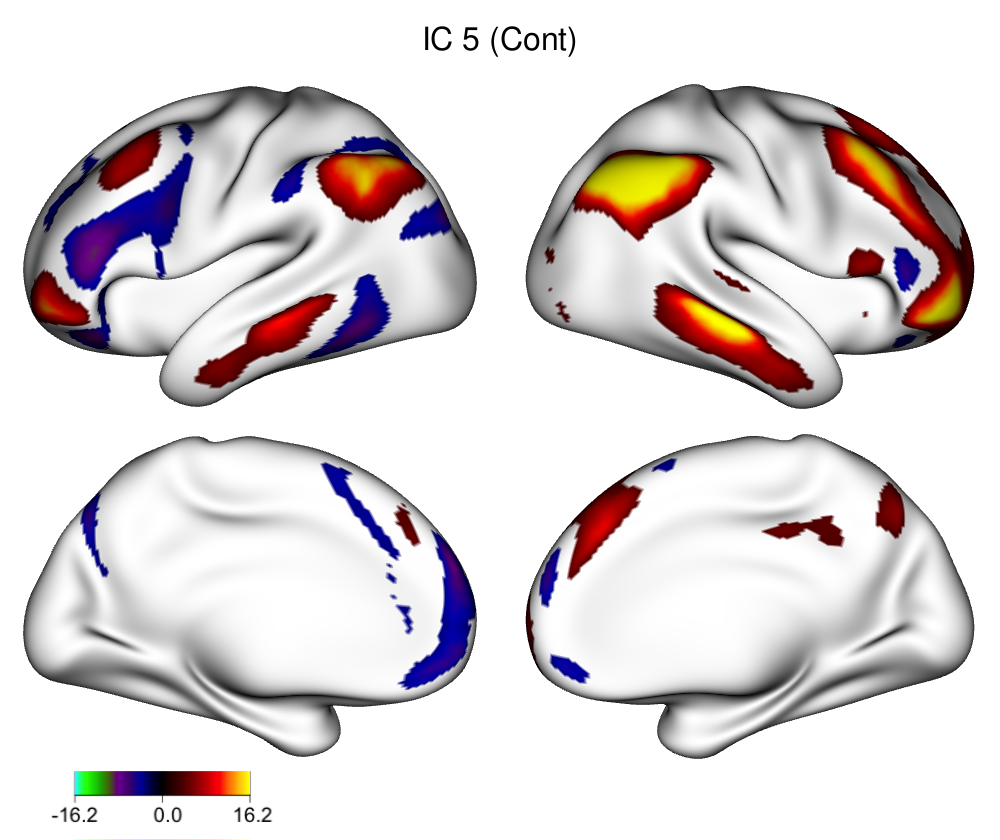}
   \includegraphics[width=0.19\textwidth]{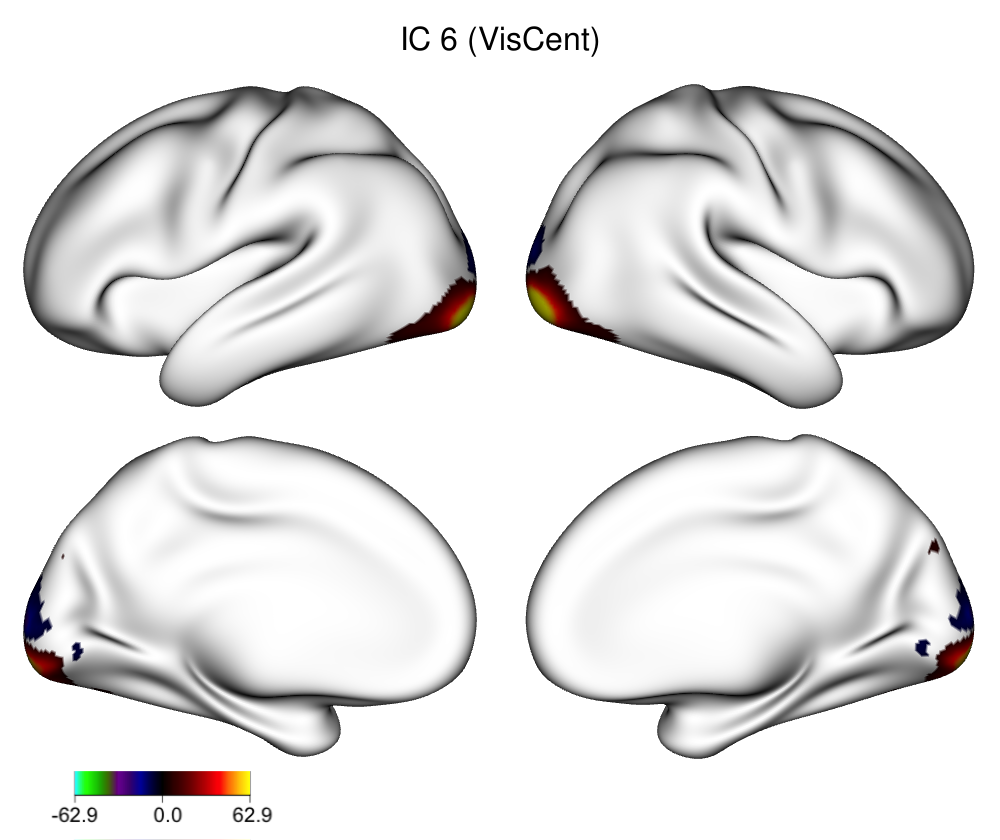}
    \includegraphics[width=0.19\textwidth]{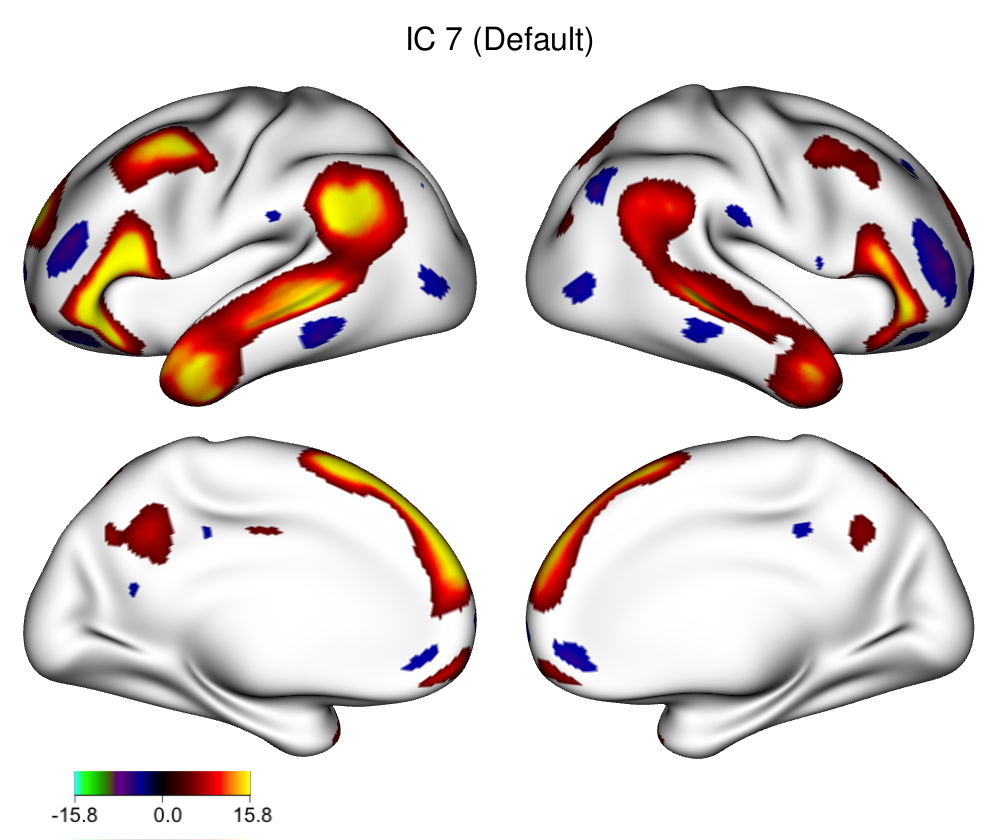}
    \includegraphics[width=0.19\textwidth]{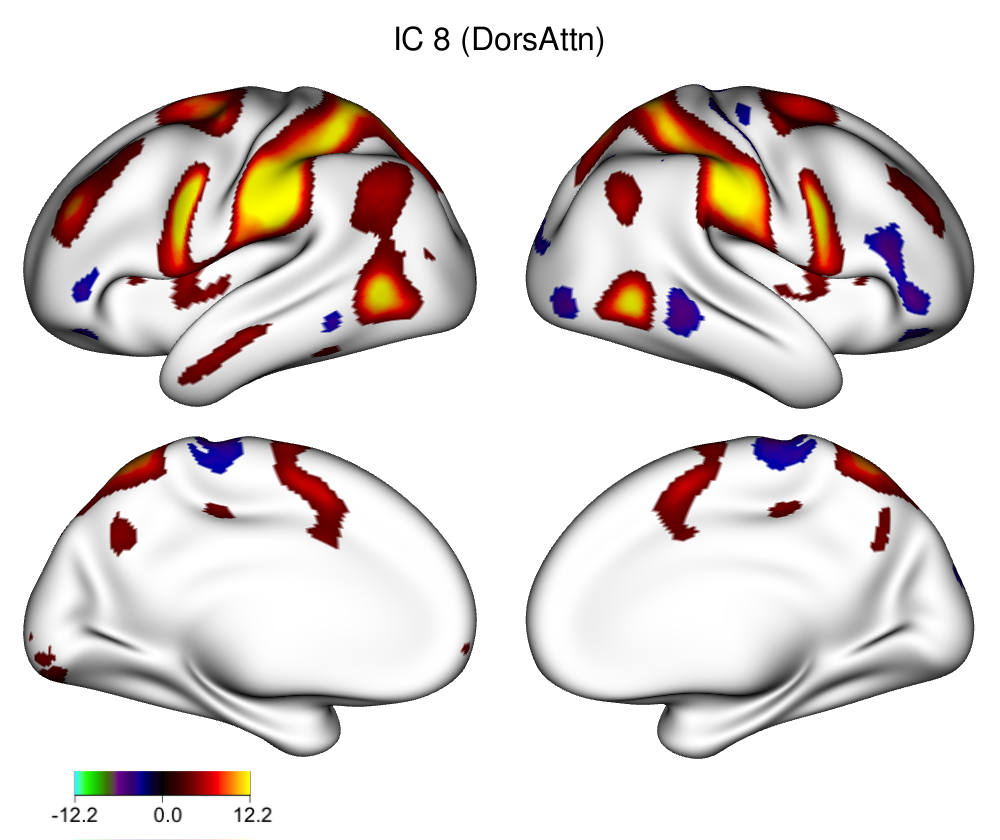}
    \includegraphics[width=0.19\textwidth]{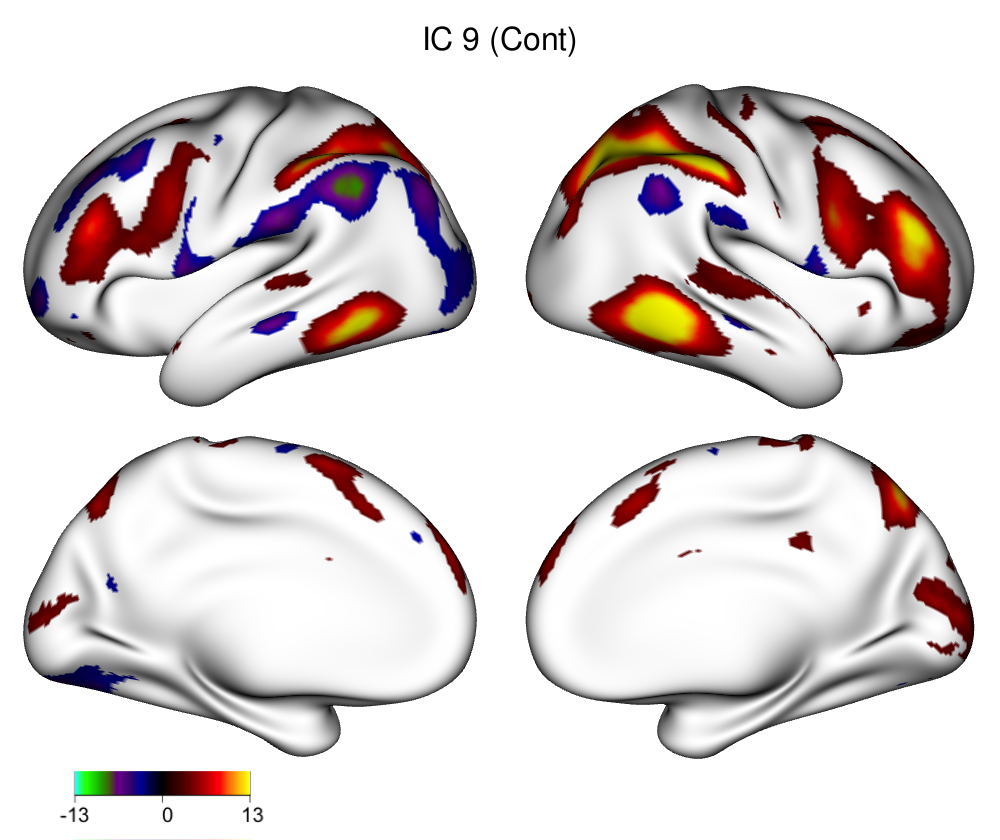}
    \includegraphics[width=0.19\textwidth]{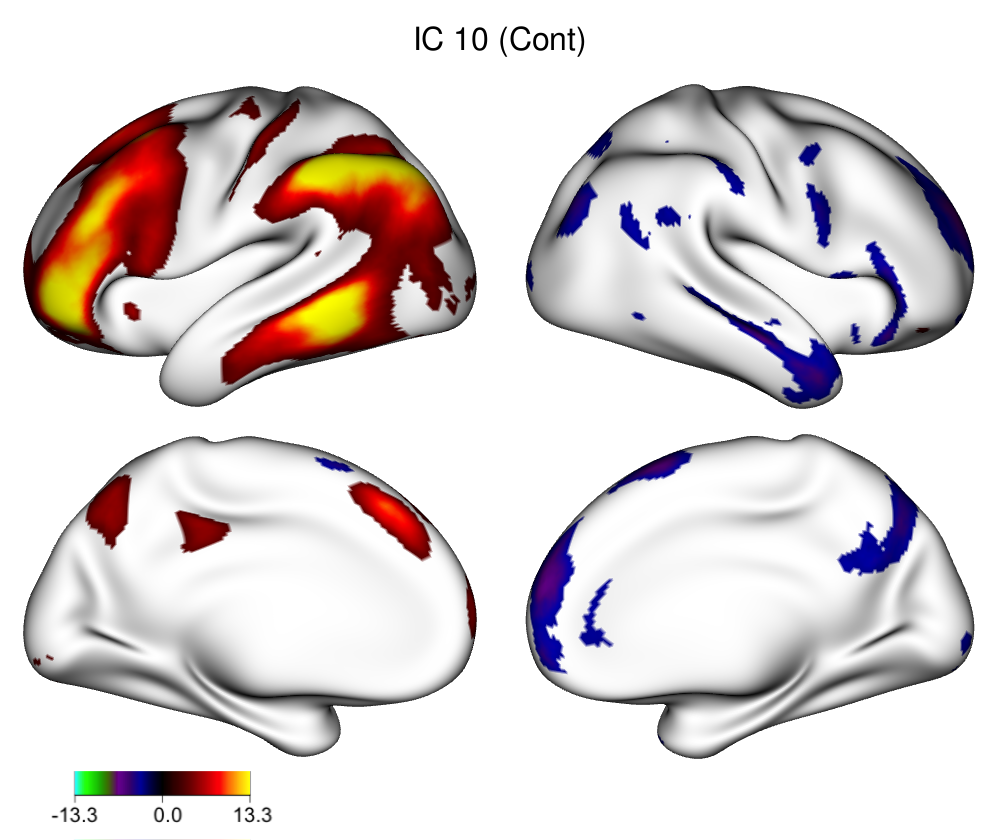}
   \includegraphics[width=0.19\textwidth]{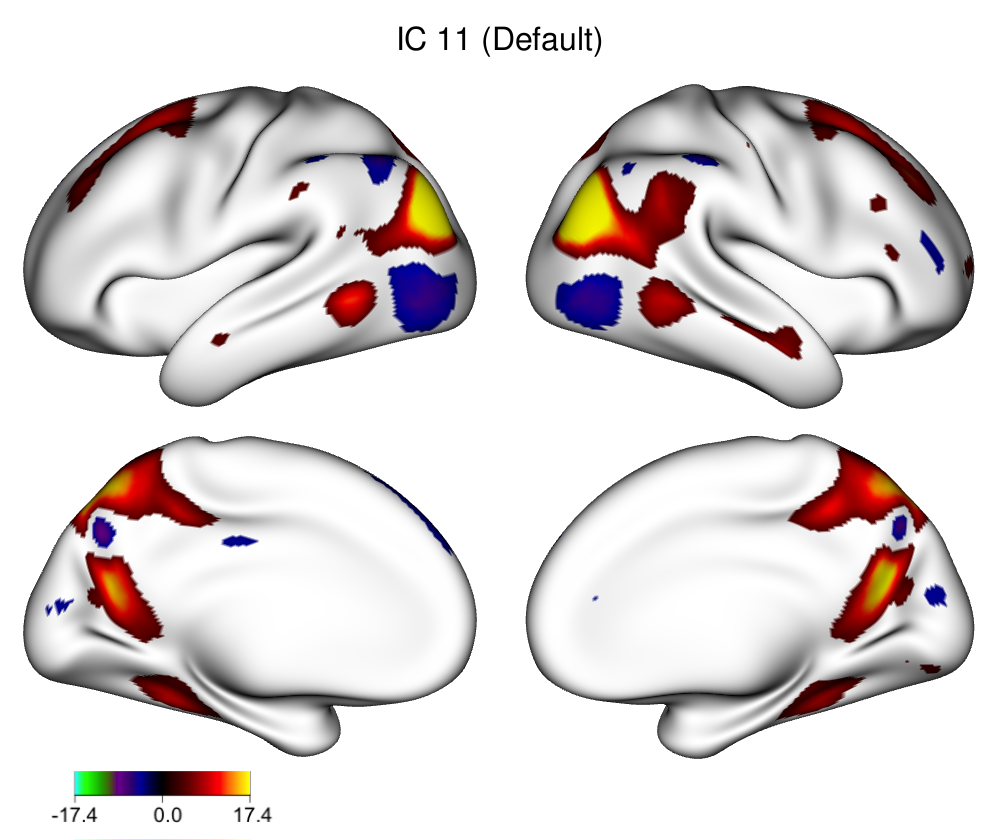}
    \includegraphics[width=0.19\textwidth]{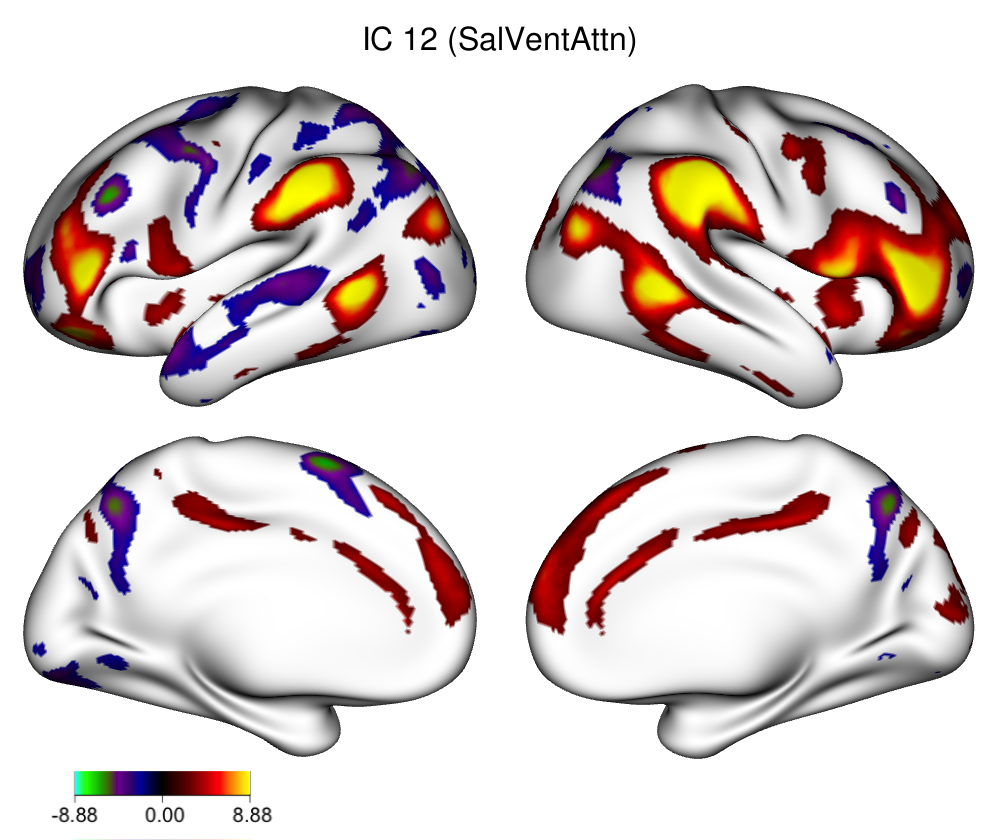}
    \includegraphics[width=0.19\textwidth]{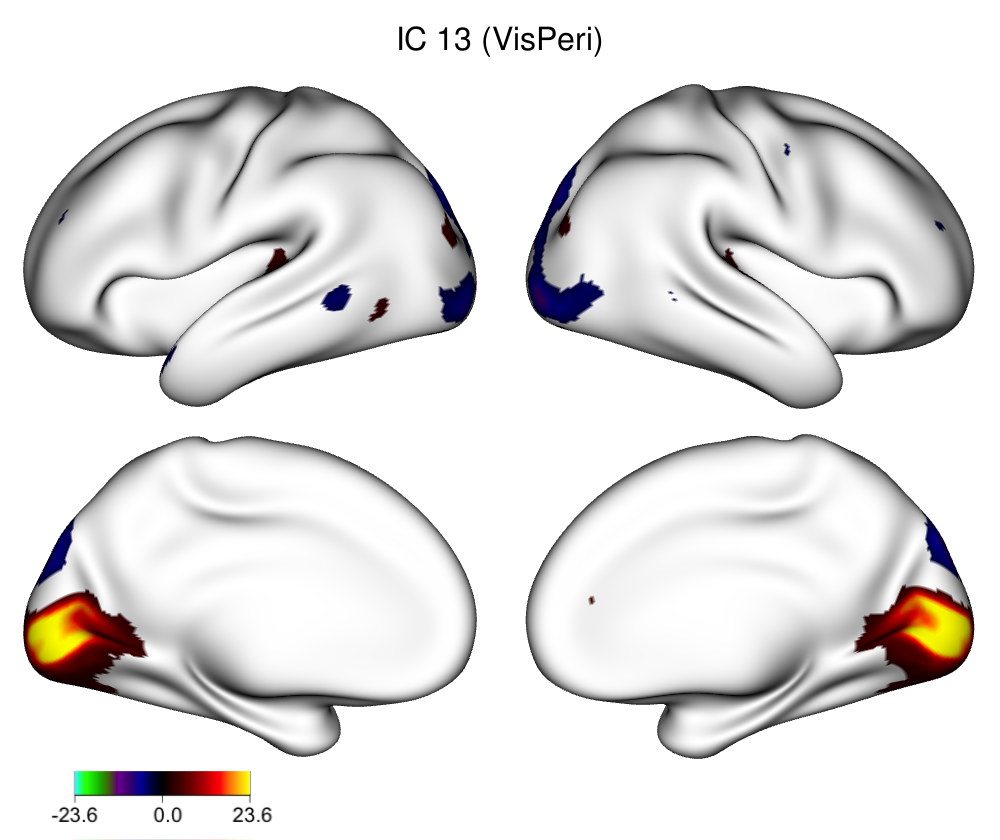}
    \includegraphics[width=0.19\textwidth]{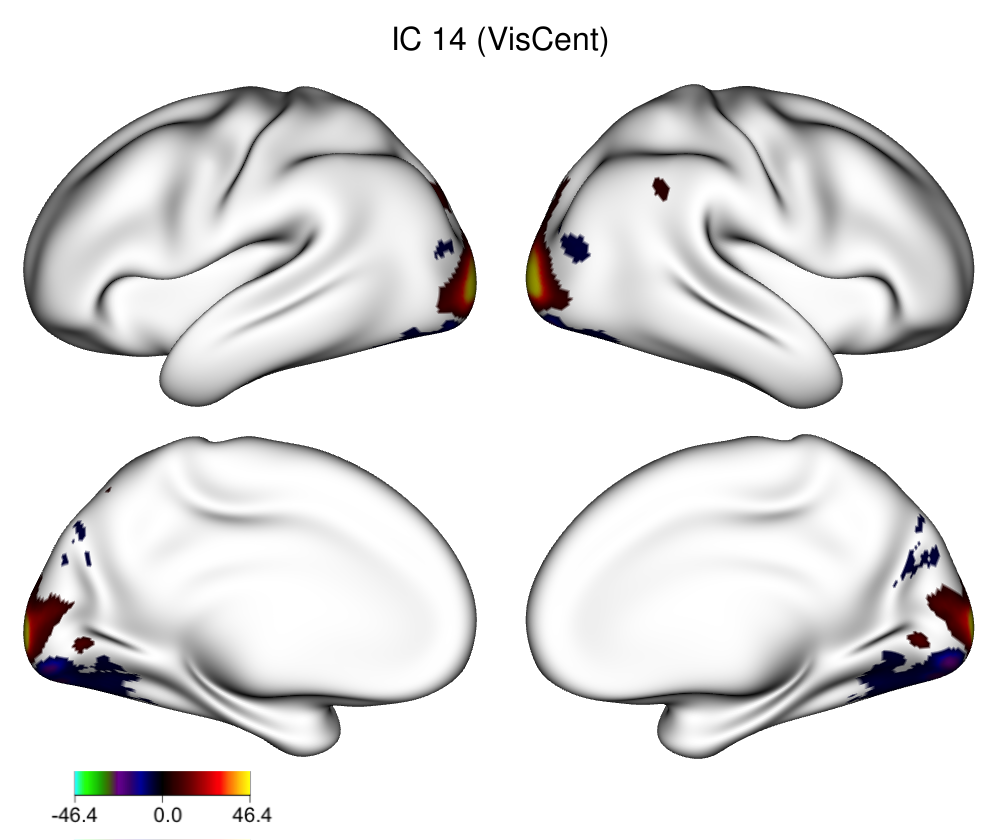}
    \includegraphics[width=0.19\textwidth]{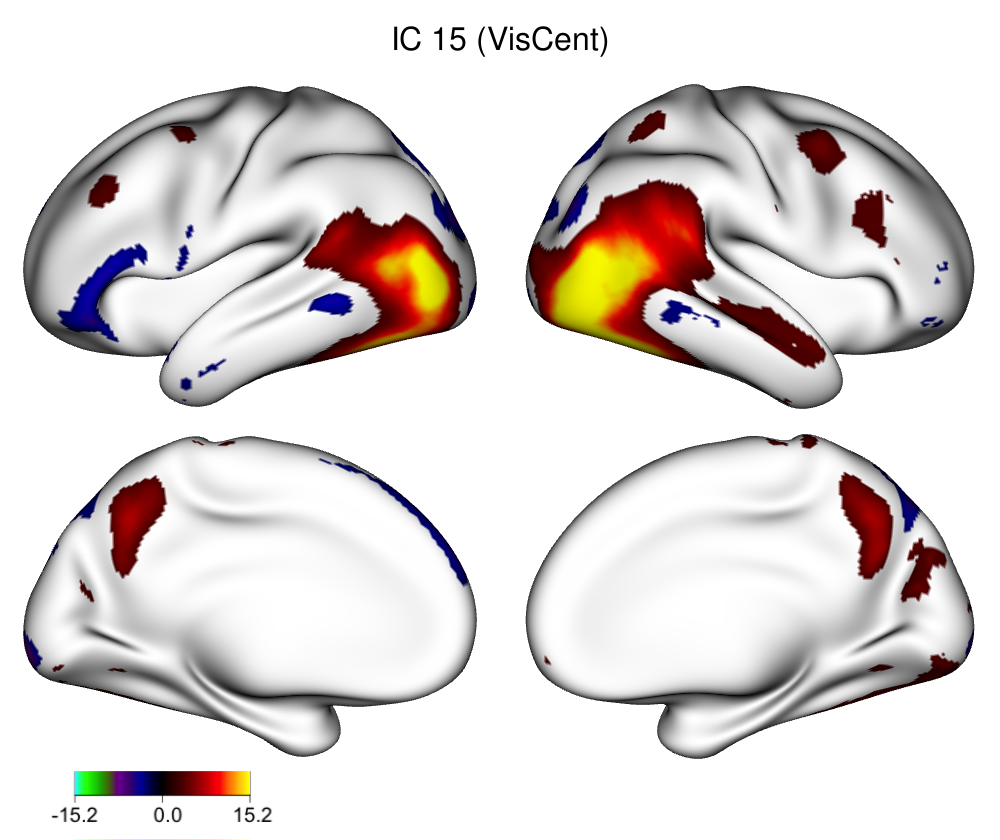}
   \includegraphics[width=0.19\textwidth]{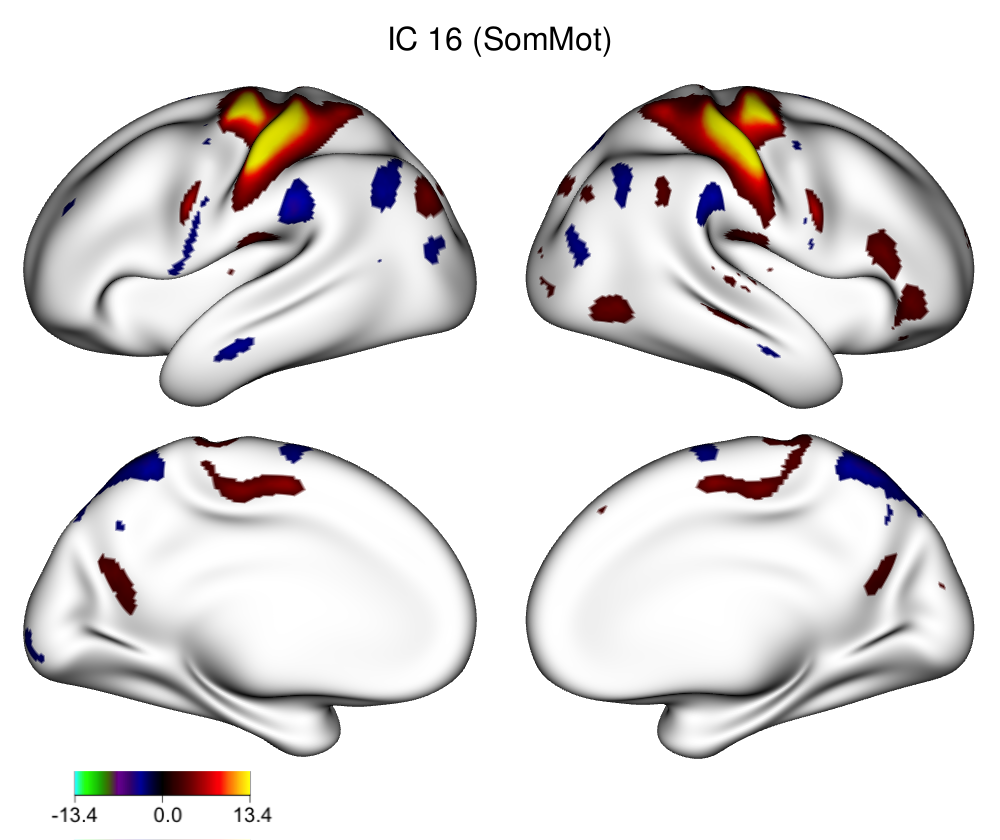}
    \includegraphics[width=0.19\textwidth]{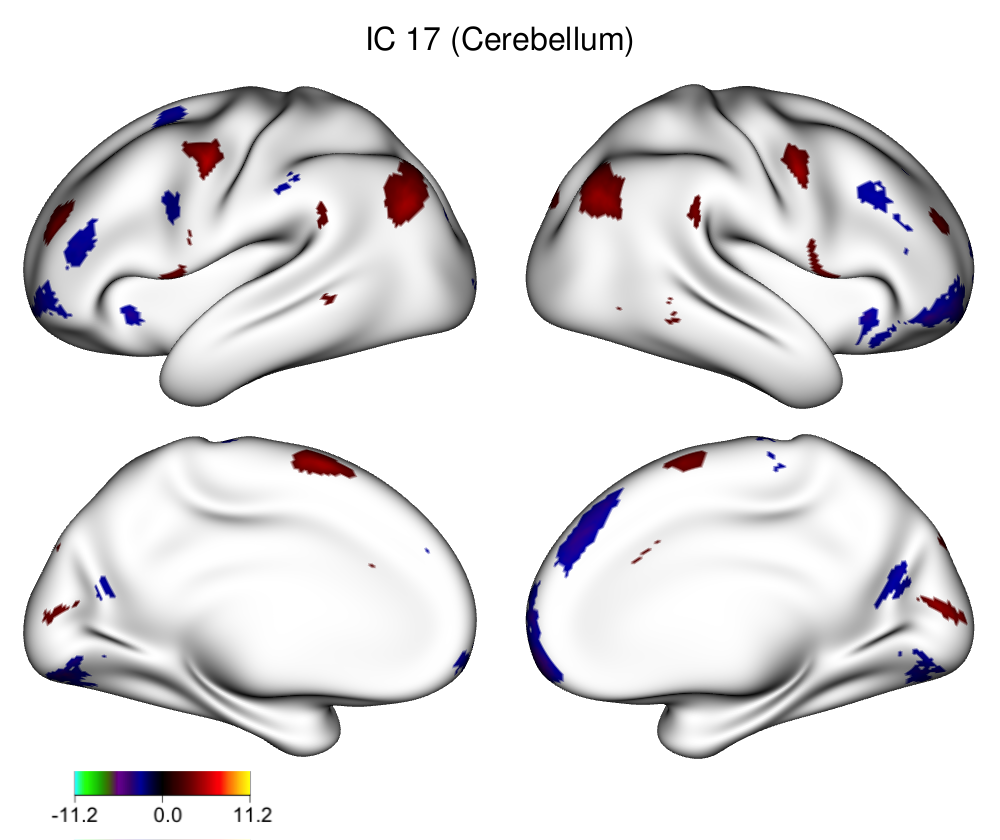}
    \includegraphics[width=0.19\textwidth]{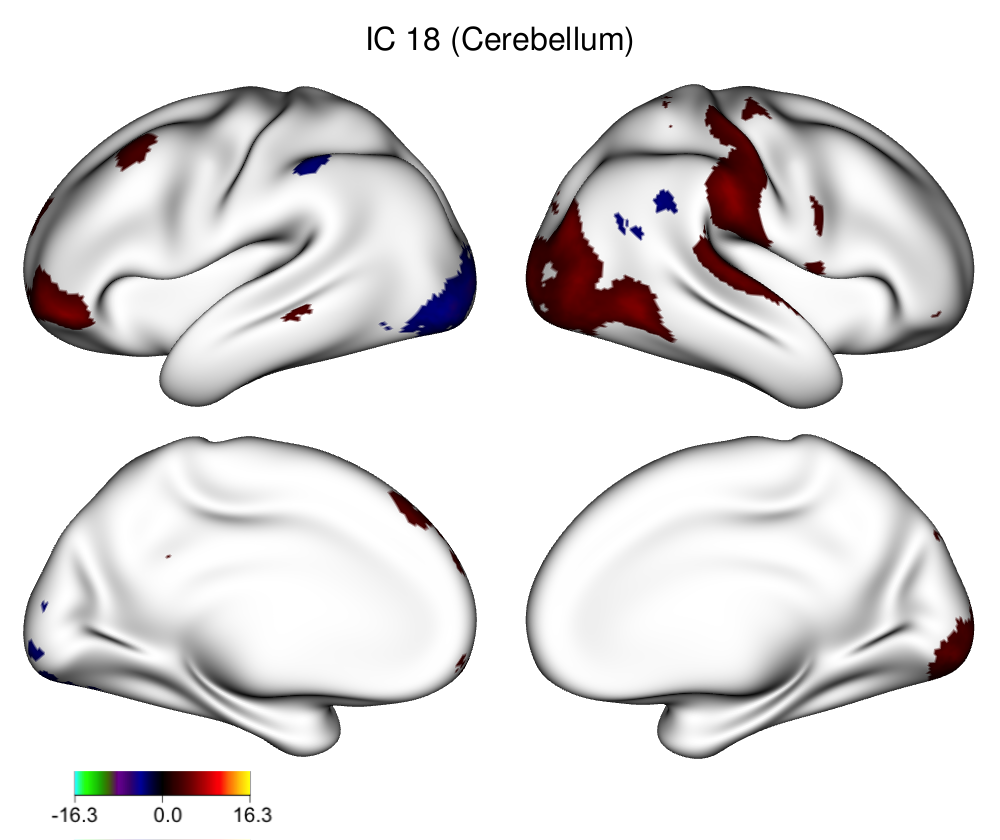}
    \includegraphics[width=0.19\textwidth]{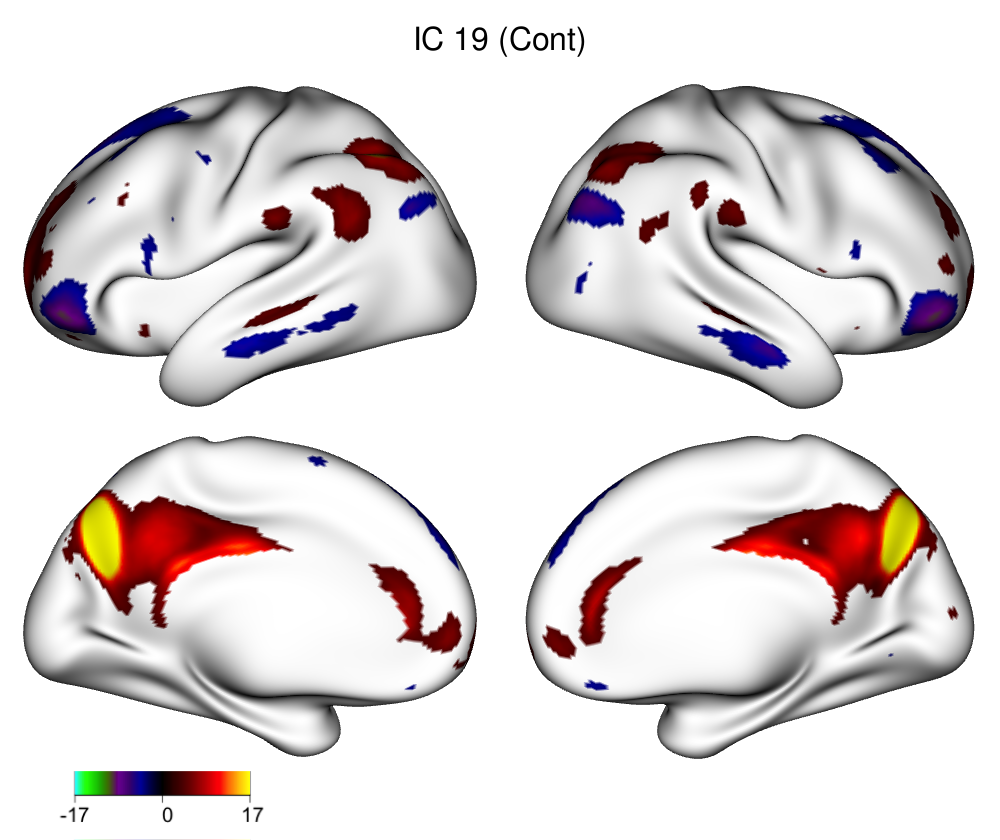}
    \includegraphics[width=0.19\textwidth]{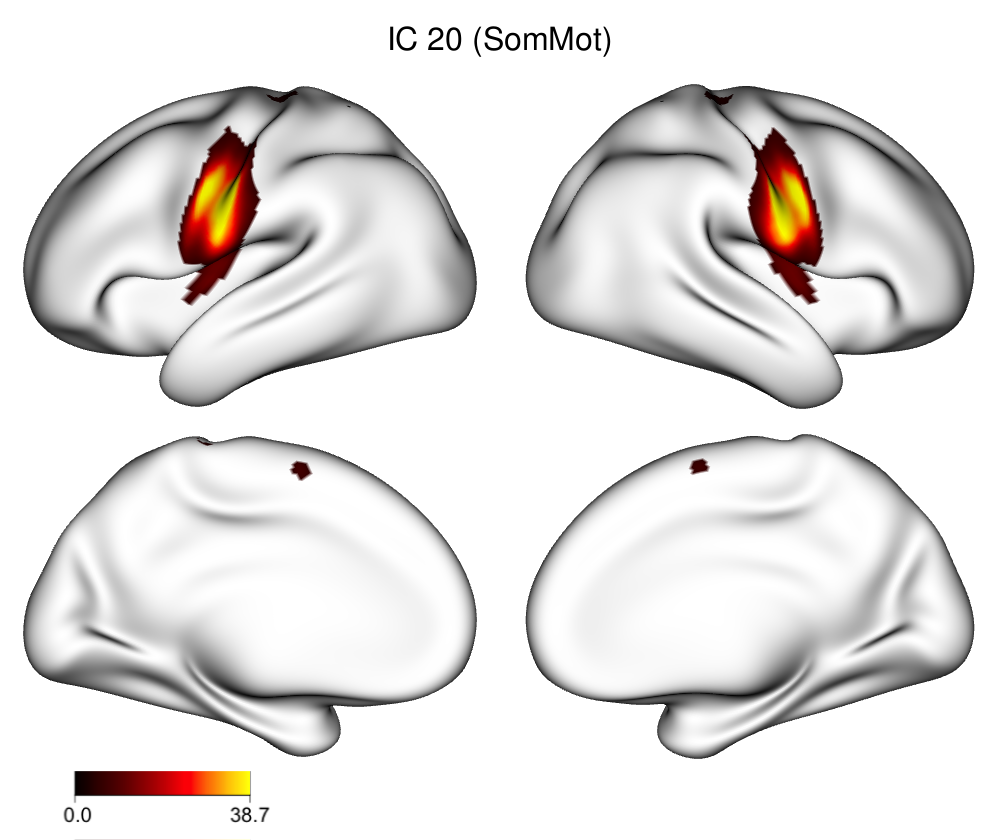}
   \includegraphics[width=0.19\textwidth]{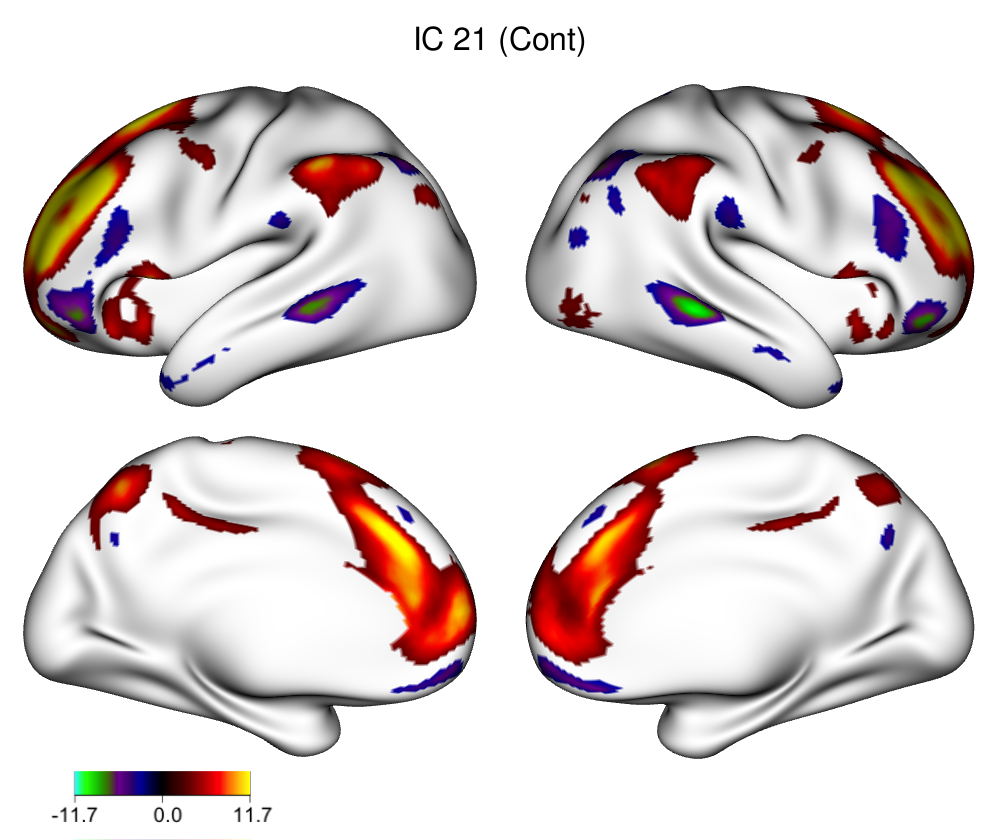}
    \includegraphics[width=0.19\textwidth]{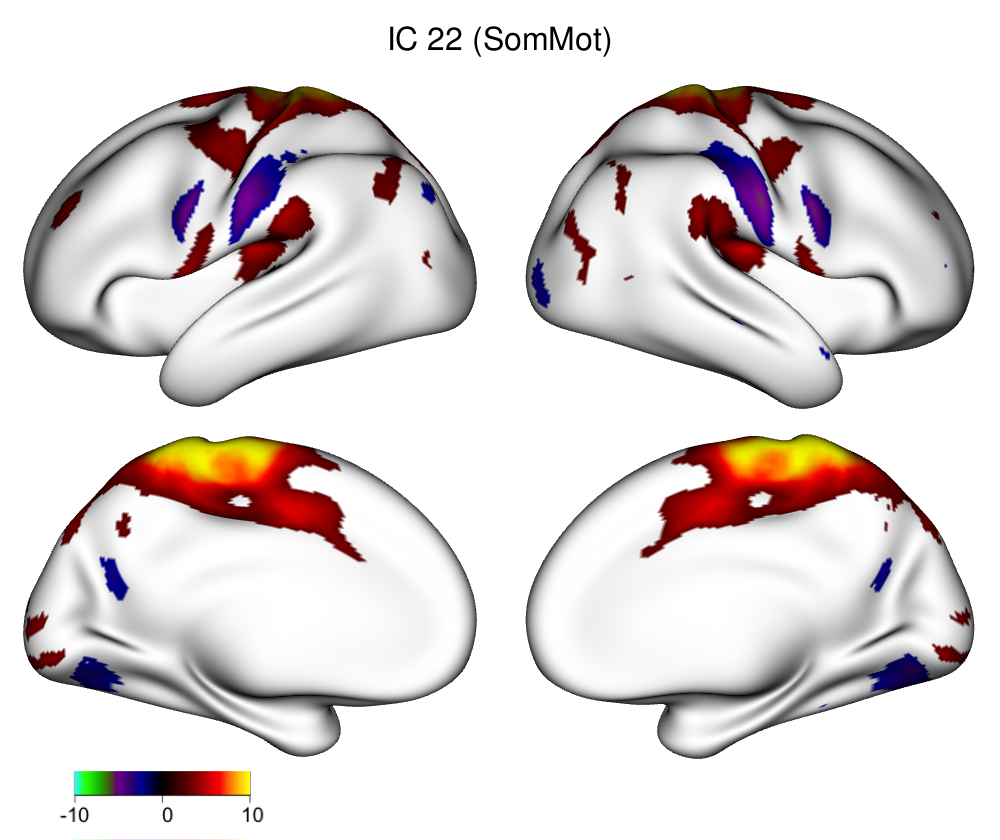}
    \includegraphics[width=0.19\textwidth]{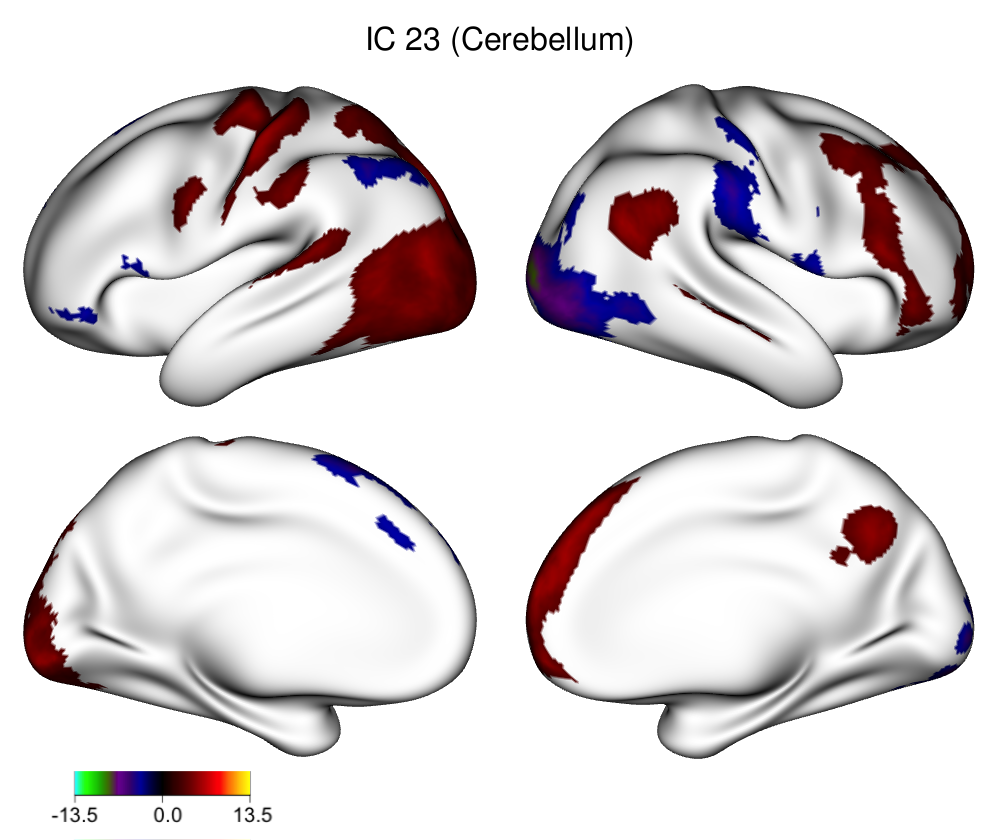}
    \includegraphics[width=0.19\textwidth]{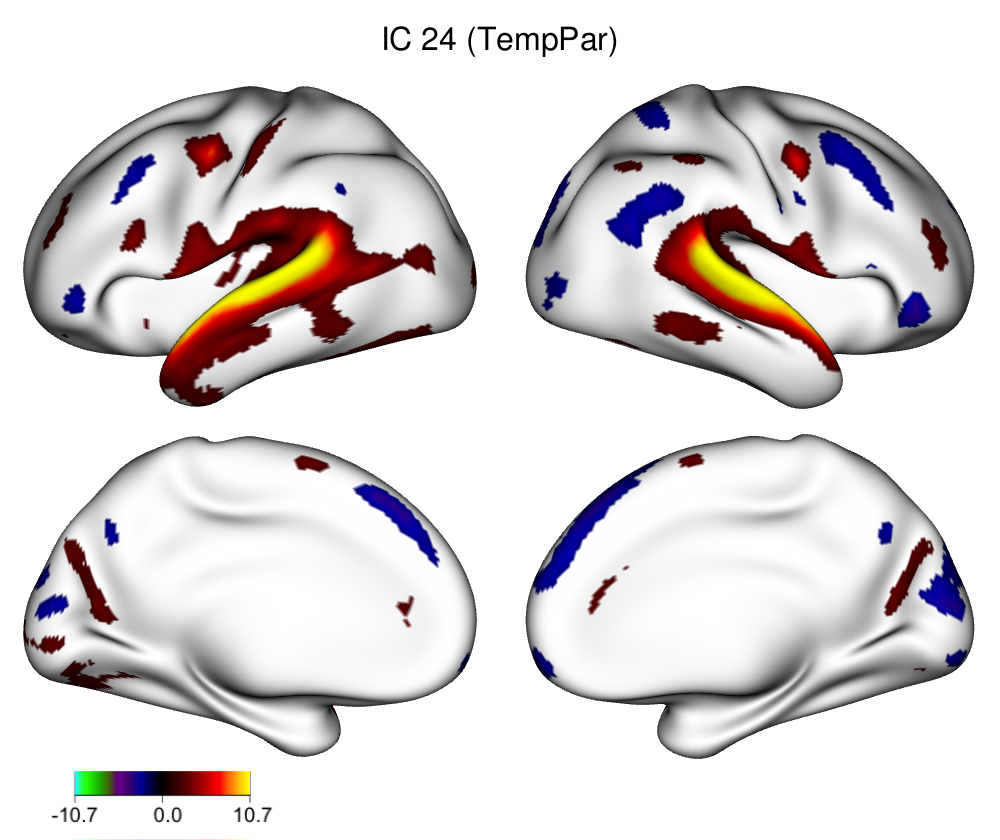}
    \includegraphics[width=0.19\textwidth]{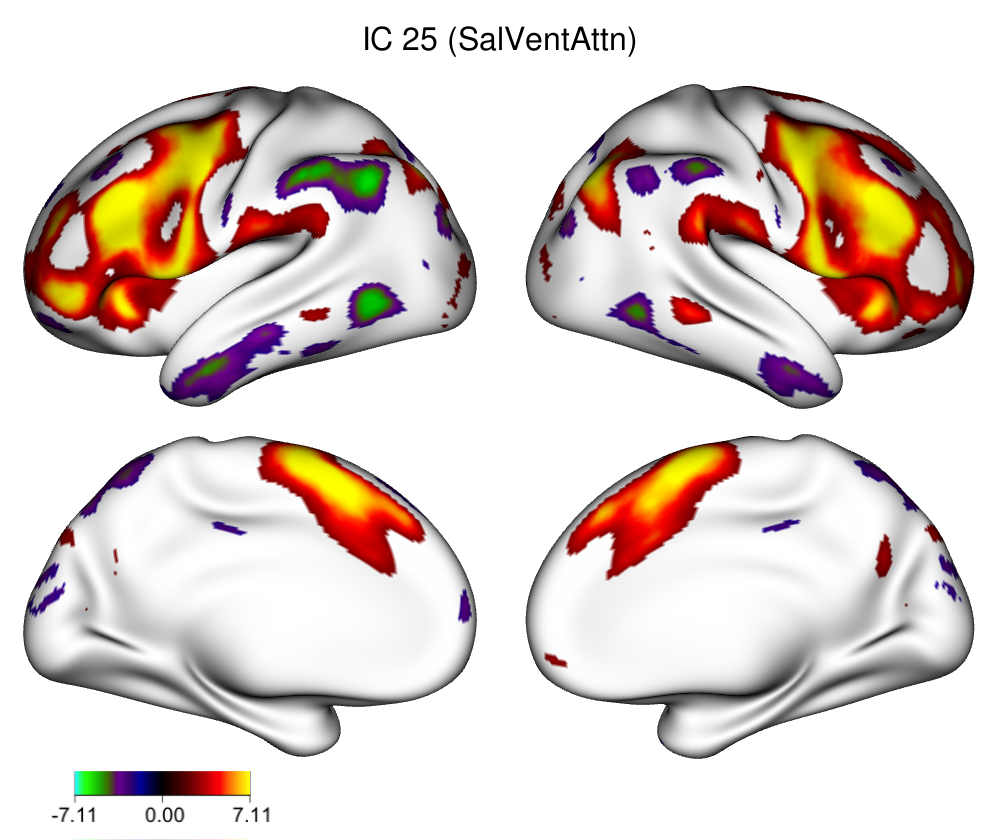}    
    \caption{Thresholded group IC maps and network assignments in HCP data analysis.}
    \label{fig:DA:groupICs_thr}
\end{figure}

For visualization of the FC matrices, it is common to group ICs by cortical resting-state network (RSN) (e.g. visual, motor) or volumetric region (e.g. cerebellum, basal ganglia). We therefore first assign each IC to a set of established network maps \citep{yeo2011organization} and Freesurfer subcortical parcels \citep{fischl2012freesurfer}. For this purpose, we first threshold each group ICA map plus or minus $x$ standard deviations from the median. The choice of $x$ is subjective and may depend on ICA resolution.  For the low-resolution HCP group ICA maps, we find $x=1$ standard deviations to capture the regions of engagement in each IC reasonably well (see Figure \ref{fig:DA:groupICs_thr}.)  After thresholding, we sum up the magnitudes of all voxels and vertices that overlap with a given parcel, then standardize by the square root of the parcel size. Each IC map is then assigned to the RSN or parcel with maximal score. In one case, a component scored highest for cerebellum but showed stronger correspondence for a cortical network with second-highest score. To avoid such cases, for ICs assigned to the cerebellum but receiving a runner up score of at least 75\% the cerebellar score, we reassign it to the runner up network. Figure \ref{fig:DA:groupICs_thr} shows the thresholded group IC maps and assignments, and Table \ref{tab:RSNs} shows the final number of ICs assigned to each network.  The full IC assignments are reported in Table \ref{tab:HCP25_networks}.


\begin{table}[H]
    \centering
    \begin{tabular}{|c|c|c|}
    \hline
    Network/Region Name & Group Name & \# of ICs \\[-2pt]
    \hline
  Visual (Central) & V & 5  \\ 
  Visual (Peripheral) & V & 2 \\ 
  Somatomotor & M & 3 \\ 
  Dorsal Attention & A & 1 \\ 
  Ventral Attention & A & 2 \\ 
  Limbic & L & 0 \\
  Control & C & 5 \\ 
  Default & D & 3 \\ 
  Temporal Parietal & TP & 1 \\ 
  Cerebellum & CB & 3 \\   
  \hline
     \end{tabular}
    \caption{\textit{Number of group ICs assigned to each functional brain network or subcortical region in HCP data analysis.} The second column shows the short group name, used to order and label the FC matrices shown in subsequent figures. }
    \label{tab:RSNs}
\end{table}


\begin{table}[H]
\centering
\begin{tabular}{| r | c |}
  \hline
IC & RSN or Parcel \\ 
  \hline
3 & VisCent \\ 
  4 & VisCent \\ 
  6 & VisCent \\ 
  14 & VisCent \\ 
  15 & VisCent \\ 
  1 & VisPeri \\ 
  13 & VisPeri \\ 
  16 & SomMot \\ 
  20 & SomMot \\ 
  22 & SomMot \\ 
  8 & DorsAttn \\ 
  12 & SalVentAttn \\ 
  25 & SalVentAttn \\ 
  5 & Cont \\ 
  9 & Cont \\ 
  10 & Cont \\ 
  19 & Cont \\ 
  21 & Cont \\ 
  2 & Default \\ 
  7 & Default \\ 
  11 & Default \\ 
  24 & TempPar \\ 
  17 & Cerebellum \\ 
  18 & Cerebellum \\ 
  23 & Cerebellum \\ 
   \hline
\end{tabular}
\caption{Group ICs assigned to each resting-state network (RSN) in HCP data analysis.}
\label{tab:HCP25_networks}
\end{table}

\newpage
\section{Additional Data Analysis Results}
\label{app:DA_results}


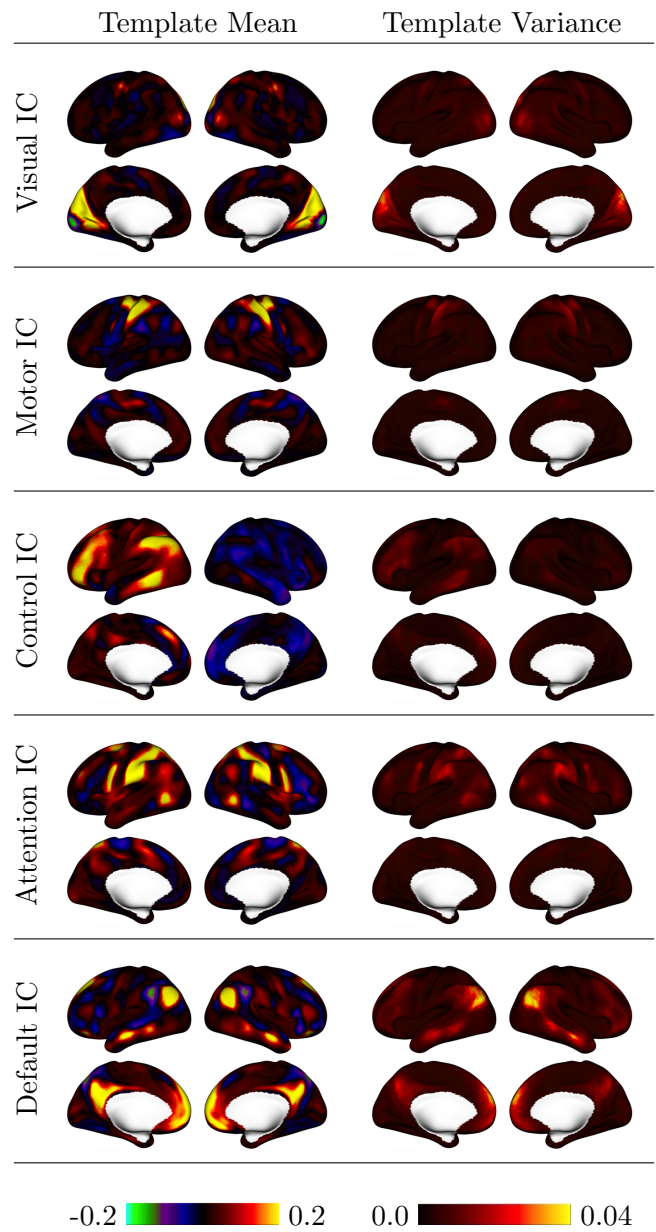
\begin{figure}[H]
    \centering
    \begin{tabular}{ccc}
    & {{Template Mean}} & {{Template Variance}} \\
    \hline
    \begin{picture}(0,80)\put(-5,35){\rotatebox[origin=c]{90}{Visual IC}}\end{picture} &
    \includegraphics[height=25mm, trim=0 3cm 0 3cm, clip]{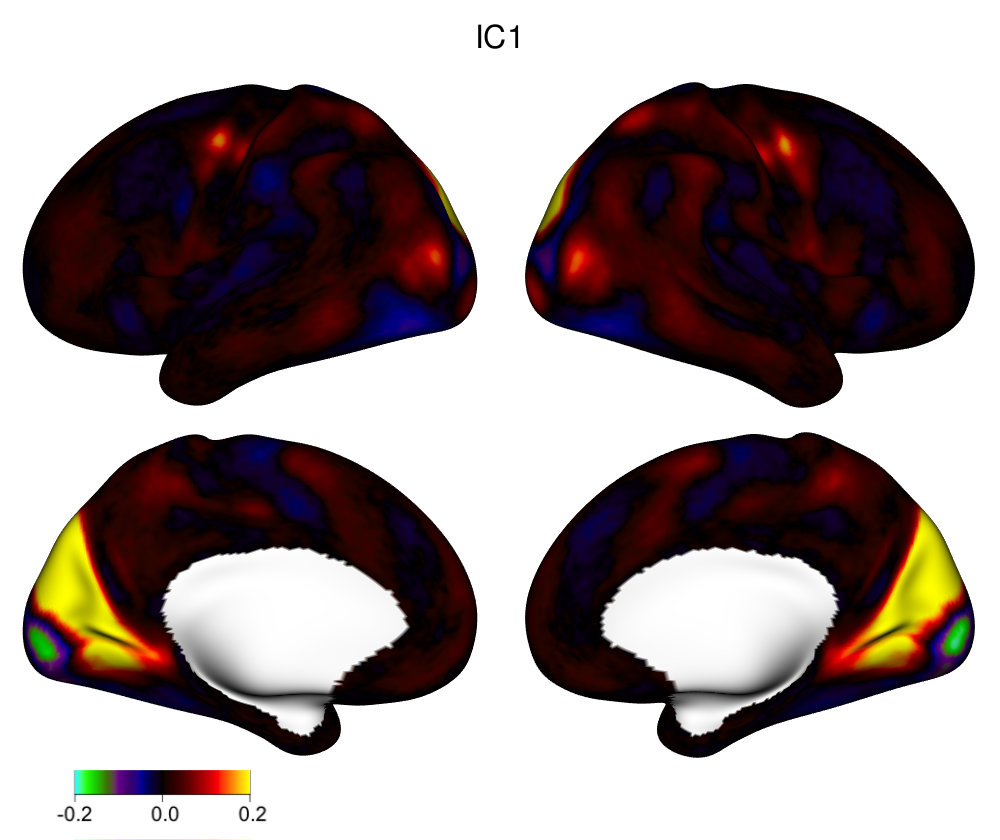} &
    \includegraphics[height=25mm, trim=0 3cm 0 3cm, clip]{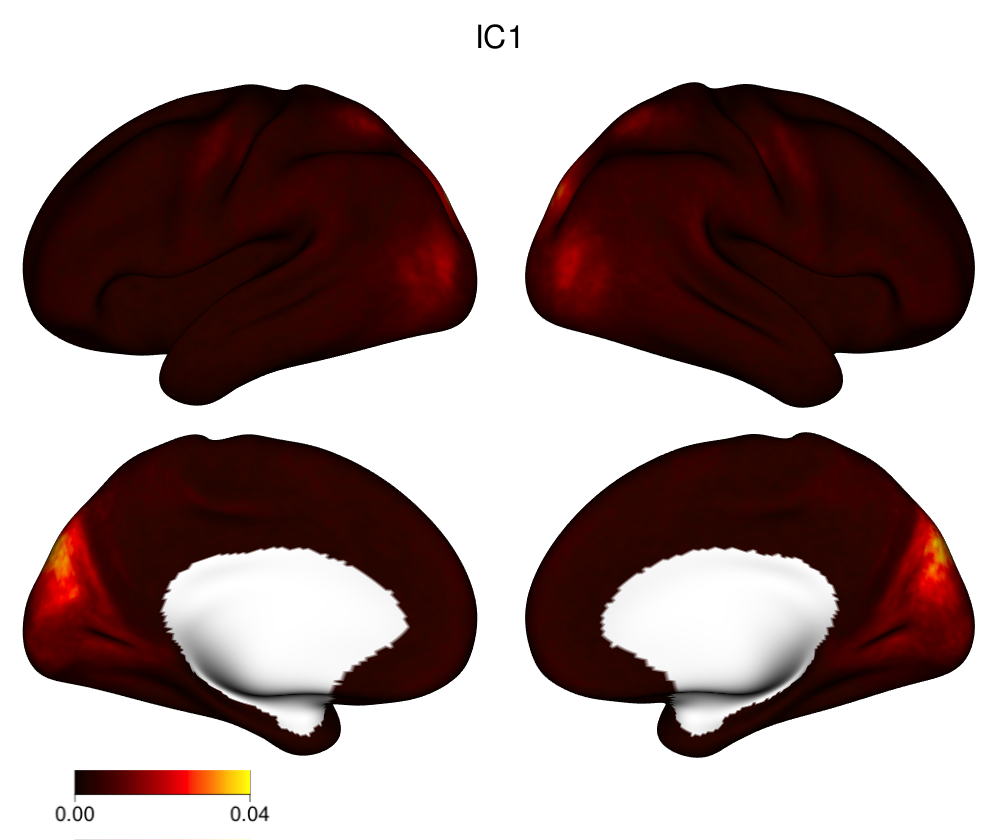} \\
    \hline
    \begin{picture}(0,80)\put(-5,35){\rotatebox[origin=c]{90}{Motor IC}}\end{picture} & 
    \includegraphics[height=25mm, trim=0 3cm 0 3cm, clip]{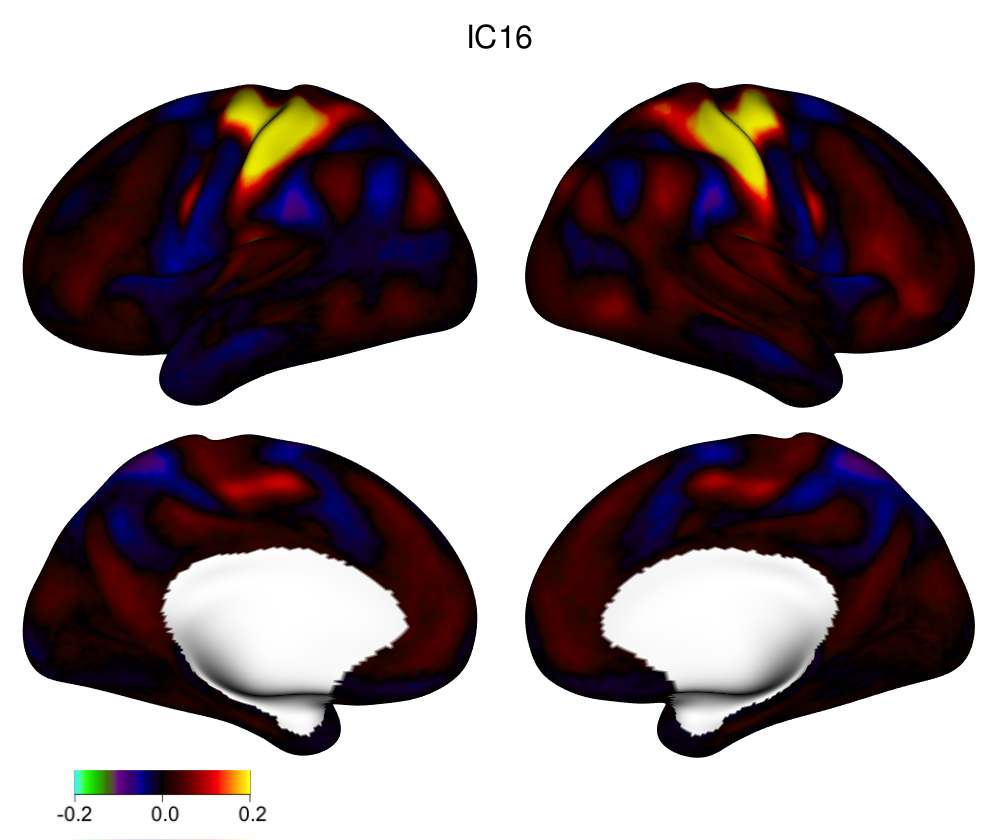} &
    \includegraphics[height=25mm, trim=0 3cm 0 3cm, clip]{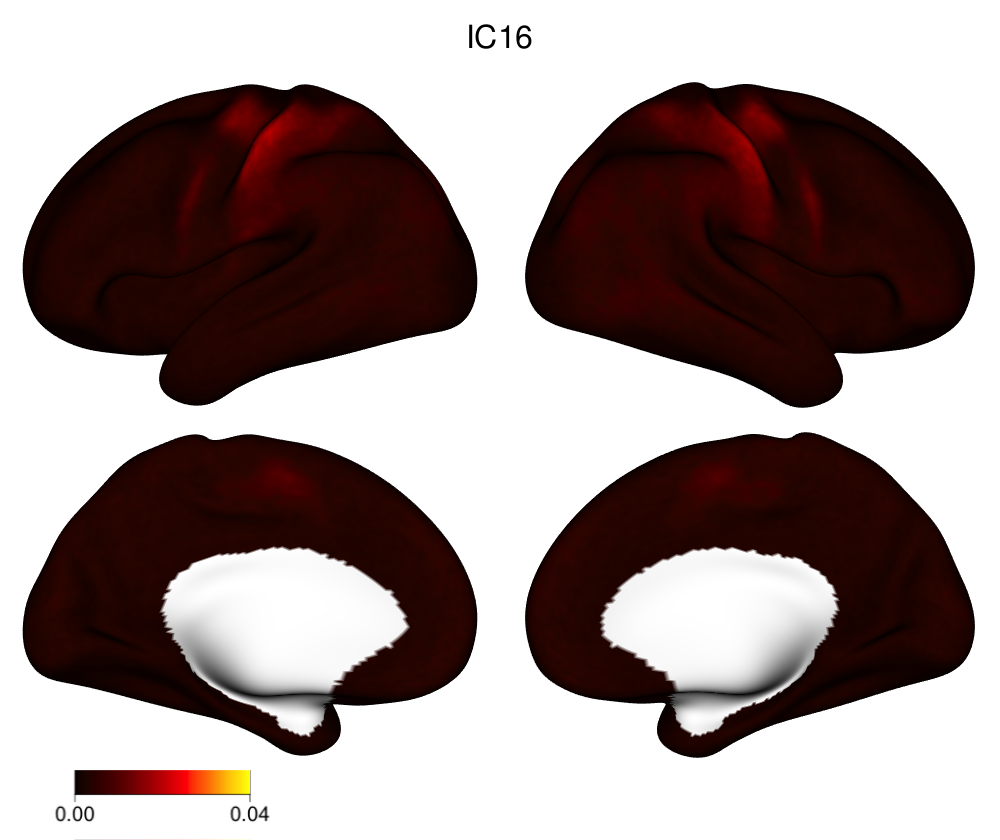} \\
    \hline
    \begin{picture}(0,80)\put(-5,35){\rotatebox[origin=c]{90}{Control IC}}\end{picture} & 
    \includegraphics[height=25mm, trim=0 3cm 0 3cm, clip]{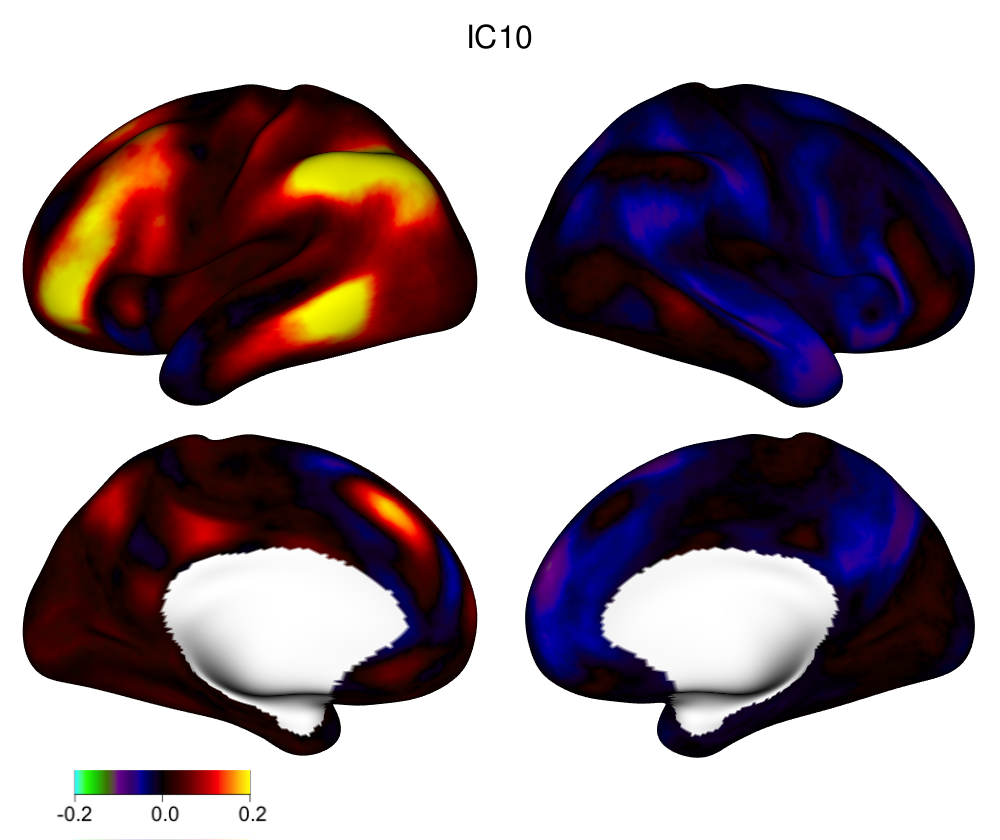} &
    \includegraphics[height=25mm, trim=0 3cm 0 3cm, clip]{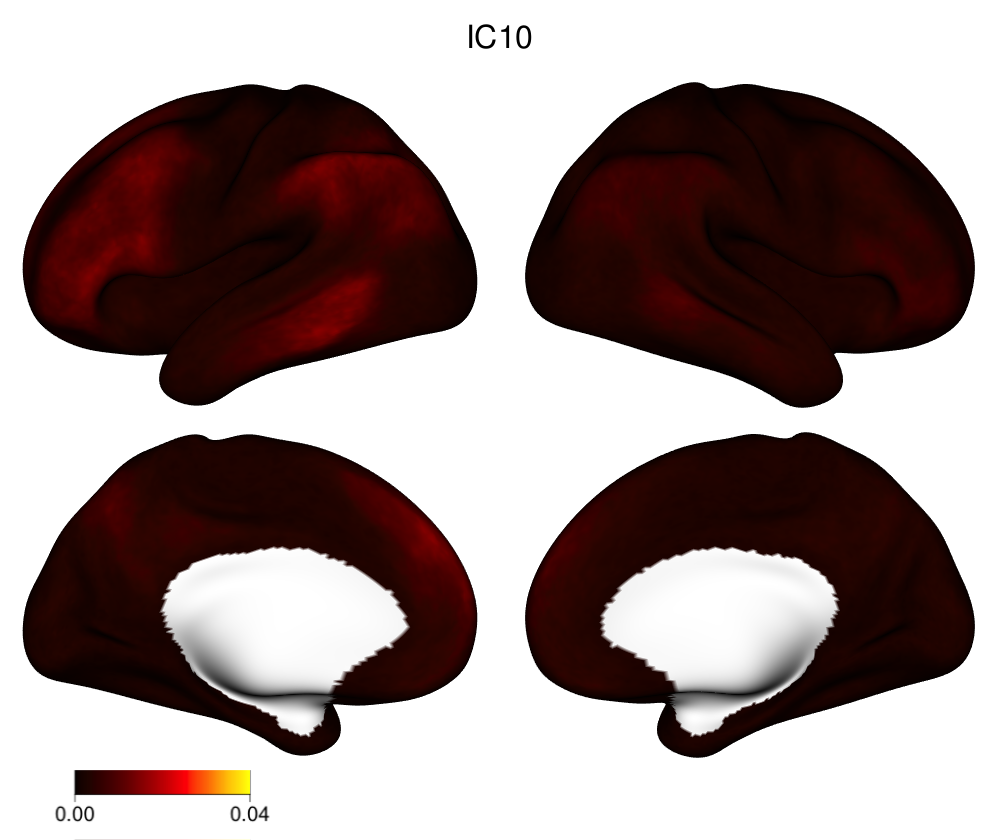} \\
    \hline
    \begin{picture}(0,80)\put(-5,35){\rotatebox[origin=c]{90}{Attention IC}}\end{picture} & 
    \includegraphics[height=25mm, trim=0 3cm 0 3cm, clip]{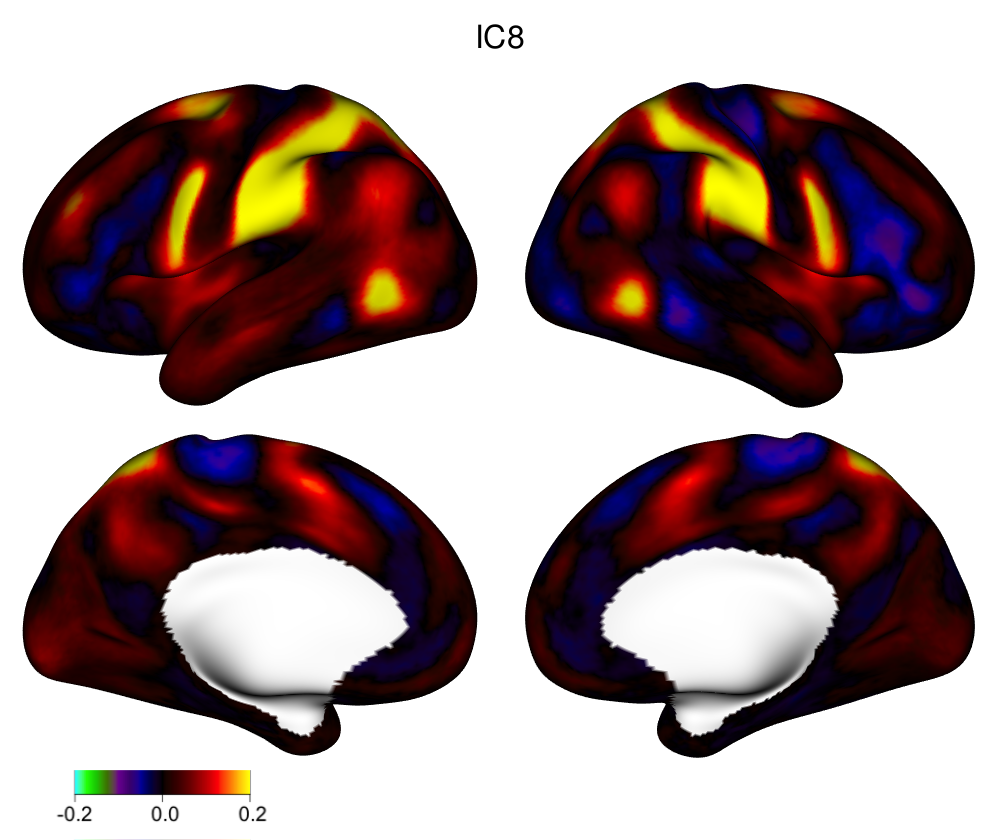} &
    \includegraphics[height=25mm, trim=0 3cm 0 3cm, clip]{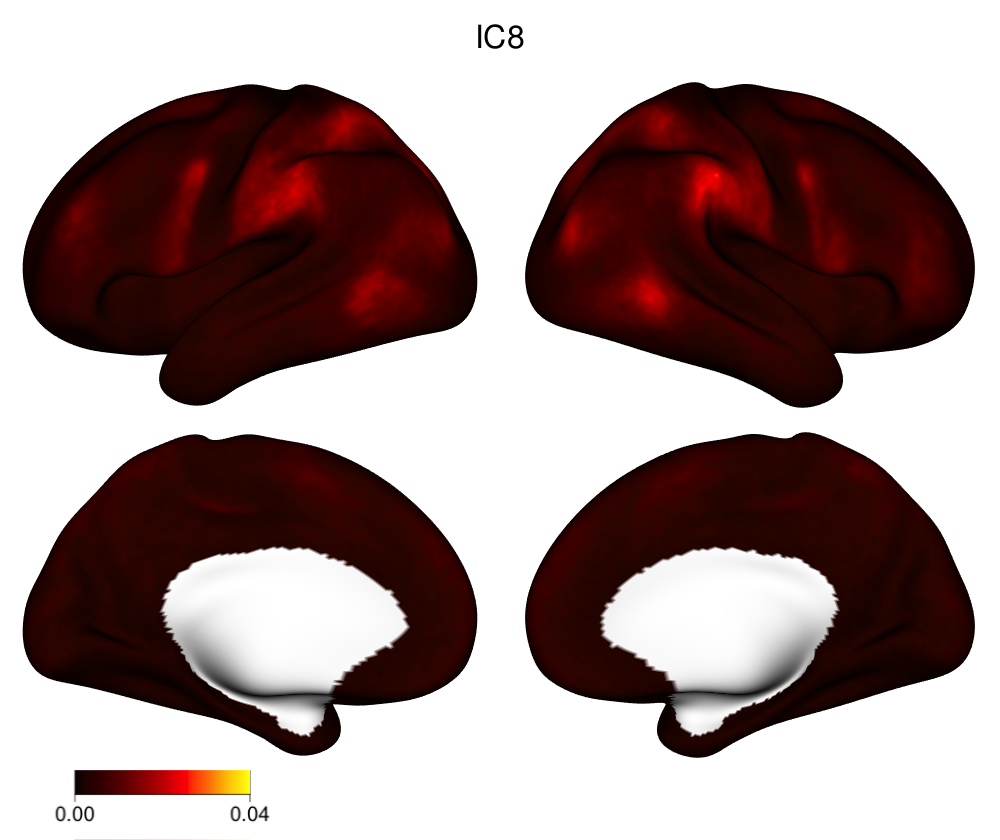} \\
    \hline
    \begin{picture}(0,80)\put(-5,35){\rotatebox[origin=c]{90}{Default IC}}\end{picture} & 
    \includegraphics[height=25mm, trim=0 3cm 0 3cm, clip]{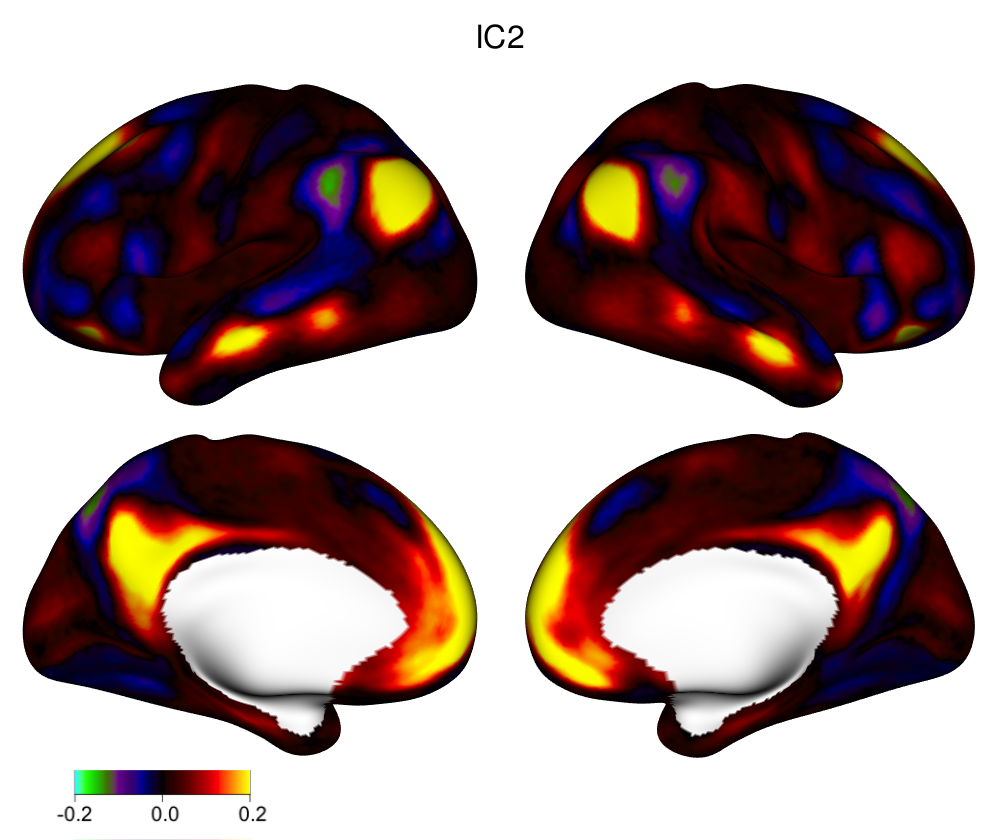} &
    \includegraphics[height=25mm, trim=0 3cm 0 3cm, clip]{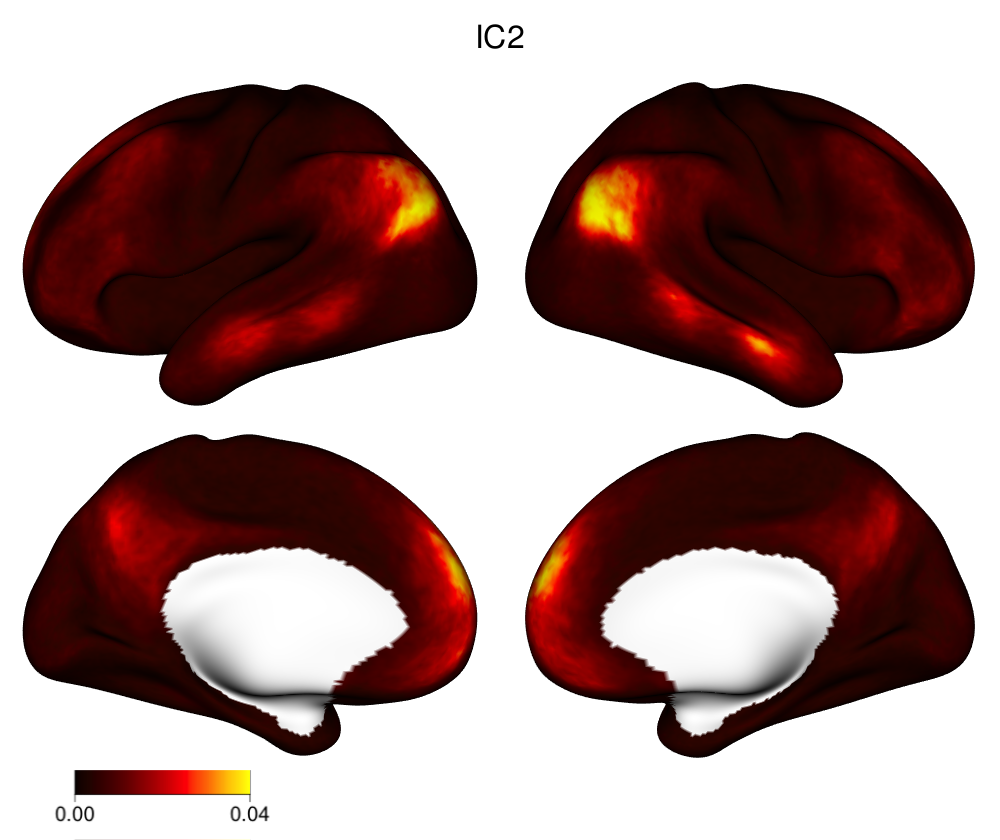} \\
    \hline \\
    & {-0.2 \includegraphics[width=2cm]{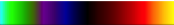} 0.2} 
    & {0.0 \includegraphics[width=2cm]{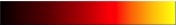} 0.04} \\
    \end{tabular}
    \caption{Prior spatial mean and variance maps for several ICs assigned to different resting-state networks in HCP data analysis.}
    \label{fig:DA:template_ICs}
\end{figure}


\begin{figure}[!p]
    \centering
    \begin{tabular}{ccccc}
    & FC-tICA (VB1) & FC-tICA (VB2) & tICA & DR \\[4pt]
    \hline\\
    \begin{picture}(10,60)\put(5,30){\rotatebox[origin=c]{90}{Subject 1}}\end{picture} &
    \includegraphics[height=20mm, trim=18cm 15cm 0 3cm, clip]{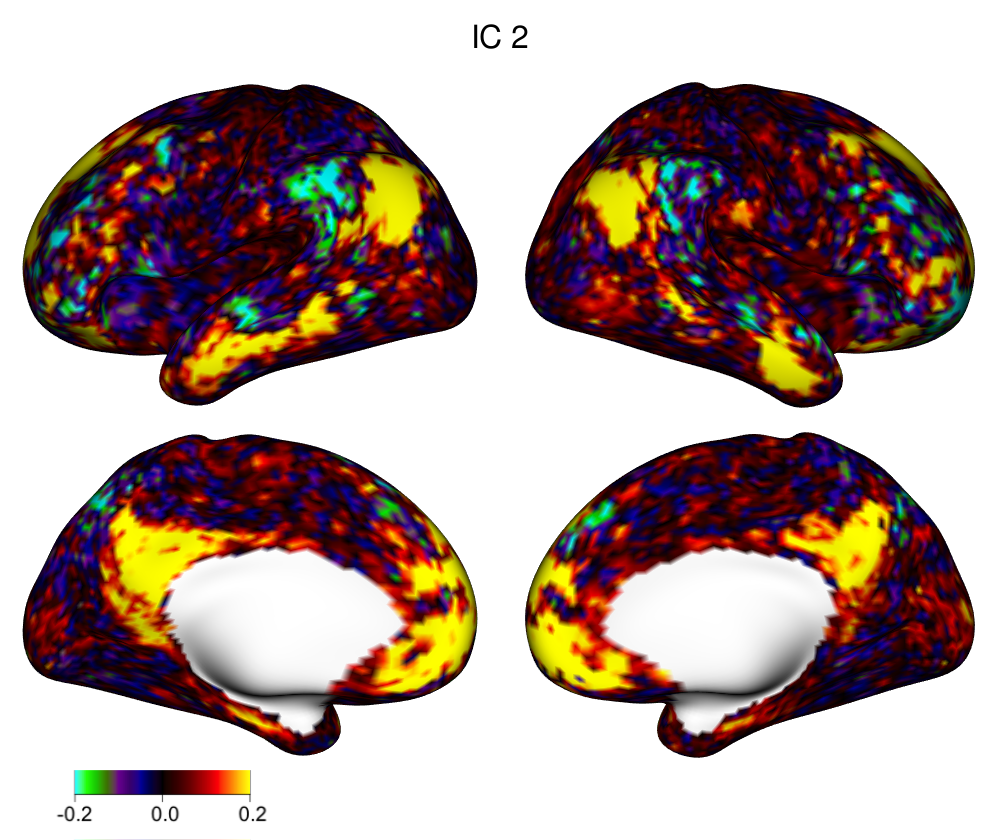} &
    \includegraphics[height=20mm, trim=18cm 15cm 0 3cm, clip]{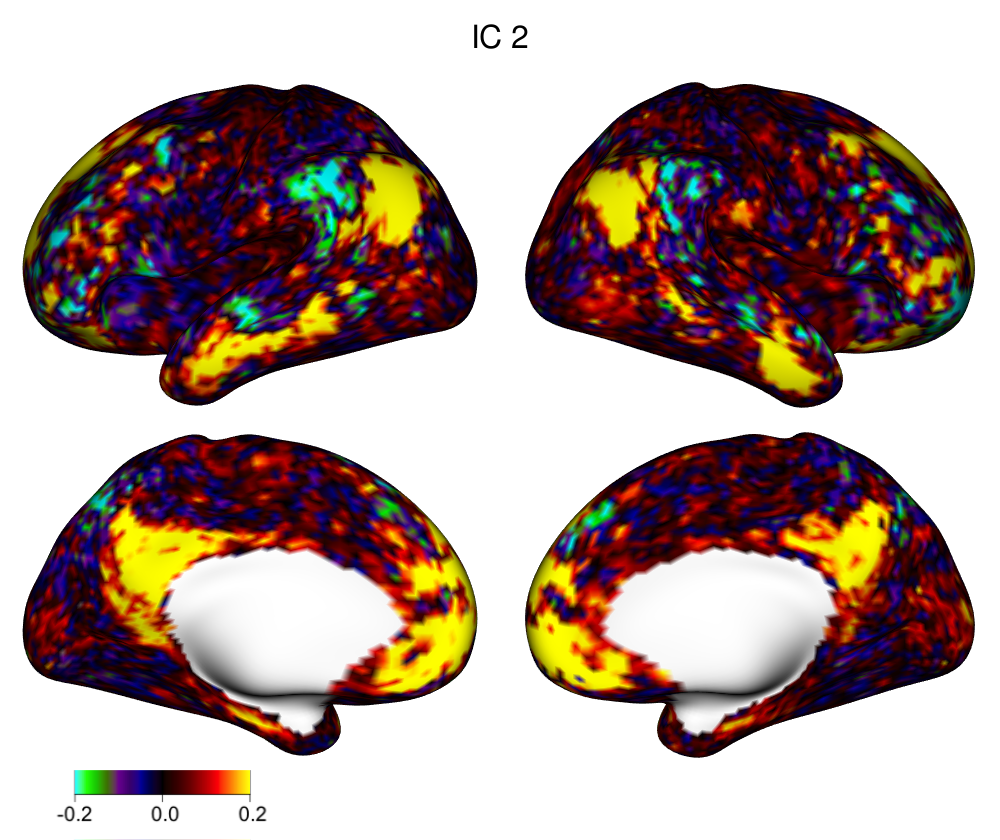} &
    \includegraphics[height=20mm, trim=18cm 15cm 0 3cm, clip]{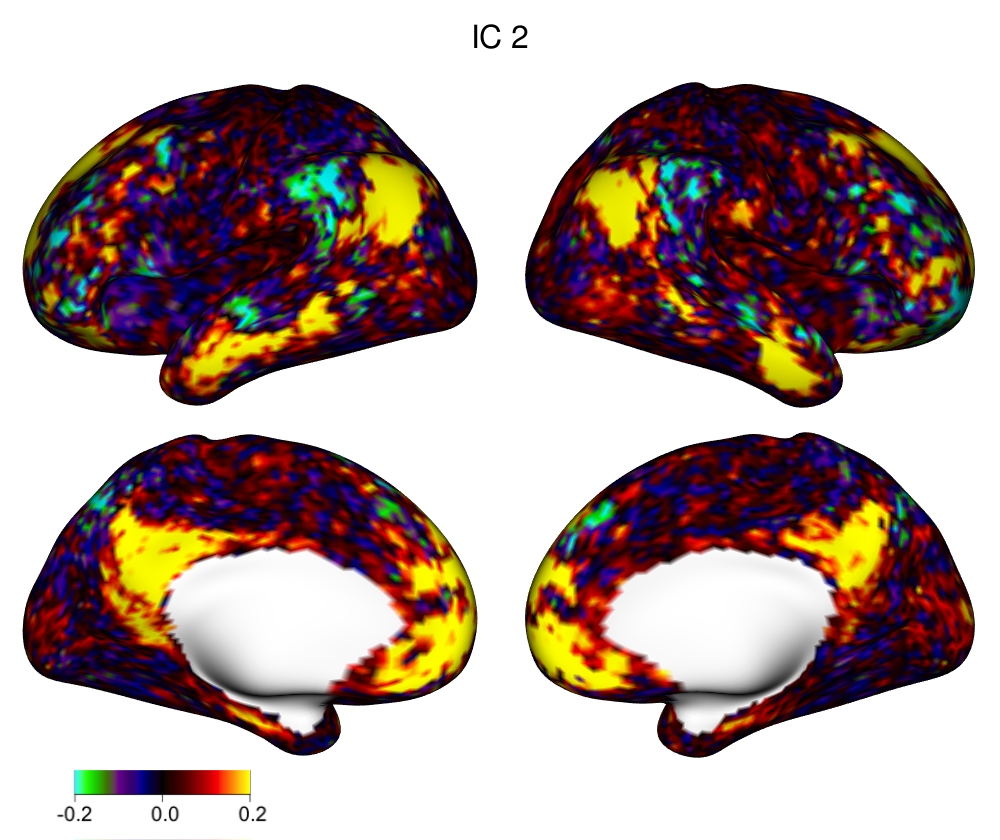} &
    \includegraphics[height=20mm, trim=18cm 15cm 0 3cm, clip]{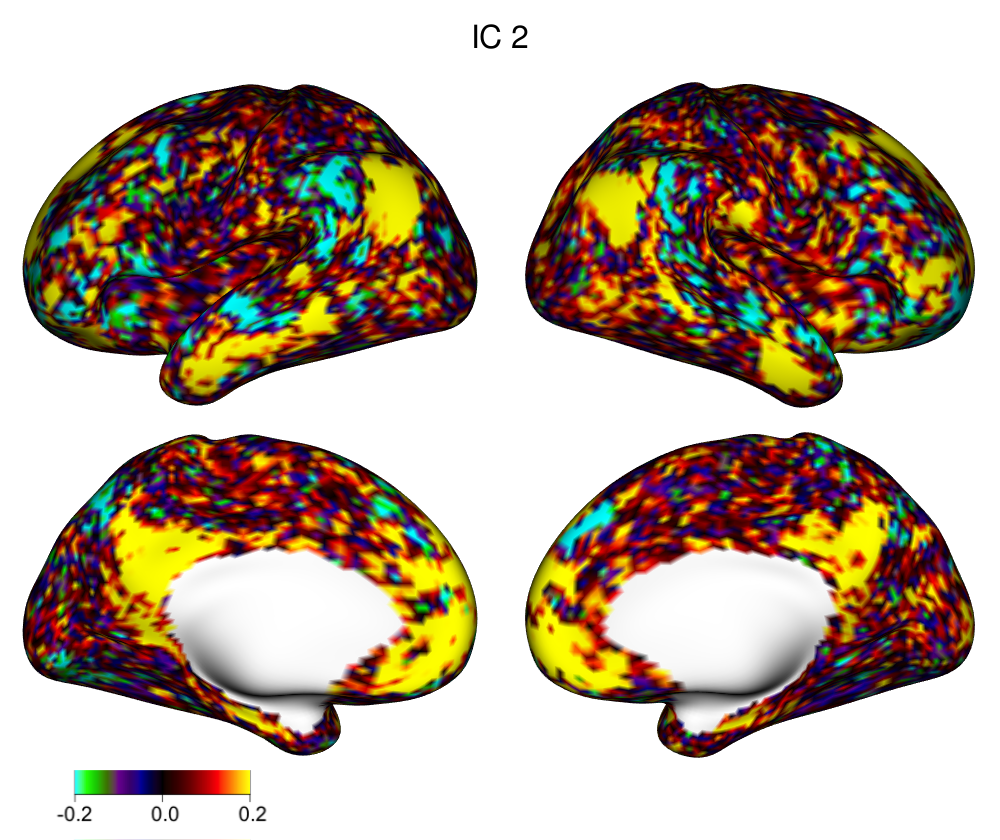} \\[4pt]
    \begin{picture}(10,60)\put(5,30){\rotatebox[origin=c]{90}{Subject 2}}\end{picture} &
    \includegraphics[height=20mm, trim=18cm 15cm 0 3cm, clip]{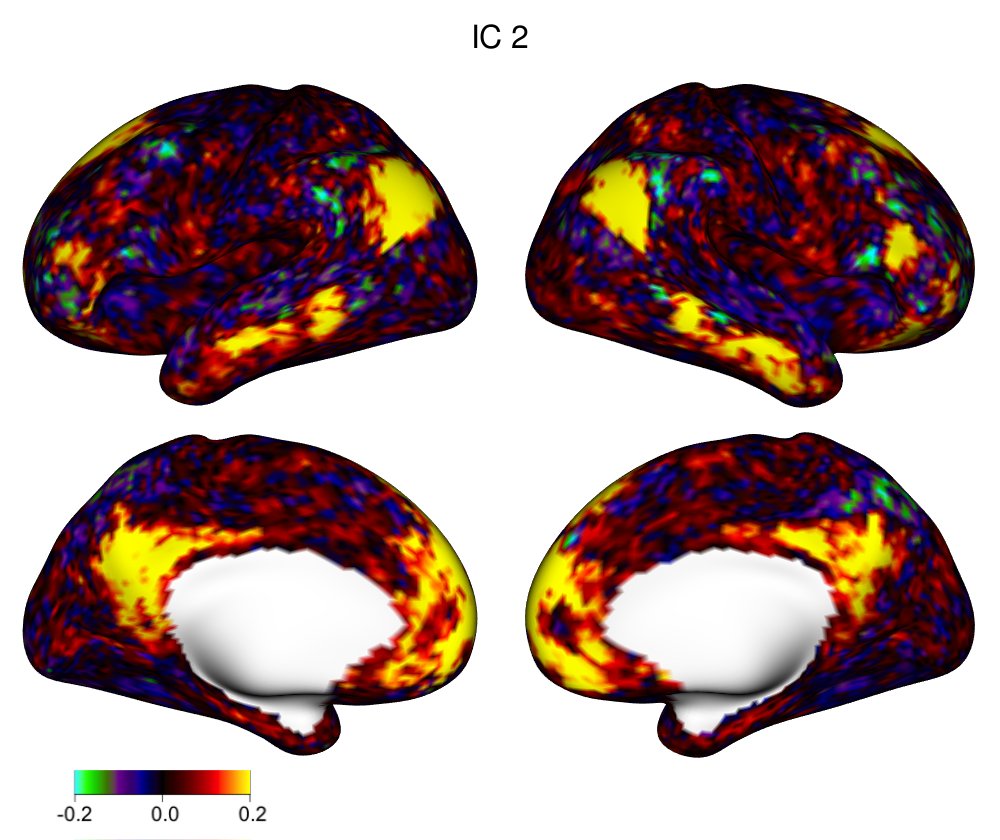} &
    \includegraphics[height=20mm, trim=18cm 15cm 0 3cm, clip]{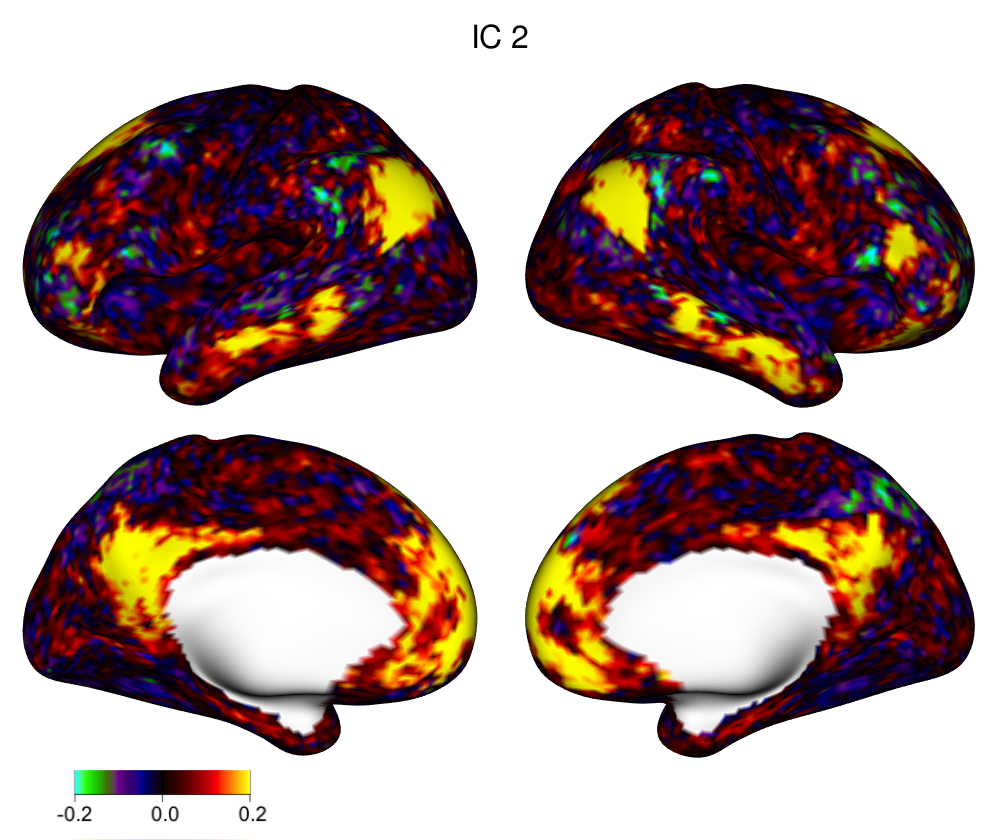} &
    \includegraphics[height=20mm, trim=18cm 15cm 0 3cm, clip]{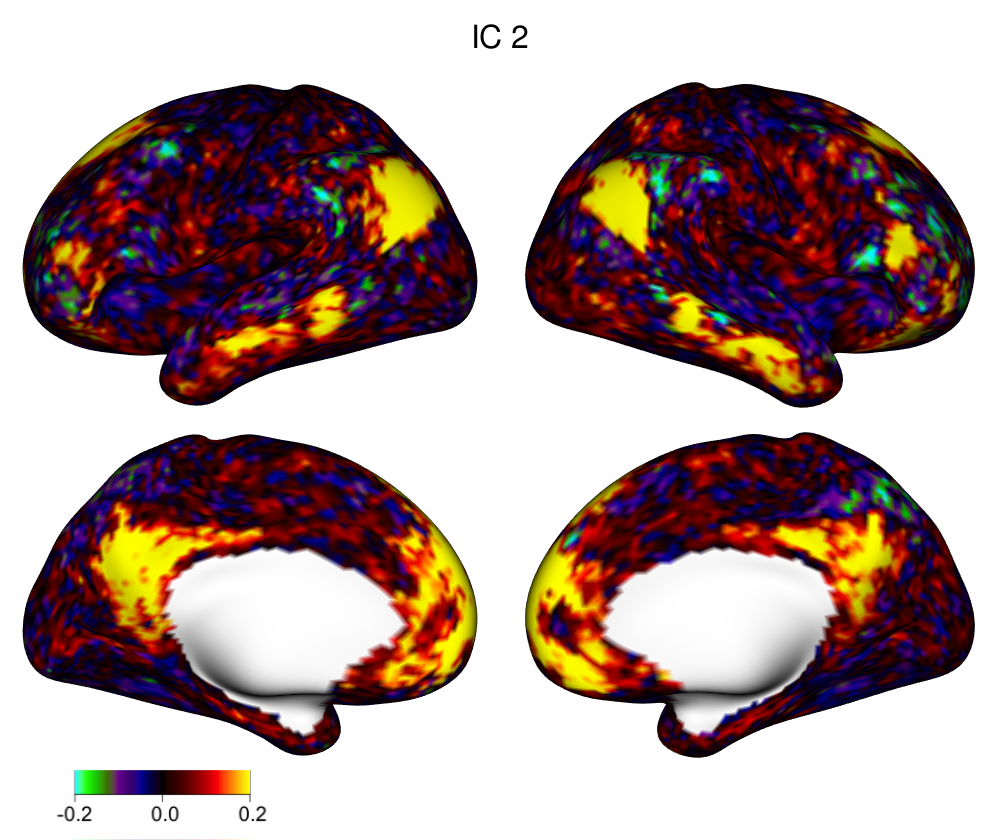} &
    \includegraphics[height=20mm, trim=18cm 15cm 0 3cm, clip]{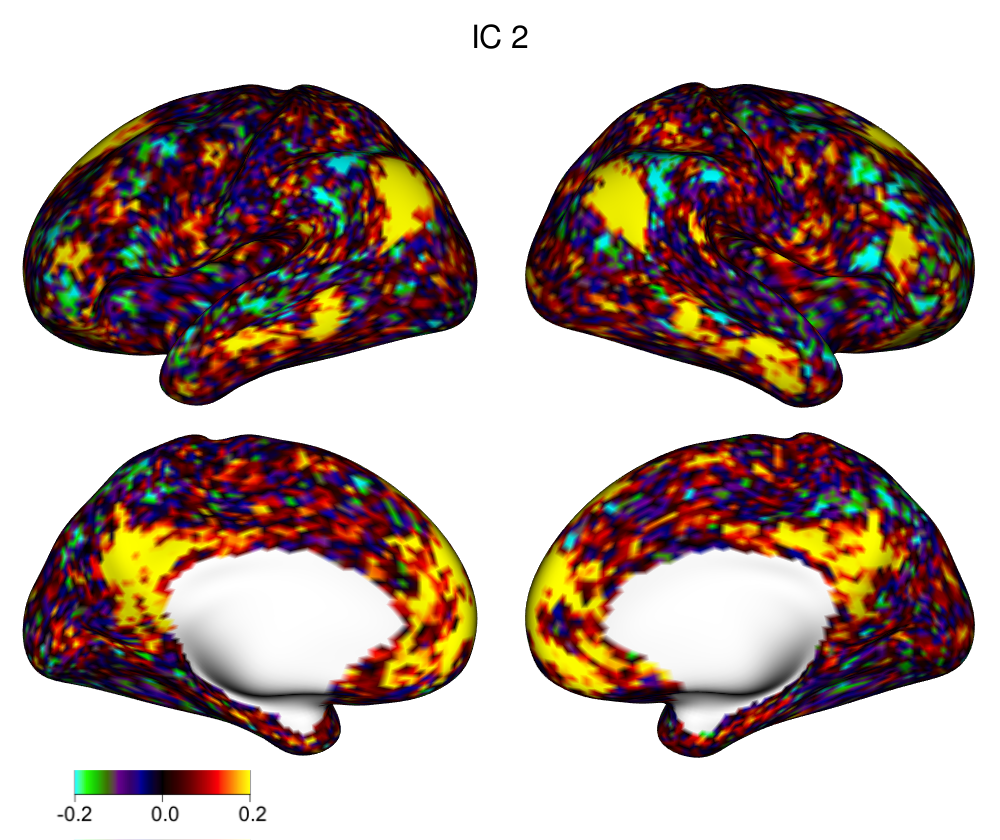} \\[4pt]
    & \multicolumn{4}{c}{-0.2 \includegraphics[width=2cm]{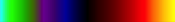} 0.2} \\[10pt]
    & \multicolumn{4}{c}{(a) IC Estimates}  \\[10pt]
    \hline\\
    \begin{picture}(10,60)\put(5,30){\rotatebox[origin=c]{90}{Subject 1}}\end{picture} &
    \includegraphics[height=20mm, trim=18cm 15cm 0 3cm, clip]{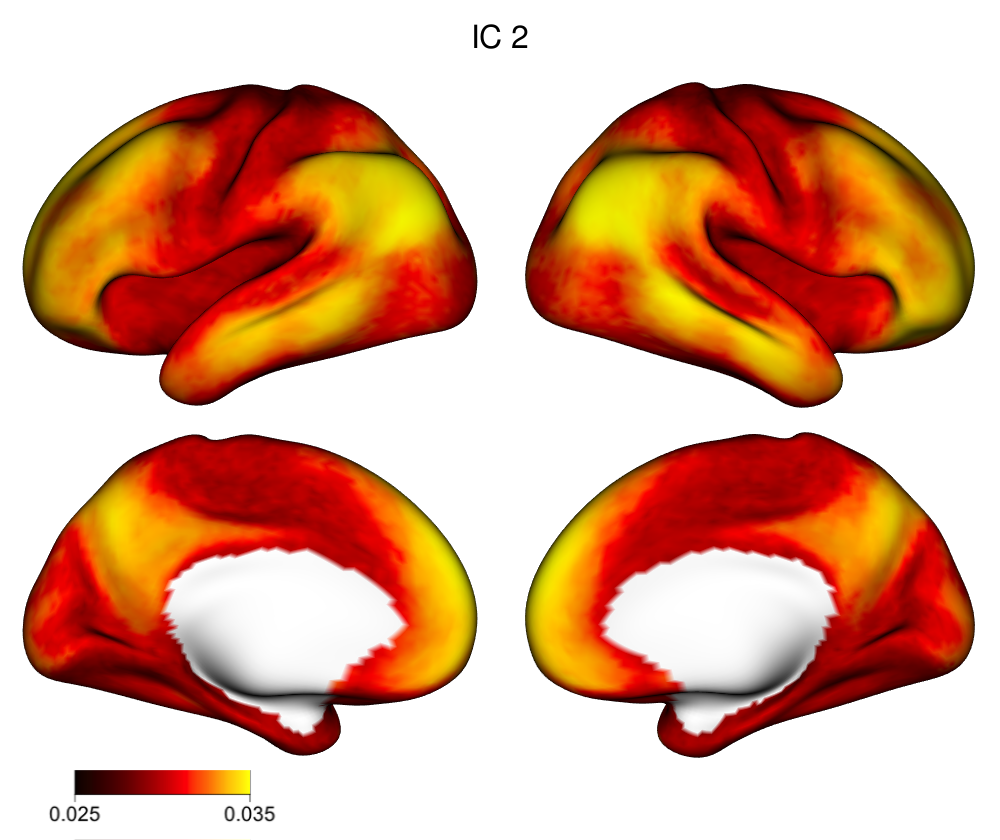} &
    \includegraphics[height=20mm, trim=18cm 15cm 0 3cm, clip]{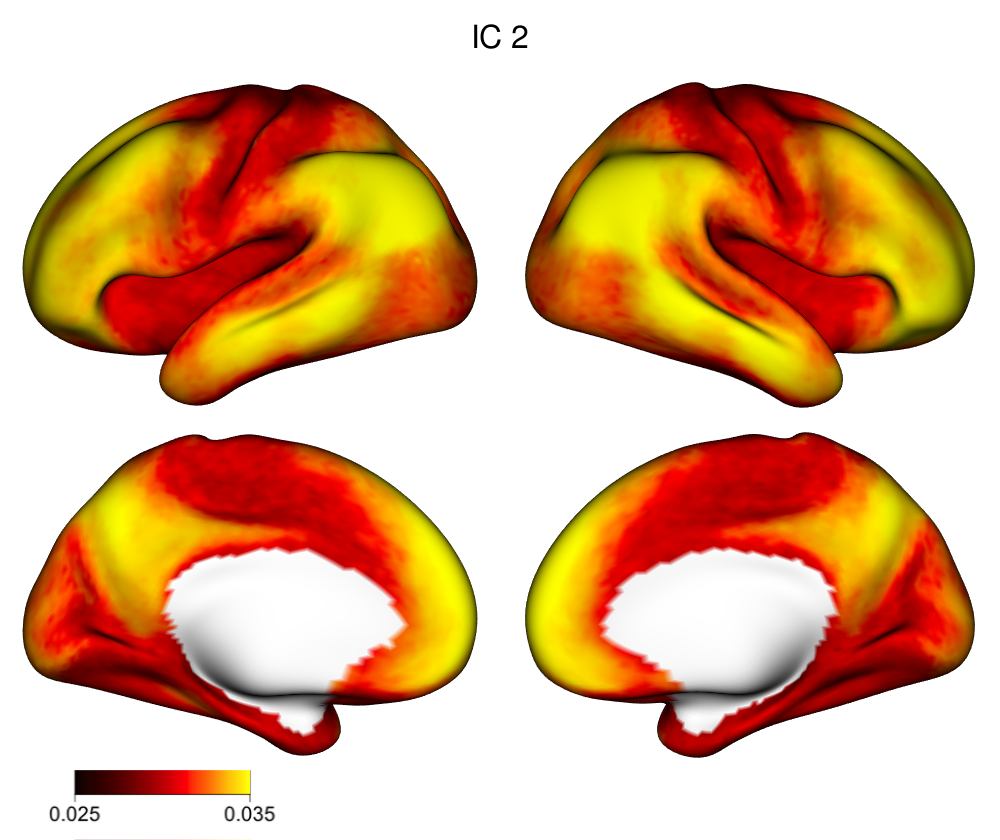} &
    \includegraphics[height=20mm, trim=18cm 15cm 0 3cm, clip]{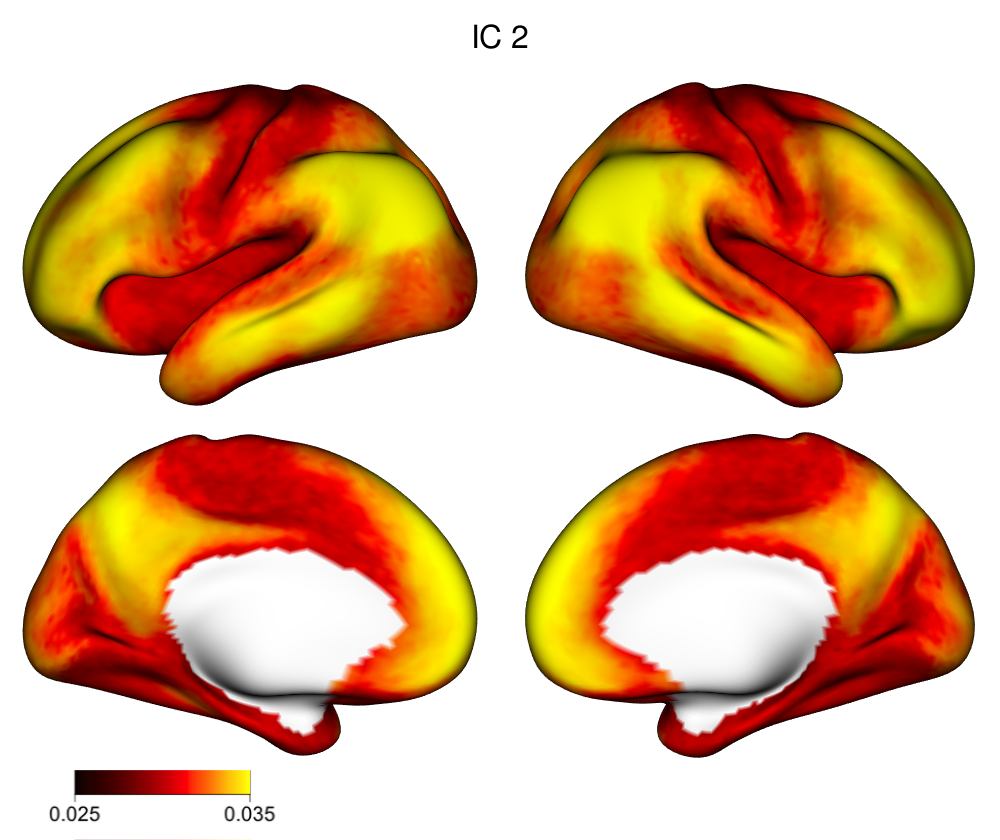}  \\[4pt]
    \begin{picture}(10,60)\put(5,30){\rotatebox[origin=c]{90}{Subject 2}}\end{picture} &
    \includegraphics[height=20mm, trim=18cm 15cm 0 3cm, clip]{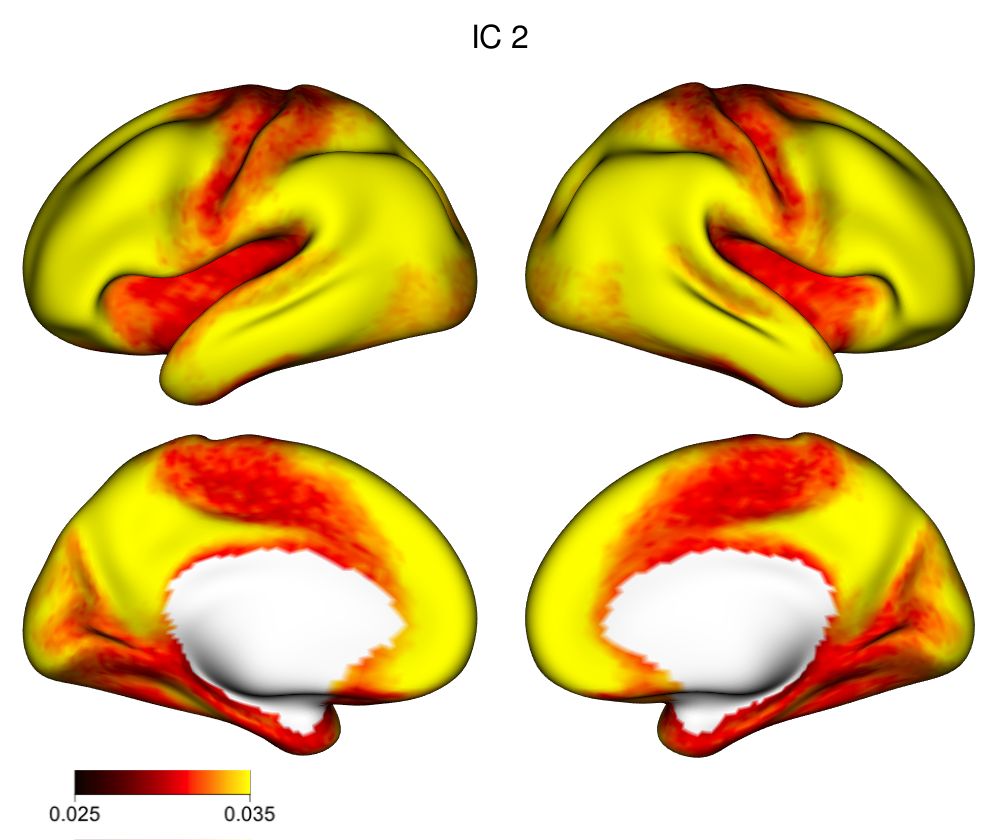} &
    \includegraphics[height=20mm, trim=18cm 15cm 0 3cm, clip]{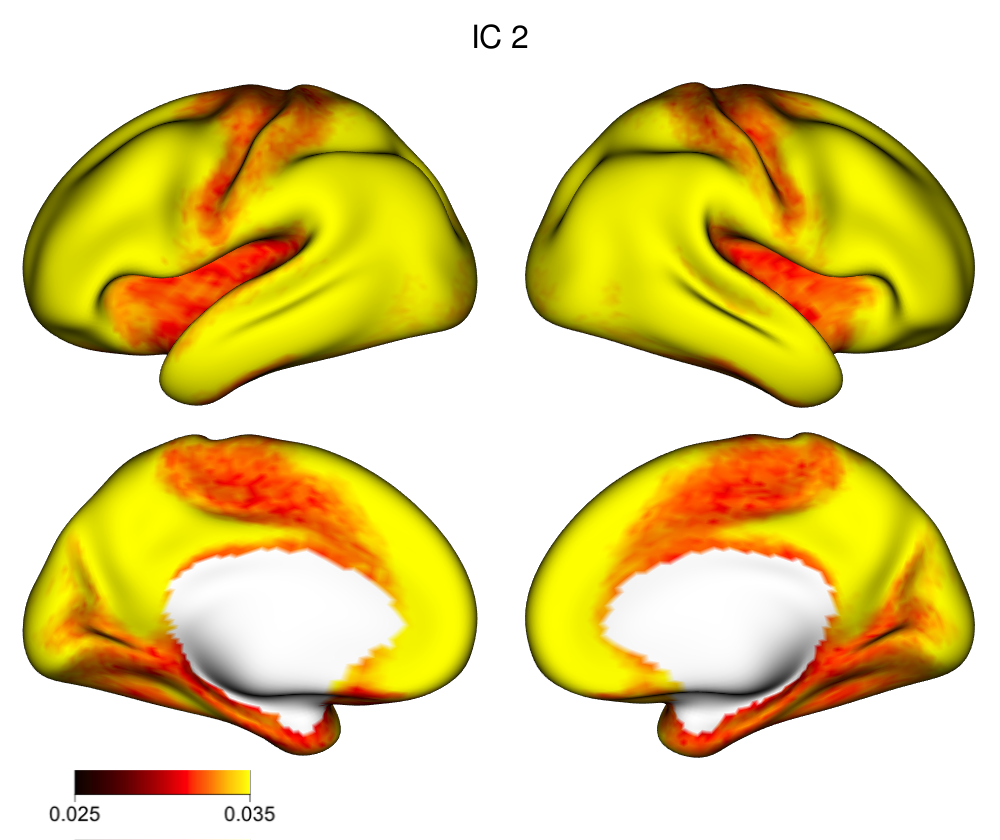} &
    \includegraphics[height=20mm, trim=18cm 15cm 0 3cm, clip]{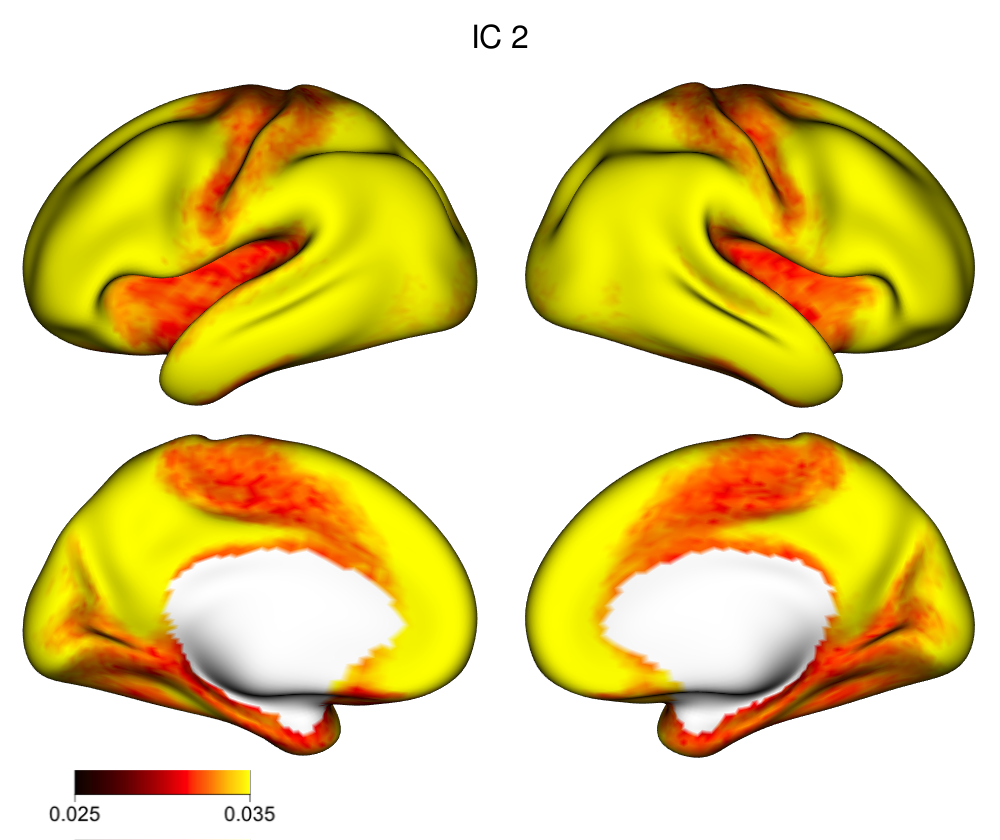}  \\[4pt]
    & \multicolumn{4}{c}{\quad 0.025 \includegraphics[width=2cm]{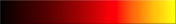} 0.035} \\[10pt]
    & \multicolumn{4}{c}{(a) Posterior SD}  \\[10pt]
    \end{tabular}
    \caption{\textit{Example IC estimates (posterior means) and posterior standard deviations.} For two example subjects, estimates of a default mode network (DMN) IC from FC template ICA (FC-tICA) and standard template ICA (tICA), along with their posterior standard deviations are shown. Estimates from the ad-hoc method dual regression (DR) are also included for comparison.}
    \label{fig:DA:IC_estimates}
\end{figure}


\begin{figure}[!p]
    \centering
    \begin{tabular}{ccccc}
    & FC-tICA (VB1) & FC-tICA (VB2) & tICA & DR \\[4pt]
        \hline
    \begin{picture}(10,70)\put(5,30){\rotatebox[origin=c]{90}{Visual IC}}\end{picture} &
    \includegraphics[height=20mm, trim=18cm 15cm 0 3cm, clip]{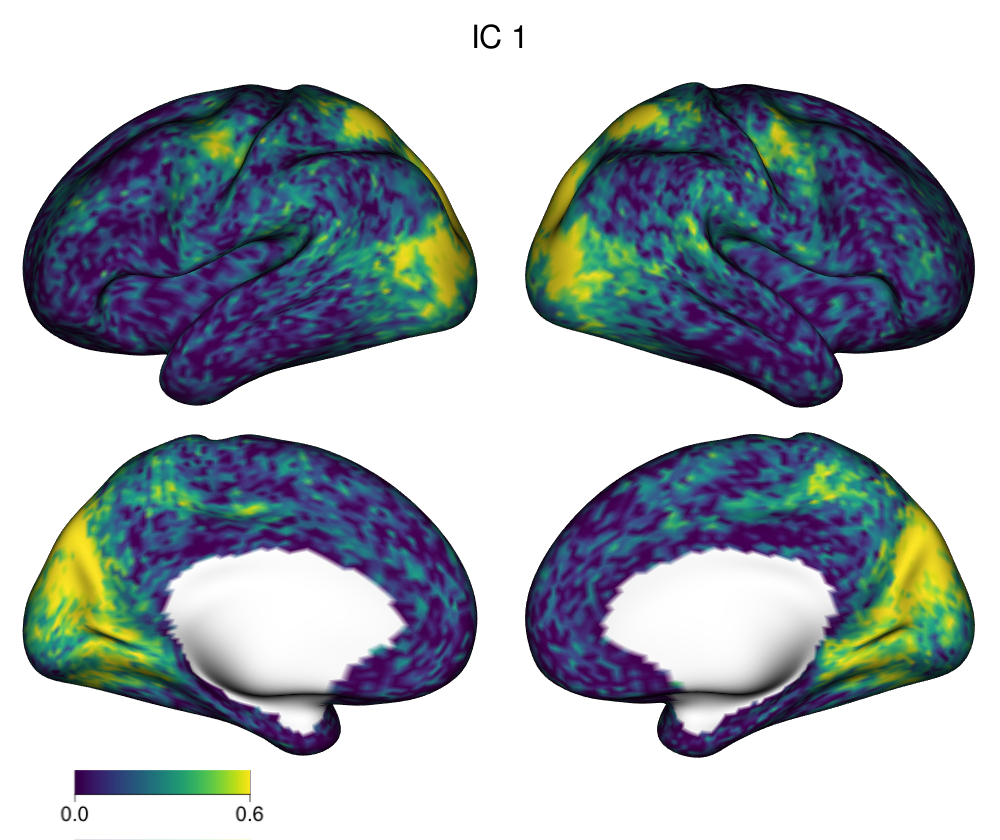} &
    \includegraphics[height=20mm, trim=18cm 15cm 0 3cm, clip]{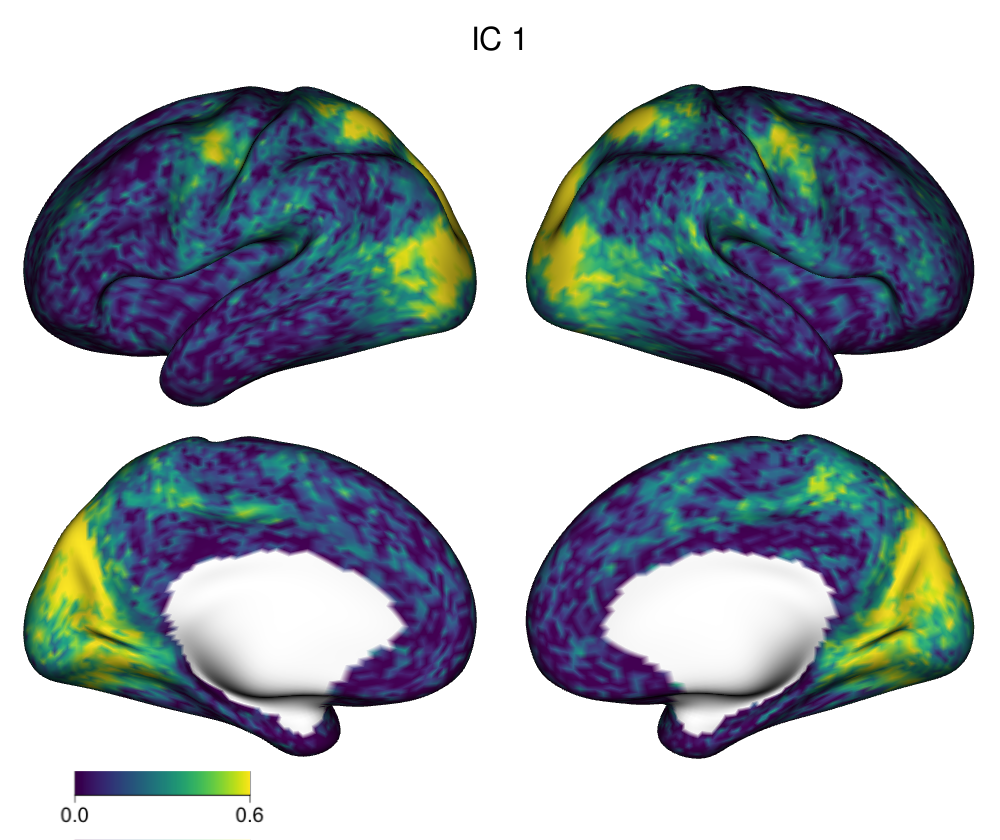} &
    \includegraphics[height=20mm, trim=18cm 15cm 0 3cm, clip]{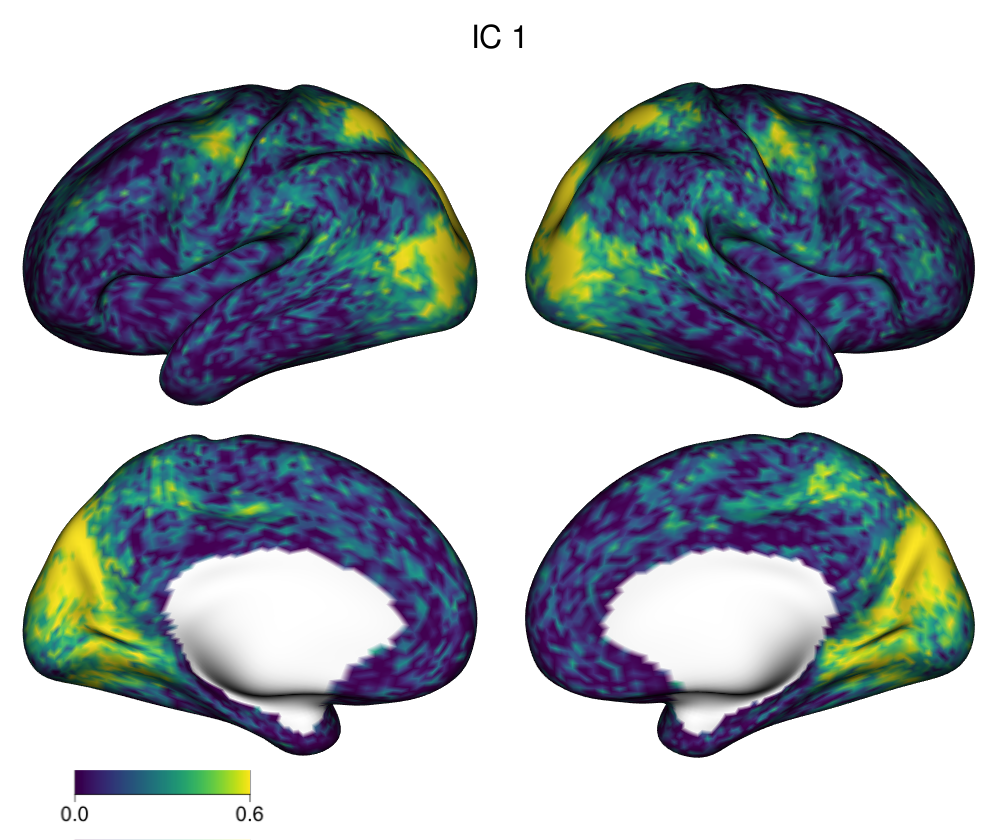} &
    \includegraphics[height=20mm, trim=18cm 15cm 0 3cm, clip]{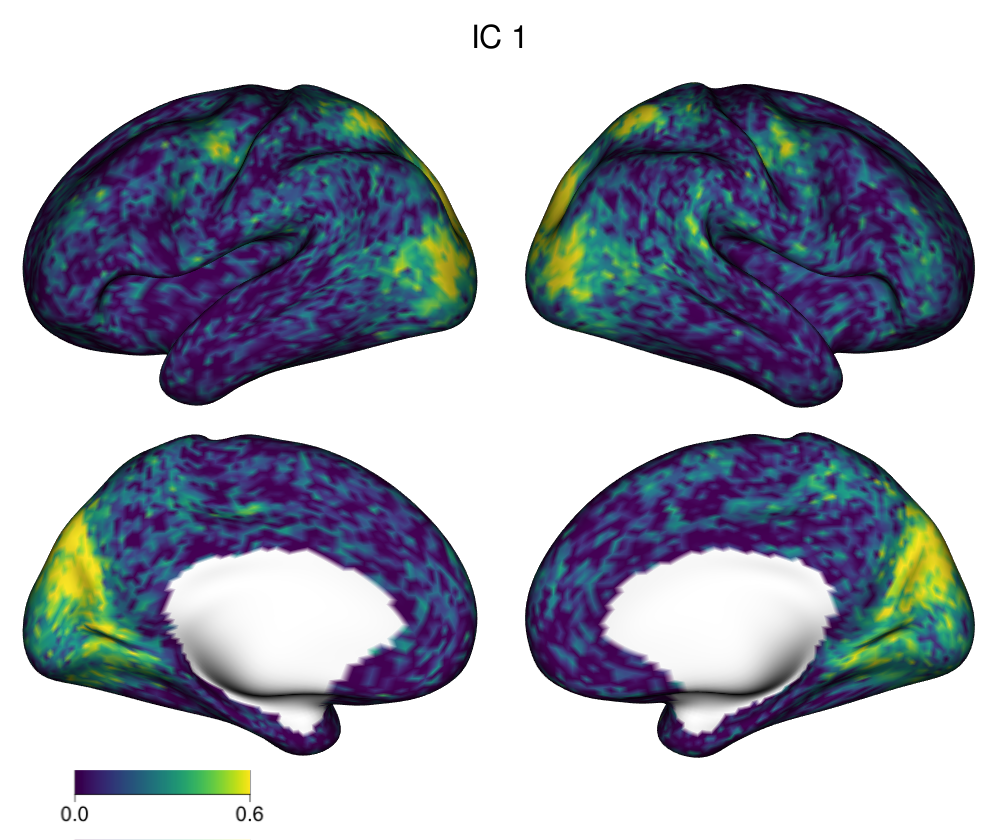} \\
    \begin{picture}(10,70)\put(5,30){\rotatebox[origin=c]{90}{Attention IC}}\end{picture} & 
    \includegraphics[height=20mm, trim=18cm 15cm 0 3cm, clip]{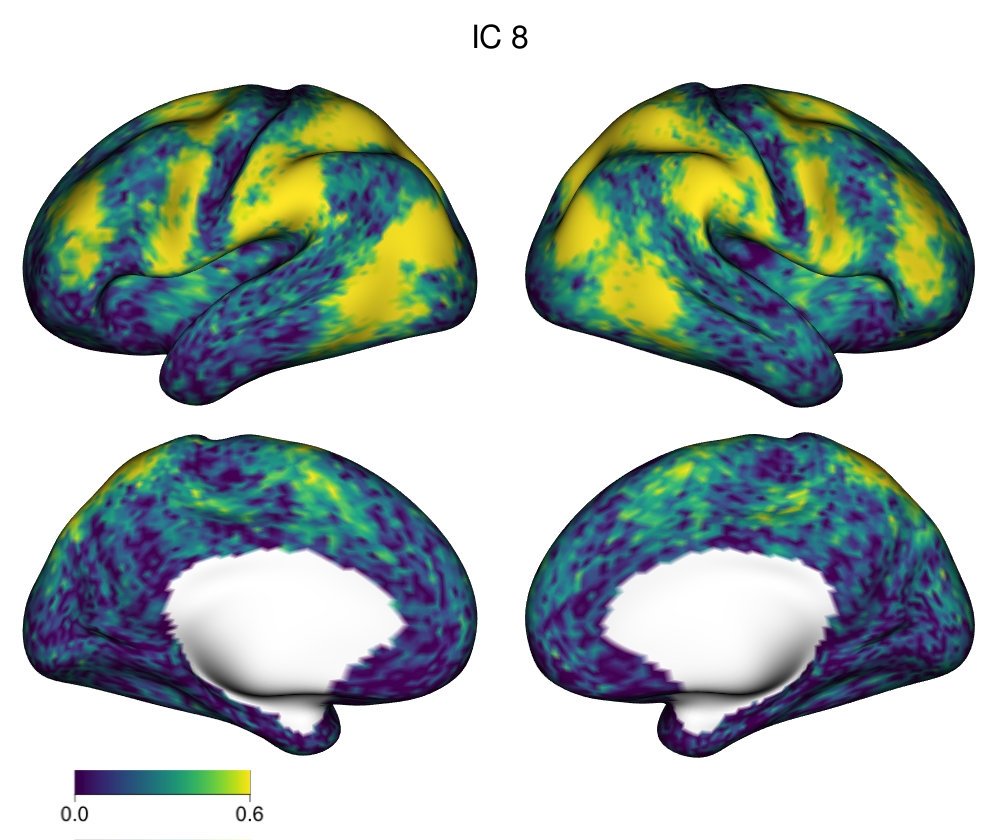} &
    \includegraphics[height=20mm, trim=18cm 15cm 0 3cm, clip]{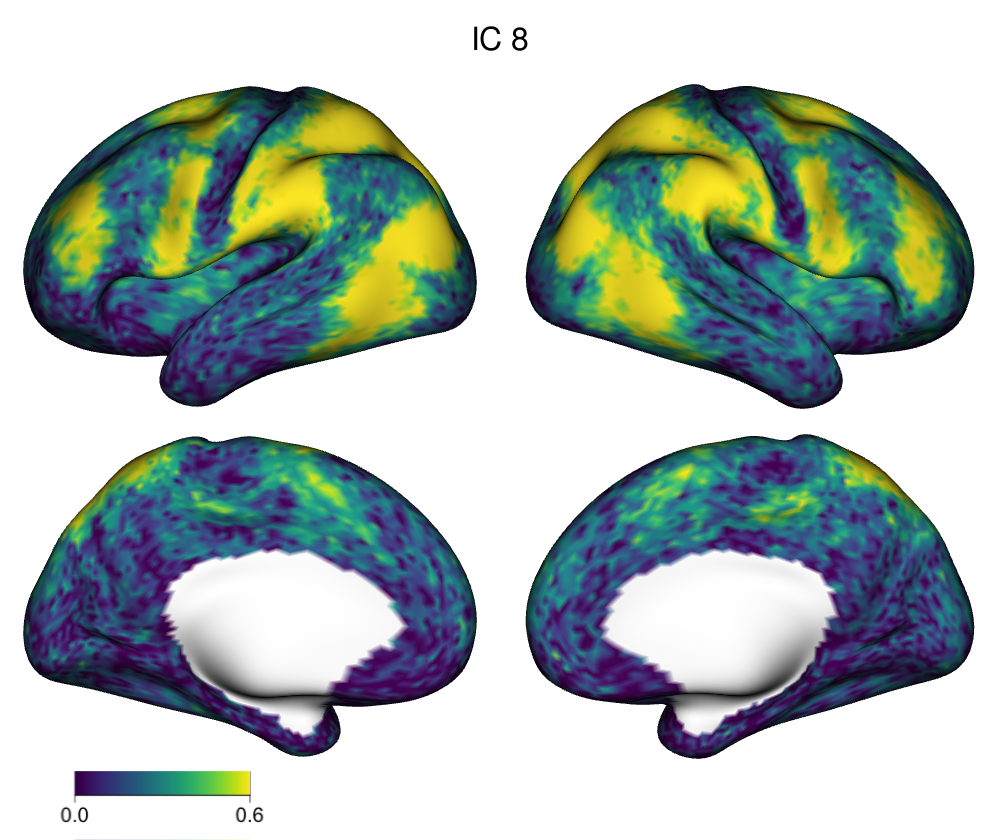} &
    \includegraphics[height=20mm, trim=18cm 15cm 0 3cm, clip]{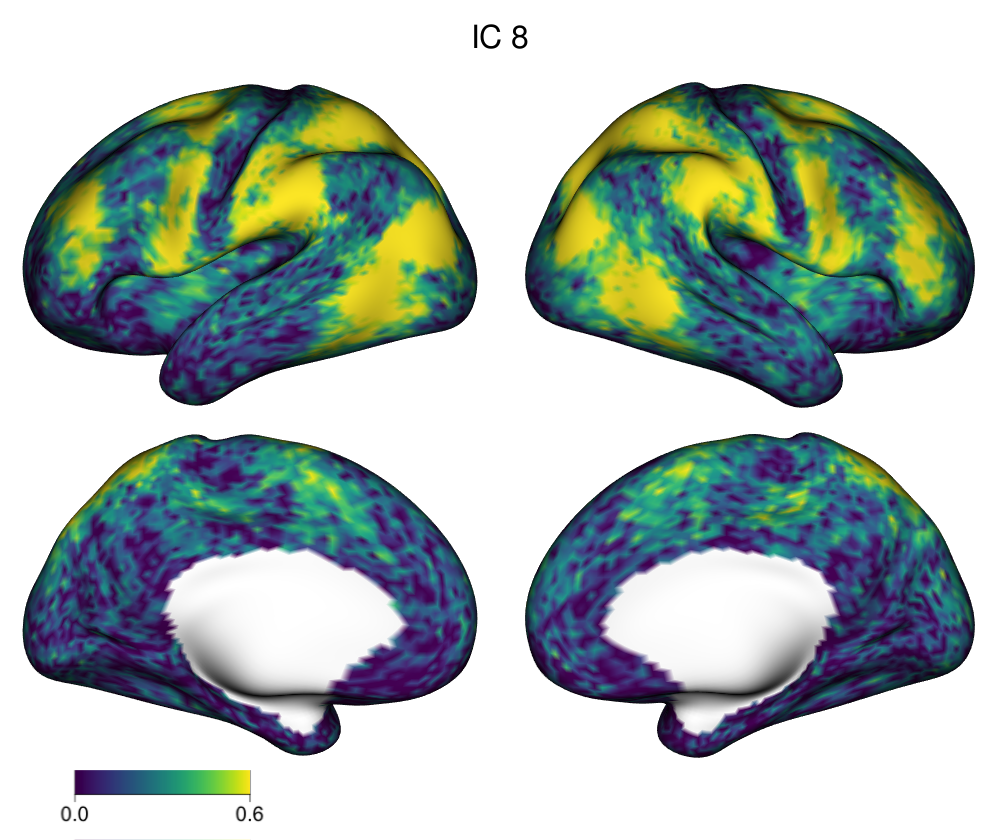} &
    \includegraphics[height=20mm, trim=18cm 15cm 0 3cm, clip]{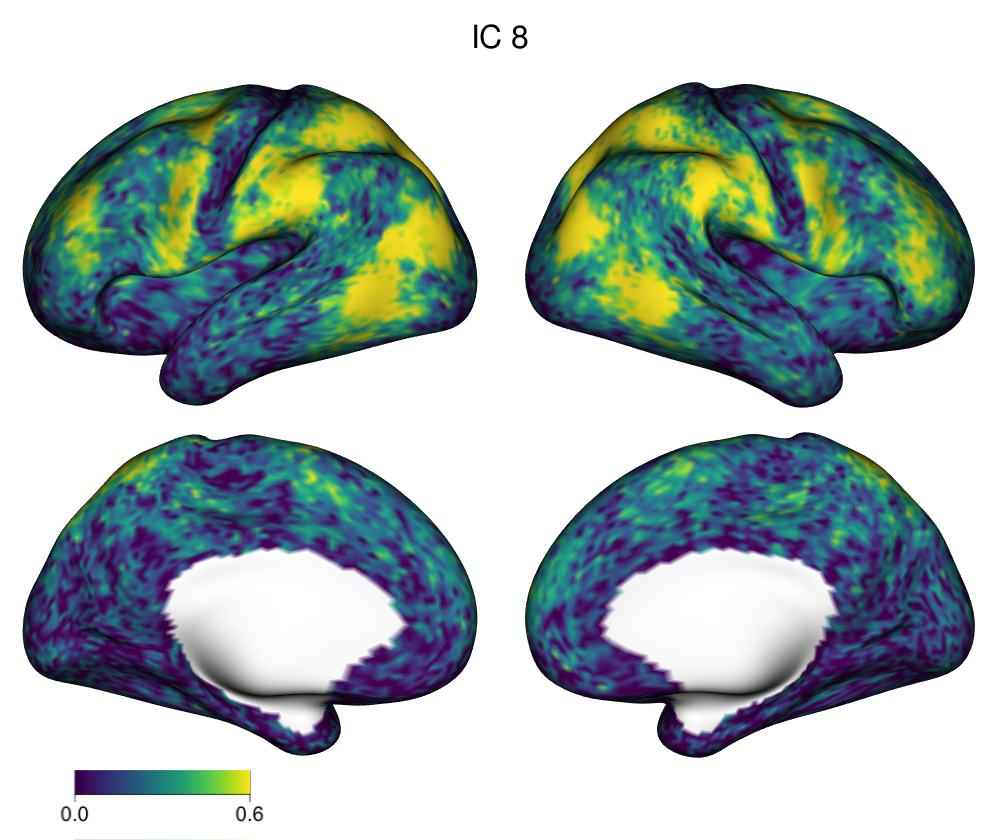} \\
     \begin{picture}(10,70)\put(5,30){\rotatebox[origin=c]{90}{Default IC}}\end{picture} & 
    \includegraphics[height=20mm, trim=18cm 15cm 0 3cm, clip]{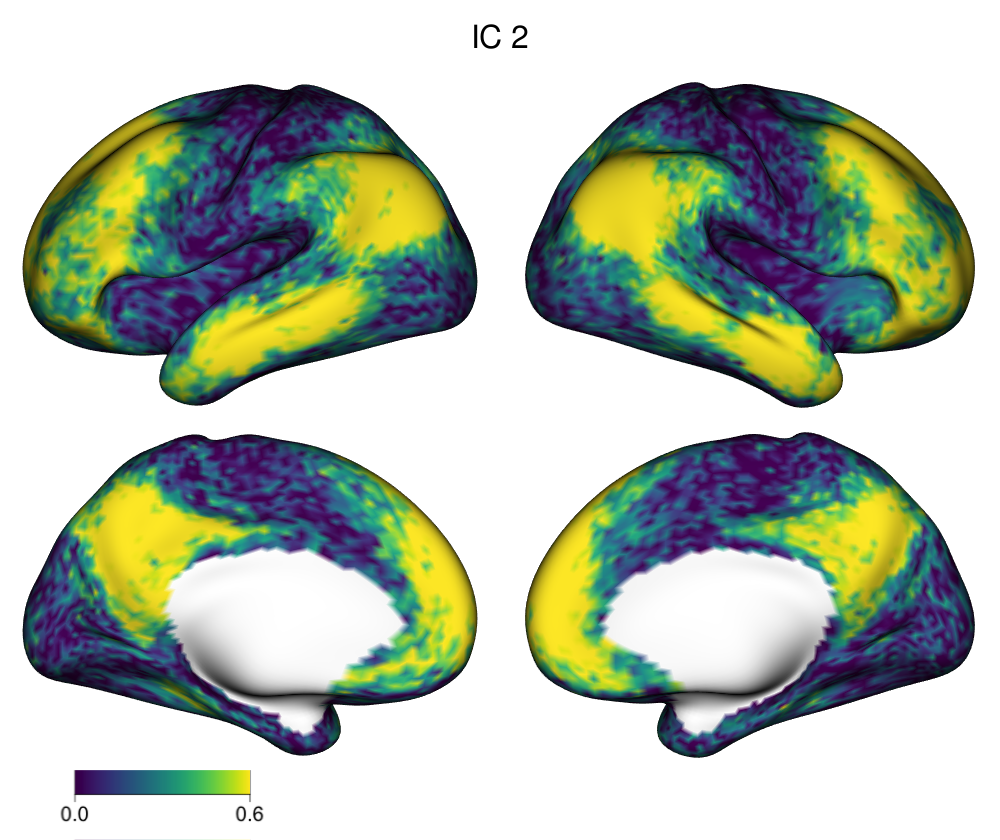} &
    \includegraphics[height=20mm, trim=18cm 15cm 0 3cm, clip]{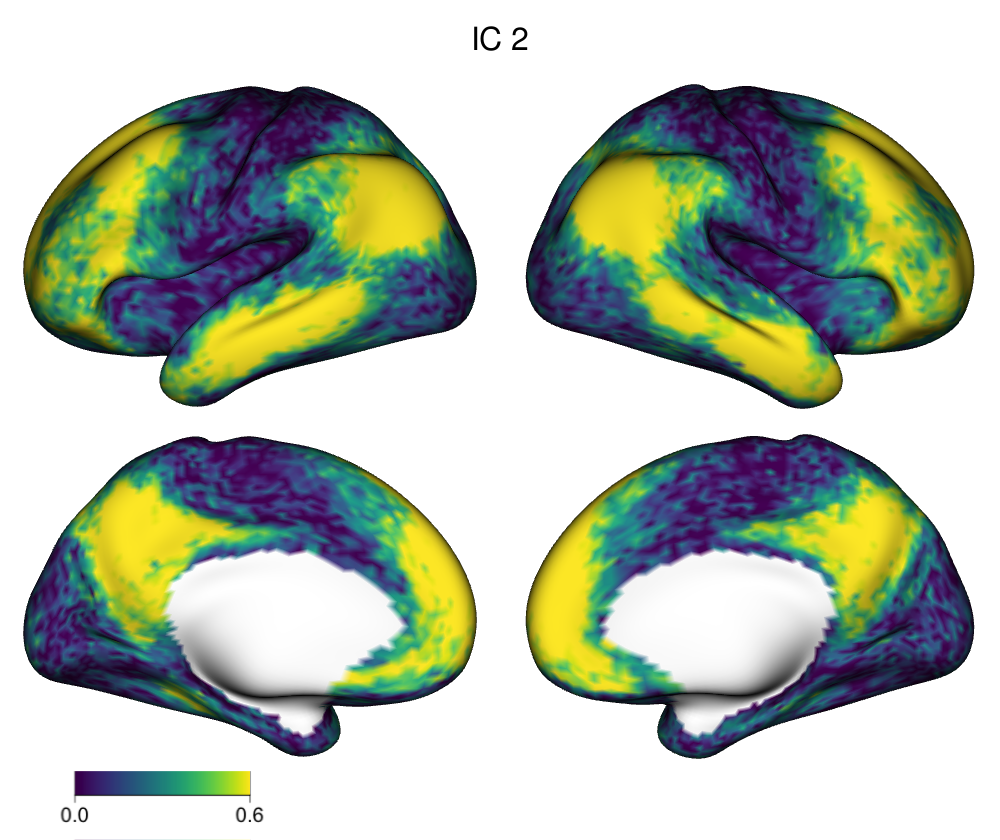} &
    \includegraphics[height=20mm, trim=18cm 15cm 0 3cm, clip]{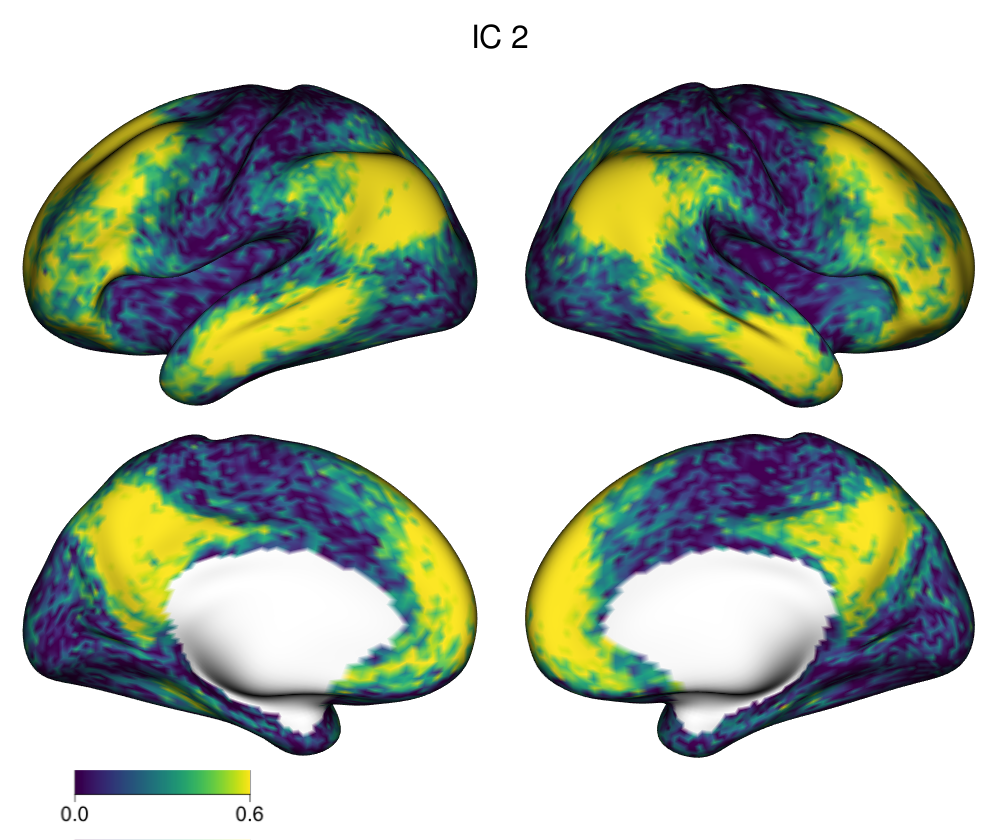} &
    \includegraphics[height=20mm, trim=18cm 15cm 0 3cm, clip]{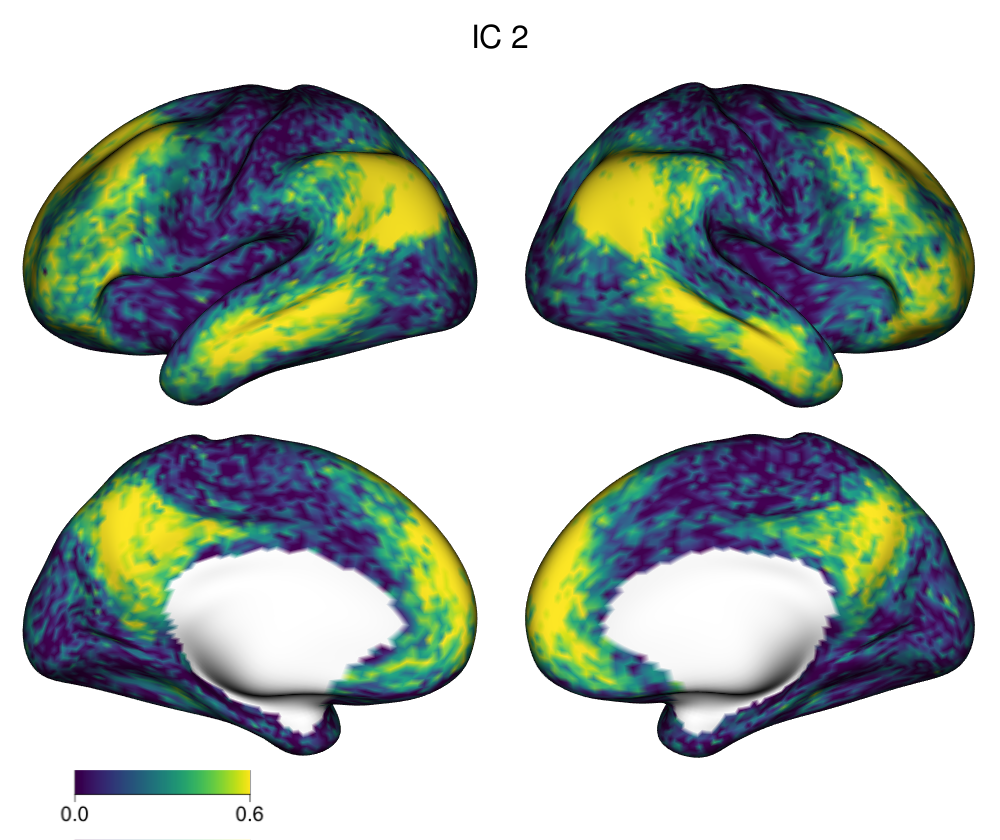} \\[10pt]
    & \multicolumn{4}{c}{ICC\quad 0.0 \includegraphics[width=2cm]{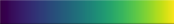} 0.6} 
    \end{tabular}
    \caption{\textit{Reliability of spatial IC maps in terms of intra-class correlation coefficient (ICC).} Three ICs are shown, but other ICs show similar patterns. As expected, ICC is generally higher within areas of engagement for each IC, since ``background'' regions exhibit little true between-subject variability.}
    \label{fig:DA:S_ICC}
\end{figure}
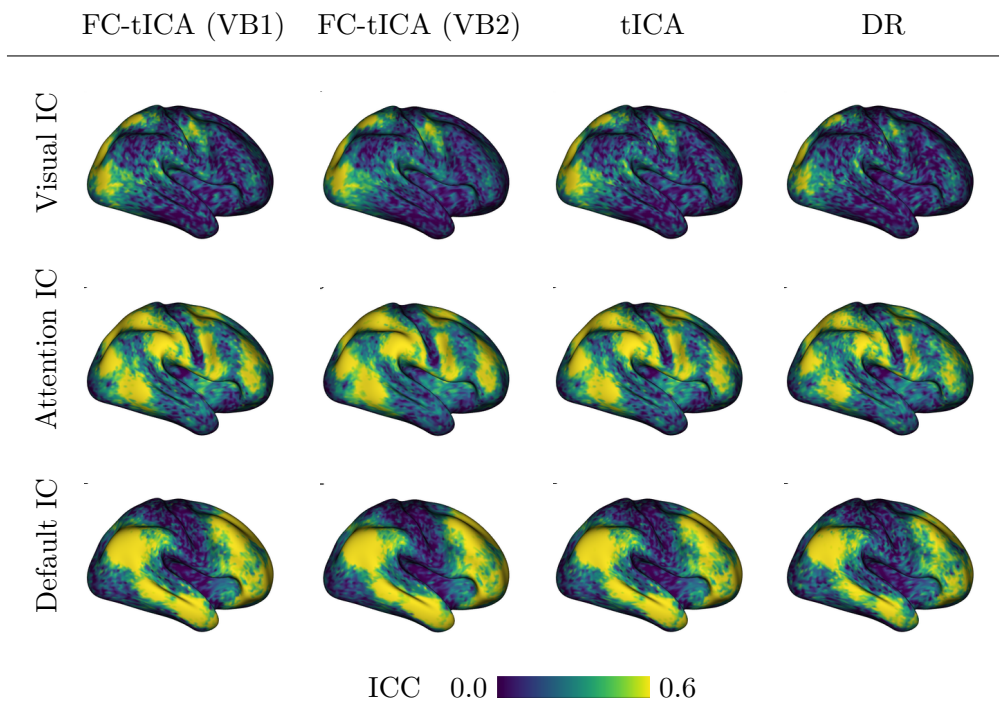

\begin{figure}[!p]
    \centering
    \includegraphics[width=4in]{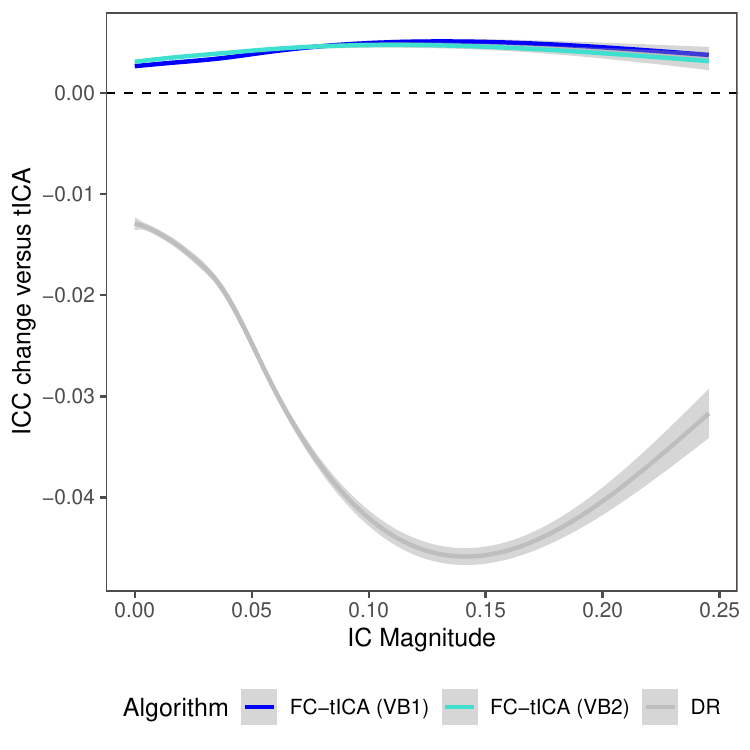}
    \caption{\textit{Change in reliability of spatial IC maps, versus standard template ICA.} The change in ICC for each IC at each vertex, compared with standard template ICA. The smoothers displayed on the plot were generated using ggplot's \texttt{geom\_smooth} function to summarize across all vertices and ICs. The x-axis shows IC magnitude (based on the template mean) and is truncated at the 99th quantile to exclude highly sparse regions from the smoother. The improvement of FC-tICA over standard tICA is seen across all IC magnitudes, suggesting that it is not isolated to background areas or high-engagement areas, but is rather represents a subtle but global reduction in noise levels.}
    \label{fig:DA:S_ICC_vs_tICA}
\end{figure}

\end{document}